\documentclass[12pt]{iopart}
\usepackage[numbers]{natbib}
\usepackage{iopams}
\usepackage{graphicx}
\usepackage{amssymb}
\usepackage{color}
\usepackage{euscript}
\usepackage{subcaption}
\usepackage[colorlinks, linkcolor={blue},
    citecolor={blue}, urlcolor={blue}]{hyperref}

\usepackage{etoolbox}
\makeatletter
\def\@mkboth#1#2{}
\newlength\appendixwidth
\preto\appendix{\addtocontents{toc}{\protect\patchl@section}}
\newcommand{\patchl@section}{%
  \settowidth{\appendixwidth}{\textbf{Appendix }}%
  \addtolength{\appendixwidth}{1.5em}%
  \patchcmd{\l@section}{1.5em}{\appendixwidth}{}{\ddt}%
}
\makeatother

\usepackage{soul}

\begin{document}
\title[Localisation and transport in bidimensional random models ...]{Localisation and transport in bidimensional random models with separable Hamiltonians}

\author{G. Corona-Patricio$^{1}$, U. Kuhl$^{2}$, F. Mortessagne$^{2}$, P. Vignolo$^{2}$, and L. Tessieri$^{1}$}
\address{$^1$ Instituto de F\'{\i}sica y Matem\'{a}ticas, Universidad Michoacana de San Nicol\'{a}s de Hidalgo, 58060, Morelia, Mexico}
\address{$^2$ Institut de Physique de Nice, Universit\'{e} C\^{o}te d'Azur, CNRS, 06100 Nice, France}
\ead{luca.tessieri@gmail.com}

\date{18th June 2019}

\begin{abstract}
We consider two bidimensional random models characterised by the
following features: a) their Hamiltonians are separable in polar
coordinates and b)~the random part of the potential depends either on the
angular coordinate or on the radial one, but not on both.
The disorder correspondingly localises the angular or the radial
part of the eigenfunctions. We analyse the analogies and the differences
which exist between the selected 2D models and their 1D counterparts.
We show how the analogies allow one to use correlated disorder to
design a localisation length with pre-defined energy dependence and to
produce directional localisation of the wavefunctions
in models with angular disorder. We also
discuss the importance of finite-size and resonance
effects in shaping the eigenfunctions of the model with angular disorder;
for the model with disorder associated to the radial variable we show
under what conditions the localisation length coincides with the
expression valid in the 1D case.
\end{abstract}


\submitto{\NJP}
\maketitle
\newpage
\tableofcontents
\newpage

\section{Introduction}

In the sixty years elapsed since the publication of the pioneering Anderson
paper~\cite{and58}, the phenomenon of Anderson localisation has been
intensively studied (see, e.g.,~\cite{tho74,lee85a,eve08,kra93} and references
therein).
Although the theory was initially conceived within the field of
condensed-matter physics, the realisation that destructive interference is
the mechanism underlying the Anderson localisation of electronic states has
led to the application of the Anderson theory to a large number of fields,
including mesoscopic physics~\cite{imr02}, propagation of electromagnetic
and acoustic waves~\cite{tat61,che67,kra92,She06,Cha00}, experiments with cold
atoms~\cite{san07,bil08,gur09,asp09,lug09,san10,mod10,pir11,sha12,Pir13}.
Even light localisation was considered~\cite{wie97,stoe06c,sch07a}, although
it should be noted that localisation of electromagnetic waves in three dimensions
was recently questioned~\cite{ski14} and that experimental reports of light
localisation in three dimensions seem to be based on incorrect interpretations
of the data~\cite{spe16}.

In the theoretical study of Anderson localisation, a special role has
been played by one-dimensional (1D) models, which are more amenable to
analytical treatment than their 2D and 3D counterparts.
The practical applicability of 1D models, however, is often reduced because
many results which can be proved for this class of systems are not valid in
higher dimensions. It has long been known, for instance, that in 1D models
any amount of uncorrelated disorder leads to the localisation of all
electronic states~\cite{mot61} so that, contrary to what happens in three
dimensions, no metal-insulator transition can occur in 1D models.
The situation is more complex in the case of correlated disorder: in
fact, specific long-range correlations of the disorder can produce an
effective metal-insulator transition even in 1D models~\cite{izr12}.
The peculiar features of 1D systems imply that a direct study of 2D and
3D models cannot be avoided in order to reach a full understanding of
Anderson localisation.

Few analytical results have been obtained for 2D and 3D models.
A crucial tool for the understanding of these models is represented by
the single-parameter scaling (SPS) theory, which was introduced
in~\cite{abr79,and80} and still provides an essential framework for the
study of Anderson localisation.
The SPS theory predicts that, for uncorrelated disorder, all states are
localised in 1D and 2D models, whereas a metal-insulator transition occurs
in 3D systems. Two is the lower critical dimension and, for the standard
Anderson model without spin-dependent terms, the scaling hypothesis
leads to the conclusion that all states are localised, although their
localisation length may be exponentially large for weak disorder~\cite{tho82}.
This is the generally accepted view, although there has always been a certain
amount of controversy concerning the possibility of a metallic phase in 2D
models even in the absence of a magnetic field (see~\cite{abr01} and
references therein).

Immediately after the publication of the original SPS paper~\cite{abr79},
diagrammatic techniques were used to show that in 2D models the conductivity
depends logarithmically on the frequency in the low-frequency
regime~\cite{gor79}.
In the early '80s the scaling hypothesis was corroborated by numerical
results obtained by MacKinnon and Kramer~\cite{mac81}. At the same time,
numerical studies of 2D models were also carried out in order to
confirm the predictions of the SPS theory~\cite{lee81}.
In the '90s Schreiber and coworkers used numerical techniques to perform
comprehensive studies of the 2D Anderson model with diagonal and off-diagonal
disorder~\cite{sch92c,tit95,eil98}. They analysed the properties of the eigenstates
and the density of states using an approach based on the transfer matrix method.
Transfer matrix methods were also used in more recent numerical studies of
the localisation length in 2D models with diagonal and off-diagonal
disorder~\cite{ung03}.
More recently, the use of a generalised form of the DPMK equation~\cite{dor82,mel88c}
was proposed to study 2D models~\cite{mar10}.

The works on 2D models mentioned so far did not consider disorder with
spatial correlations. This subject received attention in a series
of papers~\cite{mou04,dos07,mou08,mou10b} focused on models with disorder having a
power-law spectral density $S(k) \sim 1/k^{\alpha}$. For strongly correlated
disorder, these studies generally show that extended states and ballistic
diffusion processes emerge.
In the field of cold atoms, the use of correlated disorder to produce spectral
shaping of the localisation length in 2D and 3D experiments was considered
in~\cite{Plo11,Pir12}.
More recently, the effects of short-range correlations were considered for a
Bose gas confined on a 2D square lattice~\cite{cap15}. Using a 2D
generalisation of the dual dimer random model, the authors showed that
short-range correlations of the disorder can enhance quantum coherence
in a weakly interacting many-body system.

In this paper we focus our interest on two bidimensional models.
To avoid the rather intractable problems which present themselves
whenever one tries to obtain analytical results for 2D systems,
we considered models with two specific features: a) we selected models
with separable Hamiltonians in polar coordinates and b) we picked potentials
with a random part depending either on the angle or the radial variable, but
not on both coordinates at the same time, thus keeping the separability.
This choice allowed us to reduce the study of 2D models to the analysis
of associated 1D systems and thus obtain a deeper understanding of the
effects of disorder.

In particular we found that, when the random potential depends on the
angle variable, it is the angular part of the wavefunctions which is
localised; conversely, when disorder varies with the radial variable,
it is the radial part of the wavefunction which suffers localisation.
Although this might seem as a foregone conclusion, we would like to stress
that the reduction of 2D models to systems of lower dimensionality is far
from trivial. In fact, we found that the study of models with angular
disorder requires a careful analysis of finite-size effects, which
cannot be avoided since the angle variable has a necessarily finite domain;
we also discovered that resonances can play an important role in 2D
Kronig-Penney models with angular disorder.
The analysis of models with radial disorder, on the other hand, revealed
that 2D models with a central random potential can be effectively reduced
to their 1D homologues only if the potential does not decrease too
quickly away from the force centre.

Throughout the paper, we considered the case of correlated disorder. For
the model with disorder associated to the angular variable, we found that
correlated disorder can be extremely useful to enhance localisation and
compensate for the fact that the geometry of the problem sets limits
on the strength of structural disorder.
This is also an issue for experimental realisations which always involve
finite-size systems.
In the case of disorder depending on the radial coordinate, we found
that spatial correlations can be used to modulate the transmission
properties of the model within the allowed energy bands.

We think that the theoretical understanding of the relevant
features of the studied models has a twofold importance: it sheds some light
on the difficult problem of localisation in 2D systems and makes possible to
design actual devices which can be used to filter and focus waves as desired.

The paper is organised as follows.
We introduce the separable models discussed in this work in
Sec.~\ref{separable_models}. We devote Sec.~\ref{angular_disorder} to the
discussion of a 2D model with a random potential that depends only on the angle
variable. We show how a disorder of this kind produces localisation of the
angular
part of the wavefunction and we analyse in detail the effects due to the finite
size and geometrical constraints of the model.
Correlated disorder is applied to enhance the localisation in angular direction.
We show that all radial bands localise in the same direction (see
figure~\ref{wavefun_s}); this feature suggests that such models might be used
to implement directional transport.
In Sec.~\ref{radial_disorder} we shift our attention to a model with a central
random potential. In this case it is the radial component of the eigenstates
which is localised. We show how one can use spatial correlations of the disorder
to allow transmission in predefined energy windows.
We present our conclusions in Sec.~\ref{conclu}.

\section{Separable models in two dimensions}
\label{separable_models}

We consider a quantum particle in a two-dimensional plane. The particle
wavefunction obeys the Schr\"{o}dinger equation
\begin{equation}
\left[ - \nabla^{2} + U(r,\theta) \right] \psi(r,\theta) = E \psi(r,\theta)
\label{2dschr}
\end{equation}
where the symbol $\nabla^{2}$ represents the 2D Laplacian which, in polar
coordinates, can be written as
\begin{equation}
\nabla^{2} = \frac{1}{r} \frac{\partial}{\partial r} \left( r
\frac{\partial }{\partial r} \right) + \frac{1}{r^2}
\frac{\partial^{2} }{\partial \theta^{2}} .
\label{2dlap}
\end{equation}
In Eq.~(\ref{2dschr}) we used energy units such that $\hbar^{2}/2m = 1$;
we shall follow this convention throughout the rest of the paper.

As a full analytical treatment of arbitrary random potentials with spatial
correlations is out of reach, we focused our attention on systems which are separable
in $r$ and $\theta$. This choice simplifies the mathematical problem, because in
this case the wave function can be expressed in the factorised form
\begin{equation}
\psi(r,\theta) = R(r) \Theta(\theta)
\label{var_sep}
\end{equation}
and the Schr\"{o}dinger equation~(\ref{2dschr}) splits in a pair of 1D
differential equations.

From a physical point of view, we study two separable models in which randomness
is introduced either via the angular variable $\theta$ or through the radial variable $r$.
Thus we assume the random potential in Eq.~\ref{2dschr} to have the form:
\begin{equation}
U(r,\theta) =\left\{
\begin{array}{cl}
\displaystyle
\frac{1}{r^{2}} V_{1}(\theta) & \mbox{ for angular randomness,}\\
V_{2}(r) & \mbox{ for radial randomness.}\\
\end{array}
\right.
\label{2d_pot}
\end{equation}
We consider functions $V_{1}(\theta)$ and $V_{2}(r)$ constituted by sums
of random $\delta$-barriers, so that in both cases our analysis will involve
the study of variants of the aperiodic Kronig-Penney model.
We specify the exact form of the
potential~(\ref{2d_pot}) in Sec.~\ref{model_angular_disorder} for the case
of angular disorder and in Sec.~\ref{centr_pot} for the case of radial disorder.

\section{Angle-dependent disorder}
\label{angular_disorder}

\subsection{Model with angle-dependent disorder}
\label{model_angular_disorder}

We consider a quantum particle confined in an annulus on a two-dimensional
plane. We use the symbols $r_{1}$ and $r_{2}$ to denote the inner and
outer radii of the annulus, where $r_{1}>0$ and $r_{2}$ is finite but can be
arbitrarily large.
Within this domain the Schr\"{o}dinger equation~(\ref{2dschr}) holds.
We focus our attention on a potential $U(r,\theta)$ of the form
\begin{equation}
U(r,\theta) = \frac{1}{r^{2}} \sum_{n=1}^{N} U_{n}
\delta (\theta - \theta_{n}) .
\label{deltabar}
\end{equation}

Eqs.~(\ref{2dschr}) and~(\ref{deltabar}) describe a particle that
moves under the influence of a potential constituted by $N$ radial
delta-barriers. Disorder is introduced in the model via the random
strengths and positions of the barriers, which are respectively given by
\begin{equation}
U_{n} = U + u_{n}
\label{compdis}
\end{equation}
and
\begin{equation}
\theta_{n} = \alpha (n - 1) + \alpha_{n} .
\label{strucdis}
\end{equation}
In~(\ref{compdis}) the random variables $\{u_{n}\}$ represent the
fluctuations of the barrier strength around the mean value $U$.
Similarly, in~(\ref{strucdis}) $\alpha_{n}$ stands for the random
angular displacement of the $n$-th barrier with respect to the lattice
position $\alpha (n-1)$, with
\begin{displaymath}
\alpha = \frac{2 \pi}{N} .
\end{displaymath}
Note that we use the position of the first barrier as origin of the angular
coordinate, so that $\theta_{1} = 0$ and $\alpha_{1} = 0$.
We assume that the random variables $u_{n}$ and $\alpha_{n}$ have zero average
and known probability distributions.

Using the factorised form~(\ref{var_sep}), the Schr\"{o}dinger equation~(\ref{2dschr})
splits in the pair of single-variable equations
\begin{equation}
- \frac{d^{2}\Theta}{d\theta^{2}} + \sum_{n=1}^{N} U_{n} \delta
\left( \theta - \theta_{n} \right) \Theta = q^{2} \Theta
\label{angsch}
\end{equation}
and
\begin{equation}
\frac{d^{2}R}{dr^{2}} + \frac{1}{r} \frac{dR}{dr} +
\left[ E - \frac{q^{2}}{r^{2}} \right] R = 0 .
\label{radsch}
\end{equation}
Eqs.~(\ref{angsch}) and~(\ref{radsch}) define the model under study.
Note that, when units such that $\hbar^{2}/2m =1$ are chosen, the
variables $q$ and $U_{n}$ in~(\ref{angsch}) are dimensionless, while
the energy $E$ in~(\ref{radsch}) has dimensions $[L]^{-2}$.

We would like to emphasise that the model defined by Eqs.~(\ref{2dschr})
and~(\ref{deltabar}) also lends itself to the study of microwaves in a cavity
shaped as a ring cake tin. In this context, the random potential
can be mimicked with appropriate variations of the distance between the top
and bottom plates of the cavity~\cite{bar13c}.
This opens the way for the verification of the theoretical results in microwave
experiments. We express with more details the underlying idea in~\ref{app:mwc}.

We remark that our model could also find important applications in the field of
ultracold atoms, where Spatial Light Modulator (SLM) devices can be used to
generate optical random potentials of the kind analysed in this paper~\cite{barb13}.
An experimental realisation of our model with ultracold atoms in optical potentials
would provide an additional way to test our theoretical results.

\subsection{The aperiodic Kronig-Penney model and its tight-binding counterpart}
\label{angul}

In this section we analyse the Schr\"{o}dinger equation~(\ref{angsch})
for the angular part of the total wavefunction.
A crucial aspect of Eq.~(\ref{angsch}) is that it describes
a {\em finite} 1D Kronig-Penney model with compositional and structural
disorder.
The fact that $\theta$ is an angle variable implies that the domain
of the wavefunctions $\Theta(\theta)$ is restricted to the $[0,2\pi]$
interval and that periodic boundary conditions apply
\begin{displaymath}
\Theta(0) = \Theta(2\pi) .
\end{displaymath}
To analyse the solutions of the 1D model~(\ref{angsch}), it is
convenient to integrate~(\ref{angsch}) over the angular
interval $[\theta_{n}^{-}, \theta_{n+1}^{-}]$ between two barriers.
In this way one obtains the map
\begin{equation}
\left( \begin{array}{c}
\Theta_{n+1} \\
\Theta^{\prime}_{n+1} \\
\end{array}
\right)
= \mathbf{T}_{n}
\left( \begin{array}{c}
\Theta_{n} \\
\Theta^{\prime}_{n} \\
\end{array}
\right) ,
\label{map}
\end{equation}
where the transfer matrix $\mathbf{T}_{n}$ has the form
\begin{equation}\fl
{\bf T}_{n} = \left(
\begin{array}{cc}
\displaystyle
\cos \left[q \left( \alpha + \Delta_{n} \right)\right] +
\left( U + u_{n} \right) \frac{1}{q}
\sin \left[ q \left( \alpha + \Delta_{n} \right)\right] &
\displaystyle
\frac{1}{q} \sin \left[q \left( \alpha + \Delta_{n} \right) \right]\\
\displaystyle
-q \sin \left[q \left( a + \Delta_{n} \right) \right]
+\left( U + u_{n} \right)
\cos \left[ q \left( \alpha + \Delta_{n} \right) \right]&
\displaystyle
\cos \left[ q \left( \alpha + \Delta_{n} \right) \right]\\
\end{array}
\right)
\label{t_mat}
\end{equation}
and we have introduced the symbols
\begin{equation}
\begin{array}{ccc}
\Theta_{n} = \Theta(\theta_{n}^{-}) & \mbox{ and } &
\Theta_{n}^{\prime} = \Theta^{\prime}(\theta_{n}^{-})
\end{array}
\label{theta_coef}
\end{equation}
for the values of the {\em unnormalised} wavefunction $\Theta$ and of its
angular derivative $\Theta^{\prime}$ in the left neighbourhood of the $n$-th barrier.
The symbols $\Delta_{n}$ in Eq.~(\ref{t_mat}) stand for the relative
displacements of two contiguous barriers, i.e.,
\begin{equation}
\Delta_{n} = \alpha_{n+1} - \alpha_{n} .
\label{delta_n}
\end{equation}
To ensure that the periodicity conditions are satisfied, one must
have
\begin{equation}
\begin{array}{ccc}
\Theta_{N+1} & = & \Theta_{1} ,\\
\Theta_{N+1}^{\prime} & = & \Theta_{1}^{\prime} .\\
\end{array}
\label{periodic}
\end{equation}

After eliminating the derivatives $\Theta^{\prime}_{n}$ from the
map~(\ref{map}), one obtains the following recursive relation for the
$\Theta_{n}$ variables
\begin{equation}
\gamma_{n+1} \Theta_{n+1} + \gamma_{n} \Theta_{n-1} = \varepsilon_{n} \Theta_{n}
\label{tb_model}
\end{equation}
with
\begin{displaymath}
\varepsilon_{n}(q) = \frac{U_{n}}{q} +
\cot \left[ q \left( \theta_{n+1} - \theta_{n} \right) \right] +
\cot \left[ q \left( \theta_{n} - \theta_{n-1} \right) \right]
\end{displaymath}
and
\begin{displaymath}
\gamma_{n}(q) = \frac{1}{\sin \left[ q \left(\theta_{n} - \theta_{n-1} \right)
\right]} .
\end{displaymath}
Relation (\ref{tb_model}) defines the tight-binding model corresponding to the
Kronig-Penney model~(\ref{angsch}). We remark that compositional disorder
in the latter produces diagonal random terms in the former, while structural
disorder in the model~(\ref{angsch}) emerges as both diagonal and
off-diagonal disorder in the tight-binding model~(\ref{tb_model}).

The eigenfunctions $\Theta(\theta)$ of the continuous Kronig-Penney
model~(\ref{angsch}) can now be analysed in terms of the discrete
eigenstates $\{ \Theta_{n} \}$ of the tight-binding model~(\ref{tb_model})
that satisfy the periodicity conditions~(\ref{periodic}).
Note that the eigenstates $\{ \Theta_{n} \}$ completely determine
the corresponding solutions of the Schr\"{o}dinger equation~(\ref{angsch})
and viceversa.
In fact, one can easily integrate Eq.~(\ref{angsch}) within the
potential wells, using the pairs $(\Theta_{n},\Theta_{n+1})$ as boundary
conditions. Within the $n$-th well, i.e., for
$\theta \in [\theta_{n}^{+}, \theta_{n+1}^{-}]$, the angular wavefunction
has the form
\begin{equation}\fl
\Theta(\theta) = \EuScript{N} \left\{
\Theta_{n} \cos \left[ q \left( \theta - \theta_{n} \right) \right] +
\frac{\Theta_{n+1} - \Theta_{n} \cos \left[ q \left( \theta_{n+1} - \theta_{n}
\right) \right]}
{\sin \left[ q \left( \theta_{n+1} - \theta_{n} \right) \right]}
\sin \left[ q \left( \theta - \theta_{n} \right) \right] \right\}
\label{eigenstates}
\end{equation}
where $\EuScript{N}$ is a constant which can be obtained from the
normalisation condition
\begin{equation}
\int_{0}^{2 \pi} \left|\Theta(\theta) \right|^{2} \mathrm{d} \theta = 1 .
\label{norm_cond}
\end{equation}
Conversely, if the wavefunction $\Theta(\theta)$ is known, one can obtain
a solution of the tight-binding model~(\ref{tb_model}) by using~(\ref{theta_coef}) to obtain a vector $\{\Theta_{n}\}$ which can be
normalised with the condition
\begin{equation}
\sum_{n=1}^{N} \left|\Theta_{n} \right|^{2} = 1.
\label{norm_cond_2}
\end{equation}
The correspondence between {\em normalised} states of the continuous
model~(\ref{angsch}) and of its discrete counterpart~(\ref{tb_model}) breaks
down only when resonant phenomena come into play; we shall discuss this point
in Sec.~\ref{reson}.

\subsection{Vanishing disorder}

\begin{figure}[htb]
\begin{center}
\includegraphics[width=5in]{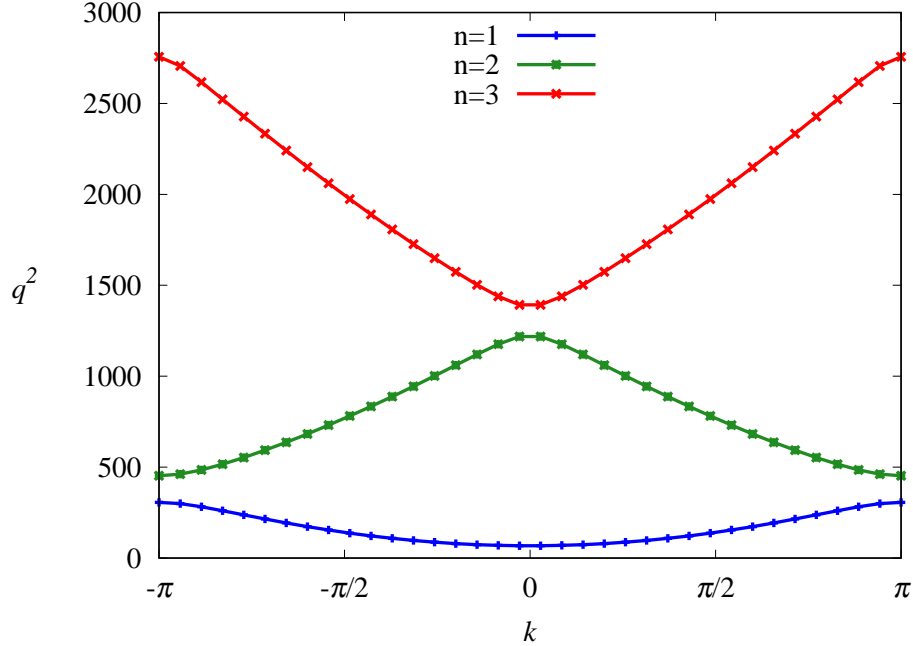}
\caption{Squared momentum $q^{2}$ versus the Bloch vector $k$ for the
first three bands. The data were obtained for a model with no disorder for $N=35$ wells
separated by barriers of strength $U = 15.0$. \label{band_structure}}
\end{center}
\end{figure}

\begin{figure}[htb]
\begin{center}
\includegraphics[width=5in]{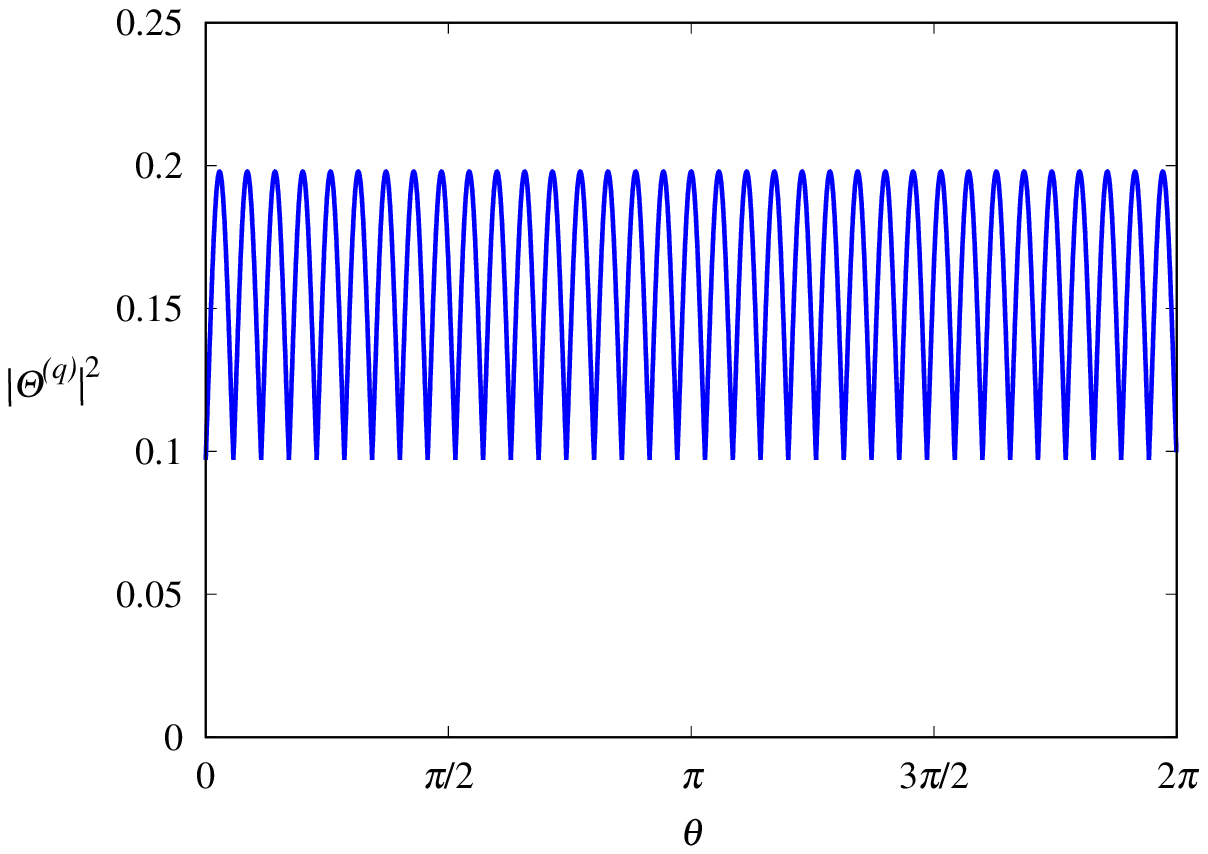}
\caption{Squared modulus of the angular eigenfunction $\Theta^{(q)}(\theta)$
versus the  angle variable $\theta$ for Bloch wavevector $k = 0.428 \pi$,
which corresponds to $q \simeq 11.04$ (i.e. $q^{2} \simeq 121.9$).
The data were obtained for a model with $N=35$ wells
separated by barriers of strength $U = 15.0$. \label{bloch_wave}}
\end{center}
\end{figure}

In the limit case of vanishing disorder, (\ref{tb_model}) reduces to
\begin{displaymath}
\Theta_{n+1} + \Theta_{n-1} = 2 \left[ \cos (q \alpha) + \frac{U}{2q}
\sin (q \alpha) \right] \Theta_{n} .
\end{displaymath}
The solutions of this equation are Bloch waves, identified by the
coefficients
\begin{equation}
\Theta_{n} = \exp \left( i k n \right) ;
\label{bloch_waves}
\end{equation}
the Bloch vector $k$ and the (angular) momentum $q$ are linked by the relation
\begin{equation}
\cos \left( k \right) = \cos \left( q \alpha \right) +
\frac{U}{2q} \sin \left( q \alpha \right) .
\label{rotangle}
\end{equation}
Note that, although $q$ is an angular quantum number for the
system~(\ref{2dschr}), the squared momentum $q^{2}$ can be interpreted
as the energy of the Kronig-Penney model~(\ref{angsch}): for this reason
we shall often refer to $q^{2}$ as the ``energy'' of the latter model.
The periodicity condition~(\ref{periodic}) is satisfied for
the $N$ non-equivalent values of the Bloch vector
\begin{displaymath}
\begin{array}{ccc}
\displaystyle
k = \frac{2 \pi}{N} l & \mbox{ with } & l=0,1, \ldots ,N-1 .
\end{array}
\end{displaymath}
Eqs.~(\ref{rotangle}) and~(\ref{bloch_waves}) completely solve the problem
of determining eigenvalues and eigenstates of the periodic Kronig-Penney
model. The dispersion relation~(\ref{rotangle}) gives the band structure
and can be solved numerically; the coefficients~(\ref{bloch_waves})
define the Bloch waves that are the eigenstates of the unperturbed
system.
In figure~\ref{band_structure} we show the structure of the first three bands
for a Kronig-Penney model with no disorder for $N=35$ barriers of strength $U=15.0$.
As noted in Sec.~\ref{angular_disorder}, in units such that $\hbar^{2}/2m = 1$
the squared momentum $q^{2}$ associated to the angle degree of freedom is
dimensionless.
In figure~\ref{bloch_wave} we show a typical Bloch eigenfunction obtained
for the same values of the parameters $N$ and $U$. The wavefunction
$\Theta^{(q)}(\theta)$ has Bloch wavevector $k = 0.428 \pi$ which corresponds
to a momentum $q \simeq 11.04$ and energy $q^{2} \simeq 121.9$ in the first
band.
The wavefunction was normalised with the condition~(\ref{norm_cond}).

\subsection{Infinite aperiodic Kronig-Penney model}
\label{infinite_kp}

To gain insight on the structure of the eigenstates of the
model~(\ref{angsch}) when disorder is present, one can consider the
corresponding infinite model, defined by the Schr\"{o}dinger equation
\begin{equation}
- \frac{d^{2}\Theta}{dx^{2}} + \sum_{n=-\infty}^{\infty} U_{n} \delta
\left( x - x_{n} \right) \Theta = q^{2} \Theta
\label{kpmodel}
\end{equation}
with $x \in \mathbf{R}$. The delta barriers in the model~(\ref{kpmodel})
are centred at the random positions $\{x_{n}\}$, with
$\langle x_{n} \rangle = n \alpha$.
As is well known, the eigenstates of the infinite Kronig-Penney
model~(\ref{kpmodel}) are localised. Their spatial extension is determined
by the inverse localisation length, which can be defined in terms of
the $\Theta_{n} = \Theta(x_{n})$ values as
\begin{equation}
\lambda = \lim_{N \to \infty} \frac{1}{N} \sum_{n=1}^{N}
\ln \left|\frac{\Theta_{n+1}}{\Theta_{n}} \right| .
\label{kp_lyap}
\end{equation}
Note that the inverse localisation length (or Lyapunov exponent)~(\ref{kp_lyap})
is a deterministic quantity due to its self-averaging property (see,
e.g.,~\cite{Lif88}).

An analytical expression for the inverse localisation length~(\ref{kp_lyap})
was derived for the case of weak disorder in~\cite{her08,her10b,her10c}.
As shown in~\cite{her08}, disorder can be considered weak provided that
\begin{equation}
\begin{array}{cccc}
\langle u_{n}^{2} \rangle \ll U^{2}, &
\langle \Delta_{n}^{2} \rangle q^{2} \ll 1, & \mbox{ and } &
\langle \Delta_{n}^{2} \rangle U \ll 1 .
\end{array}
\label{weakdis}
\end{equation}
If conditions~(\ref{weakdis}) are met, one obtains
\begin{equation}
\begin{array}{ccl}
\lambda & = & \displaystyle
\frac{1}{8 \alpha \sin^{2} k} \left[ \frac{\sin^{2} (q \alpha)}{q^{2}}
\langle u_{n}^{2} \rangle W_{1}(k) +
U^{2} \langle \Delta_{n}^{2} \rangle W_{2}(k) \right. \\
& - & \displaystyle \left.
2 \left| \frac{\sin (q \alpha)}{q} \right|
U \sqrt{\langle u_{n}^{2} \rangle
\langle \Delta_{n}^{2} \rangle W_{1}(k) W_{2}(k)} \;
\cos k \sin \left( 2 \eta \right) \right] .
\end{array}
\label{invloc}
\end{equation}
The functions $W_{1}(k)$ and $W_{2}(k)$ in~(\ref{invloc}) are the Fourier
transforms of the normalised binary correlators of the random variables
$u_{n}$ and $\Delta_{n}$, i.e.,
\begin{equation}
\begin{array}{ccl}
W_{1} \left( k \right) & = & \displaystyle
1 + 2 \sum_{l=1}^{\infty} \frac{\langle u_{n}u_{n+l} \rangle}
{ \langle u_{n}^{2} \rangle} \cos(2 k l) \\
W_{2} \left( k \right) & = & \displaystyle
1 + 2 \sum_{l=1}^{\infty} \frac{\langle \Delta_{n}\Delta_{n+l} \rangle}
{\langle \Delta_{n}^{2} \rangle} \cos (2 k l) .\\
\end{array}
\label{powerspectra}
\end{equation}
The parameter $\eta$ in~(\ref{invloc}) determines the degree of
cross-correlation of the $u_{n}$ and $\Delta_{n}$ variables. The values
of $\eta$ range from $\eta = \pi/4$ (which corresponds to the extreme case
of total positive cross-correlation) to $\eta = -\pi/4$ (total negative
cross-correlation); for $\eta = 0$ the cross-correlations vanish.
For details on the effect of cross-correlations see Refs.~\cite{her08,die12a}.

Expression (\ref{invloc}) is a perturbative result, valid within the second-order
approximation in the disorder strength. It shows that, within this
approximation, the localisation length diverges in any energy interval
over which the power spectra~(\ref{powerspectra}) vanish.
This implies that an effective localisation-delocalisation transition
can occur in the 1D Kronig-Penney model~(\ref{kpmodel}) provided that
the disorder exhibits the specific long-range correlations which make
the power spectra~(\ref{powerspectra}) vanish over a continuous energy
range.

It is possible to construct sequences of self- and cross-correlated
random variables $u_{n}$ and $\Delta_{n}$ such that the corresponding
power spectra~(\ref{powerspectra}) vanish over {\em pre-defined}
intervals. A way to produce such sequences was presented in~\cite{her10c};
for the sake of completeness, we outline the main steps in~\ref{random_sequences}.

As an application of the theory, we considered the Kronig-Penney
model~(\ref{kpmodel}) with barriers of average strength $U = 15.0$,
disorder intensity $\sqrt{\langle \Delta_{n}^{2} \rangle}~=~0.007$ and
$\sqrt{\langle u_{n}^{2} \rangle} = 4.0$ and we compared the case
of totally uncorrelated disorder with two cases of correlated disorder.
We used disorder self-correlations to generate effective mobility edges
by choosing compositional and structural disorders with identical power
spectra of the form
\begin{equation}
W_{1}(k) = W_{2}(k) = \left\{ \begin{array}{ccl}
\displaystyle
\frac{\pi}{2 \left(k_{2} - k_{1}\right)} &
\mbox{ if } & k \in \left[ k_{1}, k_{2} \right] \\
0 & \mbox{ if } &
k \in \left[ 0, k_{1} \right] \cup
\left[ k_{2}, \frac{\pi}{2}\right] \\
\end{array} \right.
\label{cond1}
\end{equation}
and we considered two cases: in the first case we set the mobility edges at
$k_{1} = 0.46 \pi$ and $k_{2} = 0.5 \pi$, while in the second case we selected
the values $k_{1} = 0.35 \pi$ and $k_{2} = 0.40 \pi$. The first choice
generates a single, wider, localisation window, while the second produces two
narrower localisation windows.
We analysed the cases of positive, absent, and negative cross-correlations;
in the figures the respective cases are marked with the numbers~(1), (2), and (3)
and associated to the colours blue, green, and red in this section as well as in
Secs.~\ref{finapkp} and~\ref{reson}.
Before proceeding, we would like to remark that we used power spectra of the
form~(\ref{cond1}) in all the cases of correlated disorder which are numerically
studied in this paper.

\begin{figure}[htb]
\begin{center}
\includegraphics[width=5in]{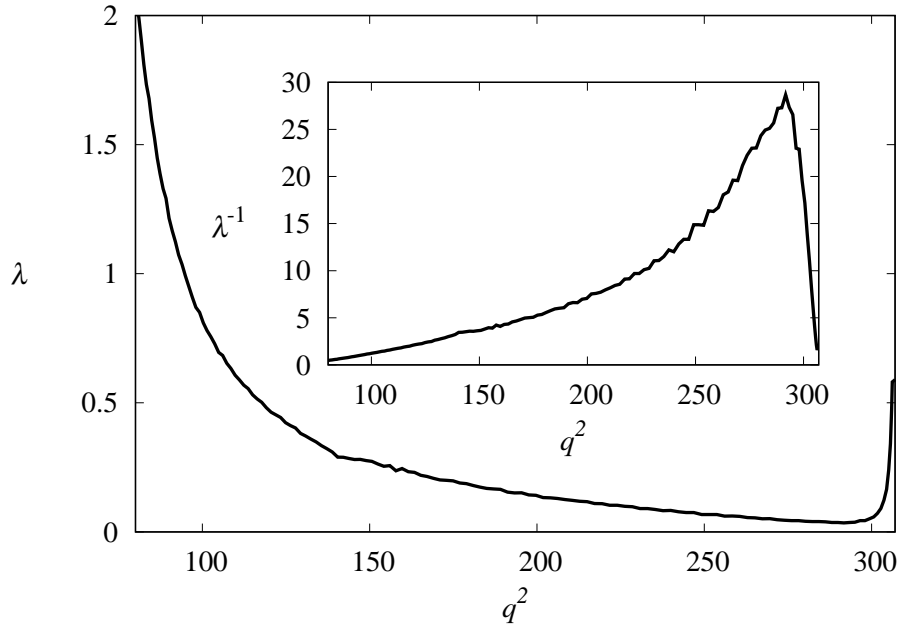}
\caption{Inverse localisation length $\lambda$ versus energy $q^{2}$
for the first energy band. The data were obtained for uncorrelated
disorder. The inset shows the corresponding localisation length.
\label{kplyap_uncor_band1}}
\end{center}
\end{figure}

\begin{figure}[htb]
\begin{center}
\includegraphics[width=5in]{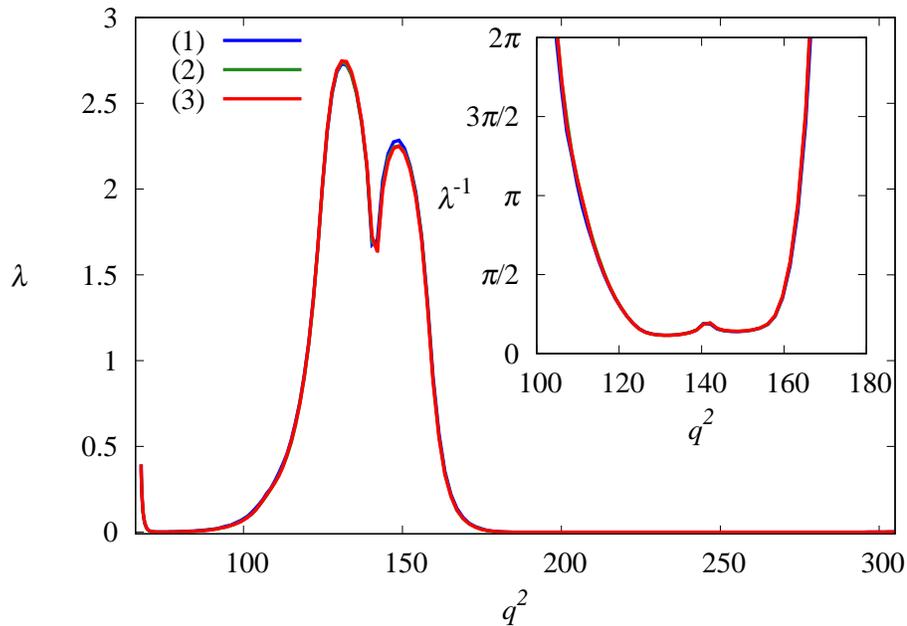}
\caption{Inverse localisation length $\lambda$ versus energy $q^{2}$
for the first energy band. Self-correlations were used to produce a single
localisation window.
The blue solid line~(1) corresponds to positive cross-correlations;
the green line~(2) and the red line~(3) to absent and negative cross-correlations, respectively.
The inset shows the corresponding localisation length.
\label{kplyap_band1_1w}}
\end{center}
\end{figure}

\begin{figure}[htb]
\begin{center}
\includegraphics[width=5in]{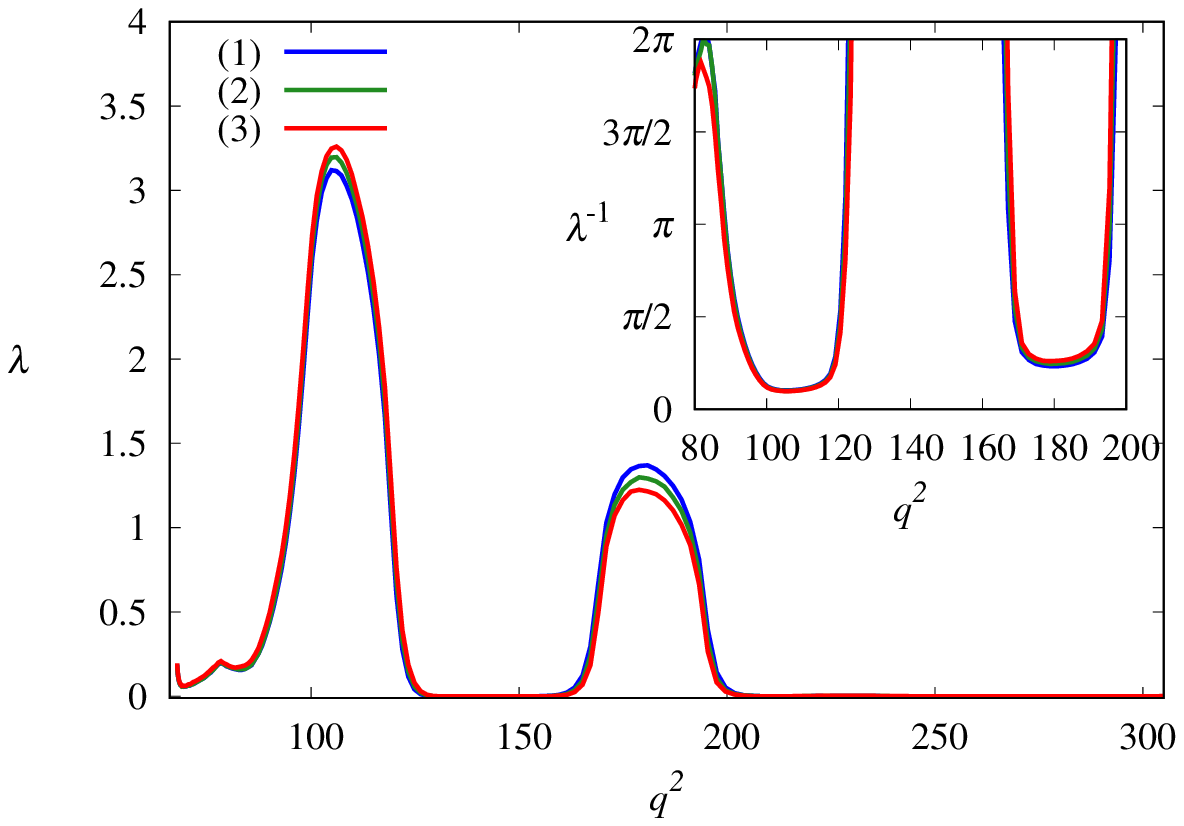}
\caption{As in figure~\ref{kplyap_band1_1w} but with different self-correlations so that two
localisation windows were generated instead of one.
\label{kplyap_band1_2w}}
\end{center}
\end{figure}

The Lyapunov exponent $\lambda$ for the infinite model~(\ref{kpmodel})
was obtained by numerically computing the right-hand side of~(\ref{kp_lyap}).
The results for the inverse localisation length are shown in
figure~\ref{kplyap_uncor_band1} for the case of uncorrelated disorder
and in figures~\ref{kplyap_band1_1w} and~\ref{kplyap_band1_2w} for the case of
correlated disorder.
In the previous figures the energy covers the first allowed band and the
inset shows the localisation length $l_{\mathrm{loc}} = \lambda^{-1}$.
For the sake of clarity, in the case of correlated disorder we considered
the localisation length only within the windows of enhanced localisation,
excluding the rest of the energy band, where $l_{\mathrm{loc}}$
is many orders of magnitude larger than $2\pi$.

By comparing figure~\ref{kplyap_uncor_band1} with figures~\ref{kplyap_band1_1w}
and~\ref{kplyap_band1_2w}, one can see that correlations of the disorder
have a twofold effect: they strongly enhance the localisation of the
eigenstates within the selected energy windows and delocalise all the other
eigenstates~\cite{kuh08a}.
This enhancement of localisation turns out to be particularly relevant for
the finite model we are interested in. In fact, the finite size of the
domain of the angle variable implies that angular localisation needs to
be sufficiently strong to be meaningful; at the same time, as discussed in
Sec.~\ref{finapkp}, in our model structural disorder is necessarily weak
because of built-in bounds, while compositional disorder is often also
reduced by physical constraints of the experimental setup. The ingenious
use of correlations to strengthen localisation, first suggested
in~\cite{kuh08a}, is therefore a crucial tool to produce angular localisation
of the wavefunctions.

\begin{figure}
\begin{subfigure}[b]{0.5\textwidth}
\includegraphics[width=\textwidth]{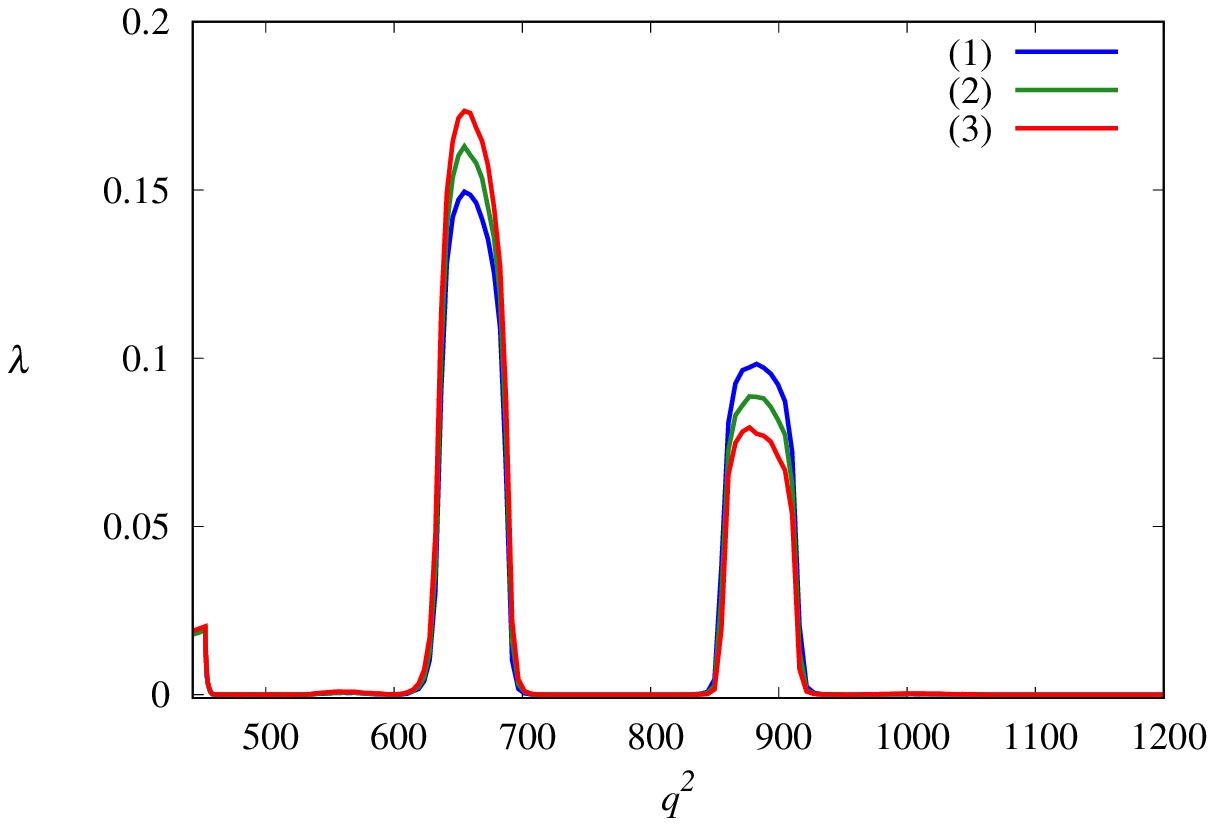}
\caption{Second energy band}
\end{subfigure}%
\begin{subfigure}[b]{0.5\textwidth}
\includegraphics[width=\textwidth]{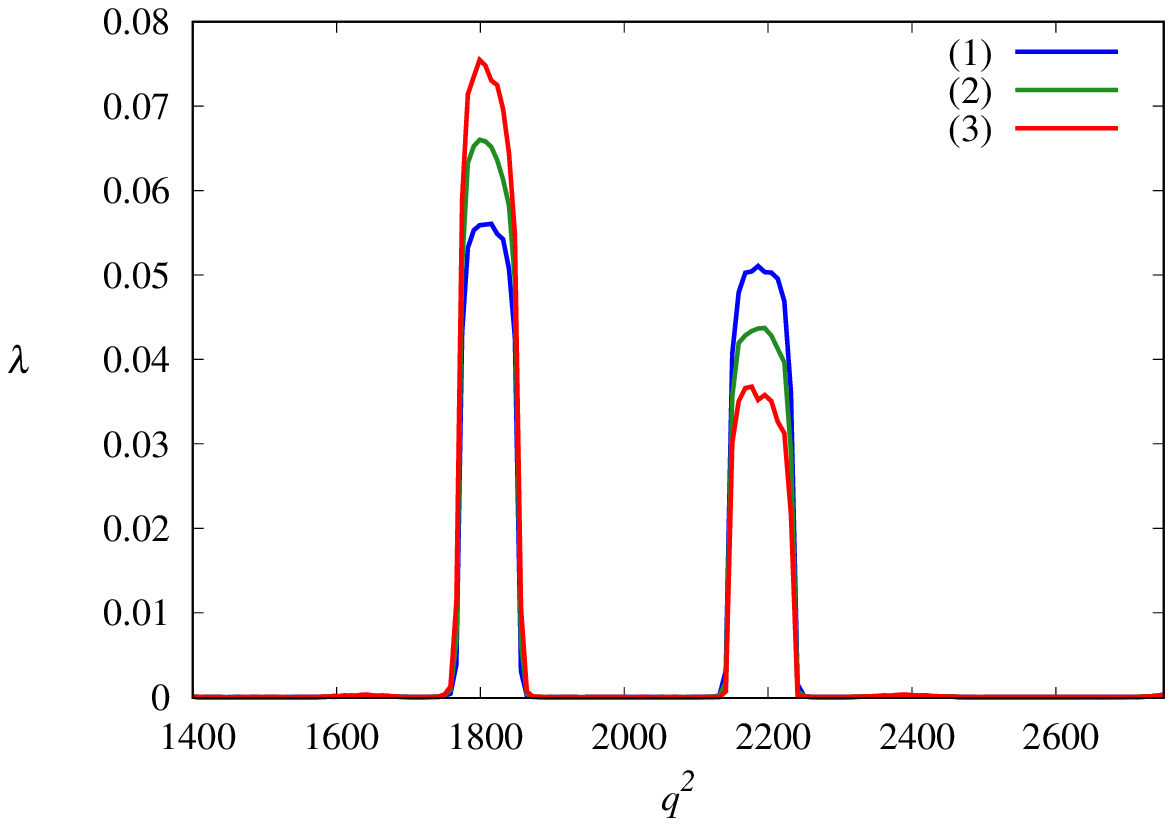}
\caption{Third energy band}
\end{subfigure}
\caption{Inverse localisation length $\lambda$ versus energy $q^{2}$ for
the second and third energy bands. The same self-correlations
used in figure~\ref{kplyap_band1_2w} were also applied here} to
produce two localisation windows.
The blue solid line~(1) corresponds to positive cross-correlation;
the green line~(2) and red line~(3) to absent and negative cross-correlations,
respectively.
\label{lyap_band_2_and_3}
\end{figure}

We also observe that in figure~\ref{kplyap_band1_1w} the Lyapunov exponent
drops in the middle of the localisation window. We interpret this decrease
as a manifestation of the anomaly which appears when the Bloch wavevector
takes the value $k = \pi/2$, as shown in~\cite{her10b}.
It is known that correlations can enhance the anomaly in the Anderson
model~\cite{tes15}; the numerical data suggest that similar effects
occur in the Kronig-Penney model.
As a final remark, we observe that cross-correlations do not change
much the value of the Lyapunov exponent when self-correlations
create a single localisation window. This is due to the position
of the selected localisation window, which is centred around the
Bloch vector $k = \pi/2$. As can be seen from~(\ref{invloc}),
cross-correlations appear in the Lyapunov exponent through a term
which is proportional to $\cos k$, and therefore their effect is
reduced if the localisation window occurs for values of the Bloch
vector which are close to $\pi/2$. Cross-correlations produce larger
differences when localisation windows occupy different positions;
as shown by figure~\ref{lyap_band_2_and_3},
their effect is more evident in higher energy bands, as greater values
of $q^{2}$ increase the relative weight of structural disorder with
respect to the compositional one.

\subsection{Finite aperiodic Kronig-Penney model}
\label{finapkp}

Some caution is required when one tries to apply the results obtained
for the {\em infinite} Kronig-Penney model~(\ref{kpmodel}) to the
{\em finite} model~(\ref{angsch}).
The fact that the angle variable is bounded within the $[0,2 \pi]$
interval  entails several differences between the two models. In the first
place, one can speak of localised states for the model~(\ref{angsch}) only
if their localisation length $l_{\mathrm loc}$ is considerably less than
the span of the angle variable, i.e., if
\begin{equation}
l_{\mathrm loc} \ll 2 \pi .
\label{strongloc}
\end{equation}

A second difference is that the finite size of the angle domain limits
the number $N$ of barriers in model~(\ref{angsch}). As already noted,
real radial barriers have a certain width and therefore $N$ cannot be
arbitrarily large.
In principle one could build very thin radial barriers;
this would make possible to increase their number. This strategy,
however, has a drawback: a higher number of barriers implies a
smaller average angular spacing $\alpha = 2 \pi/N$ and, therefore, a weaker
structural disorder. In fact, structural disorder cannot be so strong
that a barrier might step over its nearest neighbours; increasing the
number of barriers inevitably leads to barriers jumping each other even
if disorder is weak.
Barrier swaps can be prevented by confining the displacements $\Delta_{n}$
within the average disk slice allotted to each barrier; this sets the
following upper bound on the variables $\Delta_{n}$:
\begin{equation}
\left| \Delta_{n} \right| \leq \frac{\alpha}{2} = \frac{\pi}{N} .
\label{delta_bounds}
\end{equation}
The constraint~(\ref{delta_bounds}) shows that increasing the number of
barriers necessarily reduces the strength of disorder and, therefore, the
localisation of the angular wavefunction.

It is important to observe that the random displacements $\Delta_{n}$
may violate condition~(\ref{delta_bounds}) if they are generated
according to the recipe presented in~\ref{random_sequences}.
In fact, the angular displacements $\Delta_{n}$ are obtained via
sums of independent and identically distributed random variables with
finite variance, as shown by~(\ref{convol}); the central limit theorem
therefore ensures that the $\Delta_{n}$ variables have a Gaussian
distribution
\begin{equation}
p(\Delta_{n}) = \frac{1}{\sqrt{2 \pi \sigma_{\Delta}^{2}}}
\exp \left( - \frac{\Delta_{n}^{2}}{2 \sigma_{\Delta}^{2}} \right)
\label{gd}
\end{equation}
with $\sigma_{\Delta} = \sqrt{\langle \Delta_{n}^{2} \rangle}$ being
the strength of the structural disorder.
One can conclude that the probability that the $n$-th angular
displacement violates the condition~(\ref{delta_bounds}) is
\begin{equation}
\mathrm{Pr}\left( \left| \Delta_{n} \right| > \frac{\alpha}{2} \right)
= 1 - \mathrm{erf} \left( \frac{\alpha}{2 \sqrt{2} \sigma_{\Delta}} \right)
\label{prob}
\end{equation}
where $\mathrm{erf}(x)$ is the error function, defined as
\begin{displaymath}
\mathrm{erf}(x) = \frac{2}{\sqrt{\pi}} \int_{0}^{x} e^{-t^{2}} \mathrm{d} t .
\end{displaymath}

For very weak disorder $\sigma_{\Delta} \ll \alpha$ and the
probability~(\ref{prob}) is exponentially small; for stronger structural
disorder, however, the probability~(\ref{prob}) ceases to be negligeable.
In order to ensure that no violation of condition~(\ref{delta_bounds}) occurs,
one can discard the sequences of displacements in which one or more variables
$\Delta_{n}$ fail to fulfil condition~(\ref{delta_bounds}). Discarding whole
sequences rather than individual displacements preserves intact the
disorder correlations.
In mathematical terms, one should ``clip'' the tails of the Gaussian
distribution of each individual $\Delta_{n}$ variable in order to
enforce condition~(\ref{delta_bounds}).
In this way, one considers variables $\Delta_{n}$ with distribution
\begin{equation}
\begin{array}{ccc}
p(\Delta_{n}) & = & \left\{ \begin{array}{ccl}
\displaystyle
\mathcal{N} \exp \left( - \frac{\Delta_{n}^{2}}{2 \sigma_{0}^{2}} \right) &
\mbox{ for} & \displaystyle
\Delta_{n} \in \left[ -\frac{\alpha}{2},\frac{\alpha}{2} \right] \\
0 & \mbox{ for } & \displaystyle
\Delta_{n} \notin \left[ -\frac{\alpha}{2},\frac{\alpha}{2} \right] \\
\end{array} \right.
\end{array}
\label{tgd}
\end{equation}
where $\mathcal{N}$ is a normalisation constant.
Note that the parameter $\sigma_{0}^{2}$ of the distribution~(\ref{tgd})
and the variance of the disorder $\sigma_{\Delta}^{2}$ are linked by
the relation
\begin{displaymath}
\sigma^{2}_{\Delta} = \sigma_{0}^{2}
\left[ 1 - \frac{\alpha}{\sqrt{2 \pi} \sigma_{0}}
\frac{\displaystyle \exp \left( -
\frac{\alpha^{2}}{8 \sigma_{0}^{2}} \right)}
{\displaystyle \mathrm{erf} \left( \frac{\alpha}{\sqrt{8} \sigma_{0}}
\right)} \right] .
\end{displaymath}

The geometry of the problem subjects the sequences $\{\Delta_{n}\}$ of
angular displacements to an additional constraint which does not
exist in the infinite model~(\ref{kpmodel}). If the origin is set
in correspondence with the first barrier, $\theta_{1} = 0$, from
Eqs.~(\ref{strucdis}) and~(\ref{delta_n}) one obtains that the position of
the $n$-th barrier is
\begin{displaymath}
\begin{array}{ccc}
\displaystyle
\theta_{n} = \alpha (n-1) + \sum_{k=1}^{n-1} \Delta_{k} &
\mbox{ for } & n = 2,3,\ldots, N.
\end{array}
\end{displaymath}
In particular, the position of the last barrier is
\begin{displaymath}
\theta_{N} = 2 \pi \frac{N-1}{N} + \sum_{k=1}^{N-1} \Delta_{k} .
\end{displaymath}
Due to the circular nature of the problem, the $N$-th barrier cannot
be placed after the first barrier, i.e., one must have
\begin{displaymath}
\theta_{N} < 2 \pi .
\end{displaymath}
The variables~(\ref{delta_n}) must therefore satisfy the additional condition
\begin{equation}
\sum_{k=1}^{N-1} \Delta_{k} \leq \frac{2\pi}{N} .
\label{delta_bound_2}
\end{equation}
Considering weak disorder reduces but does not eliminate
the possibility of a violation of the condition~(\ref{delta_bound_2}).
In numerical simulations we resorted to ``weeding out'' the
sequences $\{\Delta_{n}\}$ for which the criterion~(\ref{delta_bound_2})
was not respected.
This introduces another difference in the statistical properties of the
variables~(\ref{delta_n}) for the finite model~(\ref{angsch}) with
respect to the infinite system~(\ref{kpmodel}).

To ascertain whether, in spite of these differences, the localisation
behaviour of the eigenstates of the infinite model~(\ref{kpmodel})
is preserved in the finite model~(\ref{angsch}), we numerically studied
the structure of the eigenstates of the latter.
To determine the eigenstates of the model~(\ref{angsch}), we wrote
the system~(\ref{tb_model}) of linear equations in matrix form:
\begin{equation}
\mathbf{M}(q)
\left( \begin{array}{c}
\Theta_{1} \\
\vdots \\
\Theta_{N} \\
\end{array}
\right) =
\left( \begin{array}{c}
0 \\
\vdots \\
0 \\
\end{array} \right)
\label{trisys}
\end{equation}
where $\mathbf{M}(q)$ is the cyclic tridiagonal matrix
\begin{equation}\fl
\mathbf{M}(q) = \left( \begin{array}{ccccccccccc}
\varepsilon_{1} & \gamma_{2} & 0 & \cdots & \cdots & \cdots & \cdots & \cdots
& \cdots & 0 & \gamma_{1} \\
\gamma_{2} & \varepsilon_{2} & \gamma_{3} & 0 & \cdots & \cdots & \cdots &
\cdots & \cdots & \cdots & 0 \\
0 & \gamma_{3} & \varepsilon_{3} & \gamma_{4} & 0 & \cdots & \cdots &
\cdots & \cdots & \cdots & 0 \\
\vdots & \ddots & \ddots & \ddots & \ddots & \ddots & & & & & \vdots \\
\vdots & & \ddots & \ddots & \ddots & \ddots & \ddots & & & & \vdots \\
\vdots & & & 0 & \gamma_{n+1} & \varepsilon_{n} & \gamma_{n+1} & 0 & & & \vdots \\
\vdots & & & & \ddots & \ddots & \ddots & \ddots & \ddots & & \vdots \\
\vdots & & & & & \ddots & \ddots & \ddots & \ddots & \ddots & \vdots \\
0 & \cdots & \cdots & \cdots &\cdots & \cdots & 0 &
\gamma_{N-2} & \varepsilon_{N-2} & \gamma_{N-1} & 0 \\
0 & \cdots & \cdots & \cdots & \cdots & \cdots & \cdots & 0 &
\gamma_{N-1} & \varepsilon_{N-1} & \gamma_{N} \\
\gamma_{1} & 0 & \cdots & \cdots & \cdots & \cdots & \cdots & \cdots &
0 & \gamma_{N} & \varepsilon_{N} \\
\end{array}
\right) .
\label{trimat}
\end{equation}
We let the momentum $q$ vary within the allowed bands and we determined
the values of $q$ for which a non-vanishing solution $\{ \Theta_{n}^{(q)} \}$
of the system~(\ref{trisys}) exists.
More specifically, for each value of $q$ we numerically diagonalised the
matrix~(\ref{trimat}), computing the eigenvalues $\{ \mu_{k}(q) \}$ and
the corresponding eigenvectors $\{ \Theta^{(q)}(k) \}$ (with
$k = 1, \ldots, N$).
We used these results to evaluate the ``density of states''
\begin{displaymath}
\rho = -\frac{1}{\pi}
\sum_{k=1}^{N} \mathrm{Im} \frac{1}{\mu^{(q)}(k) + i \varepsilon} .
\end{displaymath}
For each value of $q$ such that $\rho = O \left( 1/\varepsilon \right)$,
we identified the eigenstate $\Theta^{(q)}(\overline{k})$ with the
smallest eigenvalue $\mu_{\overline{k}}(q)$ (in absolute value) as 
a solution of the homogenous system~(\ref{trisys}).
In this way we obtained the eigenstates of the tight-binding
system~(\ref{tb_model}) and, via~(\ref{eigenstates}), the eigenfunctions
of the disordered Kronig-Penney model~(\ref{angsch}).
We then proceeded to analyse the localisation properties of both models
making use of the inverse participation ratio (IPR) to evaluate the spatial
extension of the discrete eigenvectors of the tight-binding
model~(\ref{tb_model}) and of the continuous eigenstates of the finite
Kronig-Penney model~(\ref{angsch}).
As a further check, we also computed the entropic localisation length of
the eigenvectors of the tight-binding model~(\ref{tb_model}).

The participation ratio $P^{-1}(q)$ represents a commonly used
measure of the portion of space where an eigenstate significantly differs
from zero~\cite{kra93,weg80}.
For the discrete model~(\ref{tb_model}) we defined the IPR using the
relation
\begin{equation}
P(q) = \frac{N}{2\pi} \sum_{n=1}^{N} \left| \Theta_{n}^{(q)} \right|^{4} ,
\label{ipr}
\end{equation}
with the vectors $\{ \Theta_{n}^{(q)} \}$ normalised according to~(\ref{norm_cond_2}).
We chose the prefactor in~(\ref{ipr}) so that $P^{-1} = 2 \pi$ for an
extended state. We used the definition
\begin{equation}
P(q) = \int_{0}^{2 \pi} \left| \Theta^{(q)}(\theta) \right|^{4} \mathrm{d}
\theta
\label{cont_ipr}
\end{equation}
for the eigenstates of the continuous model~(\ref{angsch}) normalised
with the condition~(\ref{norm_cond}).

The entropic localisation length was first applied in~\cite{izr88,izr89,cas90c} to
quantum chaos problems and represents a measure of the effective number of
components of eigenvectors.
After normalising the eigenvectors $\{ \Theta^{(q)}_{n} \}$ with the
condition~(\ref{norm_cond_2}), we computed the corresponding information
entropy
\begin{equation}
S_{N} (q) = - \sum_{n=1}^{N} |\Theta_{n}^{(q)}|^{2} \ln |\Theta_{n}^{(q)}|^{2}
\label{entropy}
\end{equation}
from which we obtained the entropic localisation length, defined as
\begin{equation}
l_{e}(q) =
\frac{2 \pi}{N} \exp \left[ S_{N} \left( q \right) \right] .
\label{ell}
\end{equation}
As in the previous case, the normalisation factor of the entropic localisation
length~(\ref{ell}) was chosen so that, for the extended Bloch
waves~(\ref{bloch_waves}), $l_{e} = 2 \pi$.

We observe that both the entropic localisation length $l_{e}$ and
the inverse participation ratio $P$ are functions of random
eigenvectors  and, as such, are random variables themselves whose
values fluctuate from one disorder realisation to the next.
This sets another difference between the finite models~(\ref{angsch})
and~(\ref{tb_model}) and their infinite counterpart~(\ref{kpmodel}),
which is endowed with a {\em self-averaging} inverse localisation length.
To gain insight on the statistical properties of the entropic localisation
length and of the IPR, we considered the average value and the standard
deviation of both parameters.
We defined the average entropic localisation length as
\begin{equation}
\langle l_{e}(q) \rangle = \frac{1}{N_{r}} \sum_{n=1}^{N_{r}} l_{e}^{(n)}(q)
\label{aell}
\end{equation}
where $N_{r}$ is the total number of disorder realisations and
$l_{e}^{(n)}(q)$ is the value of $l_{e}(q)$ obtained in the $n$-th
realisation.
When performing the ensemble average~(\ref{aell}), one should also
consider that the eigenvalues of the momentum $q$ suffer slight shifts
from one disorder realisation to the next. For this reason we obtained
$\langle l_{e} \rangle$ by dividing the energy band in 100 intervals
and by constructing a histogram for $l_{e}(q)$.
In the case of the inverse participation ratio, we computed its ensemble
average
\begin{equation}
\langle P(q) \rangle = \frac{1}{N_{r}} \sum_{n=1}^{N_{r}} P^{(n)}(q)
\label{aipr}
\end{equation}
and then we plotted the inverse $\langle P \rangle^{-1}$.

As observed in~\cite{izr90}, (\ref{aell}) is not the only
meaningful way to define an average entropic localisation length.
One could also consider the alternative form
\begin{equation}
\overline{l_{e}(q)} =
\frac{2 \pi}{N} \exp \left[ \langle S_{N}(q) \rangle \right]
\label{aell2}
\end{equation}
with
\begin{displaymath}
\langle S_{N}(q) \rangle = \frac{1}{N_{r}} \sum_{n=1}^{N_{r}}
S_{N}^{(n)}(q) .
\end{displaymath}
Our numerical data, however, showed that in the present problem
formulae~(\ref{aell}) and~(\ref{aell2}) produce strikingly similar
results, the main difference being that the entropic localisation
length defined by~(\ref{aell2}) usually turns out to be a few
percents shorter than its counterpart~(\ref{aell}), in agreement with
the results found in~\cite{izr90}.

\begin{figure}
\begin{center}
\includegraphics[width=0.49\textwidth]{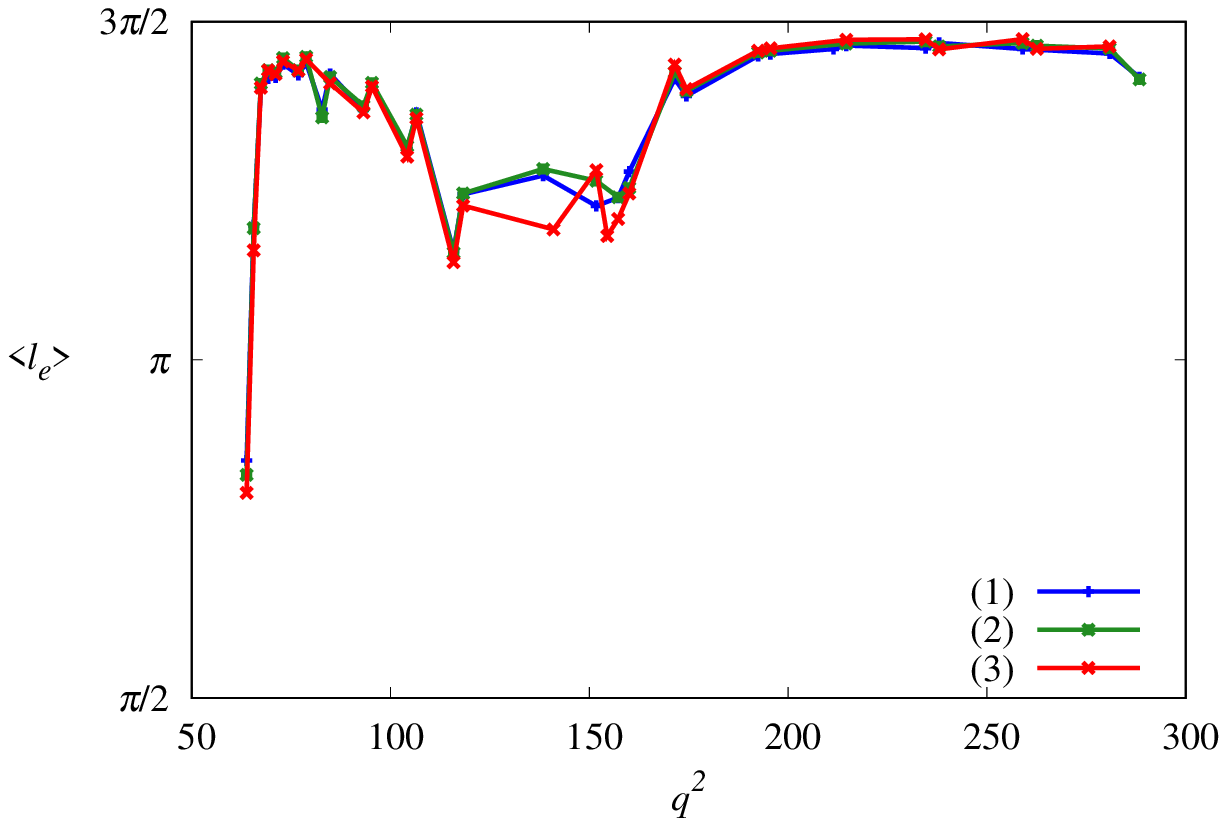}
\includegraphics[width=0.49\textwidth]{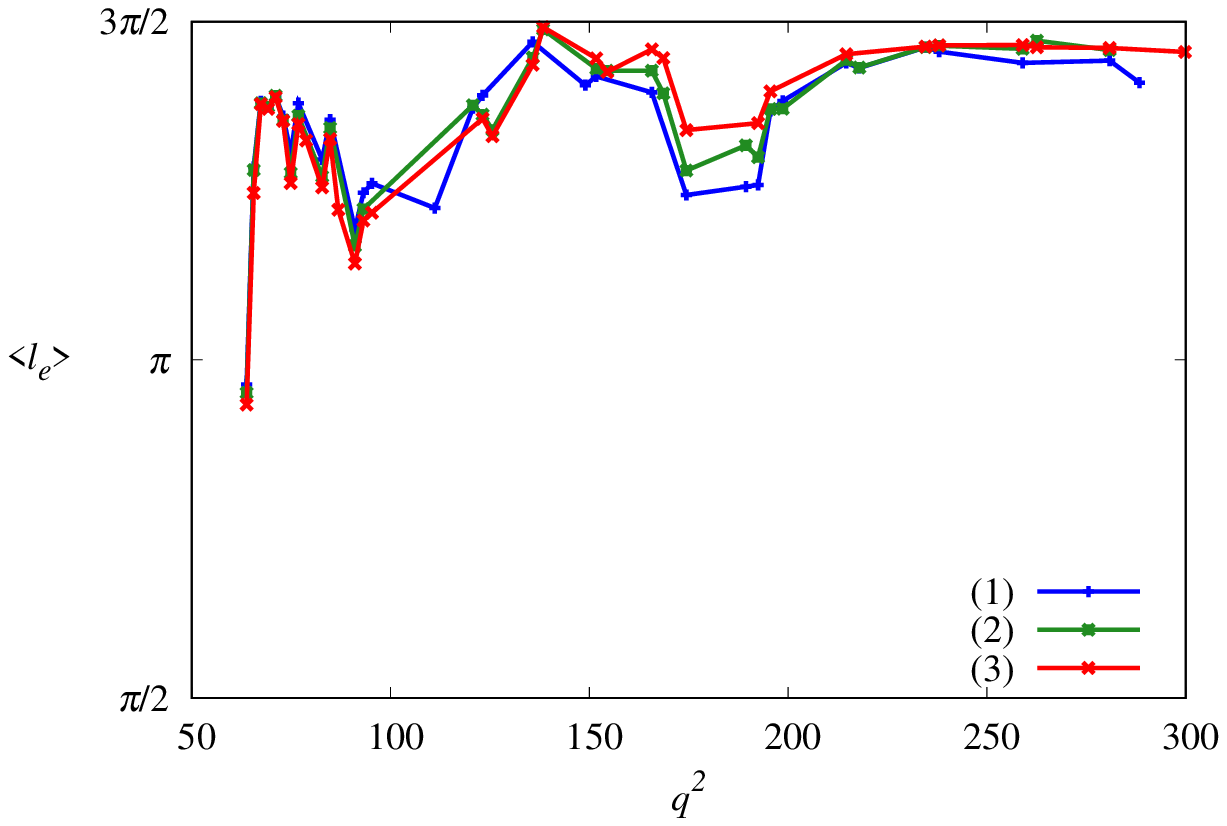}\\
\includegraphics[width=0.49\textwidth]{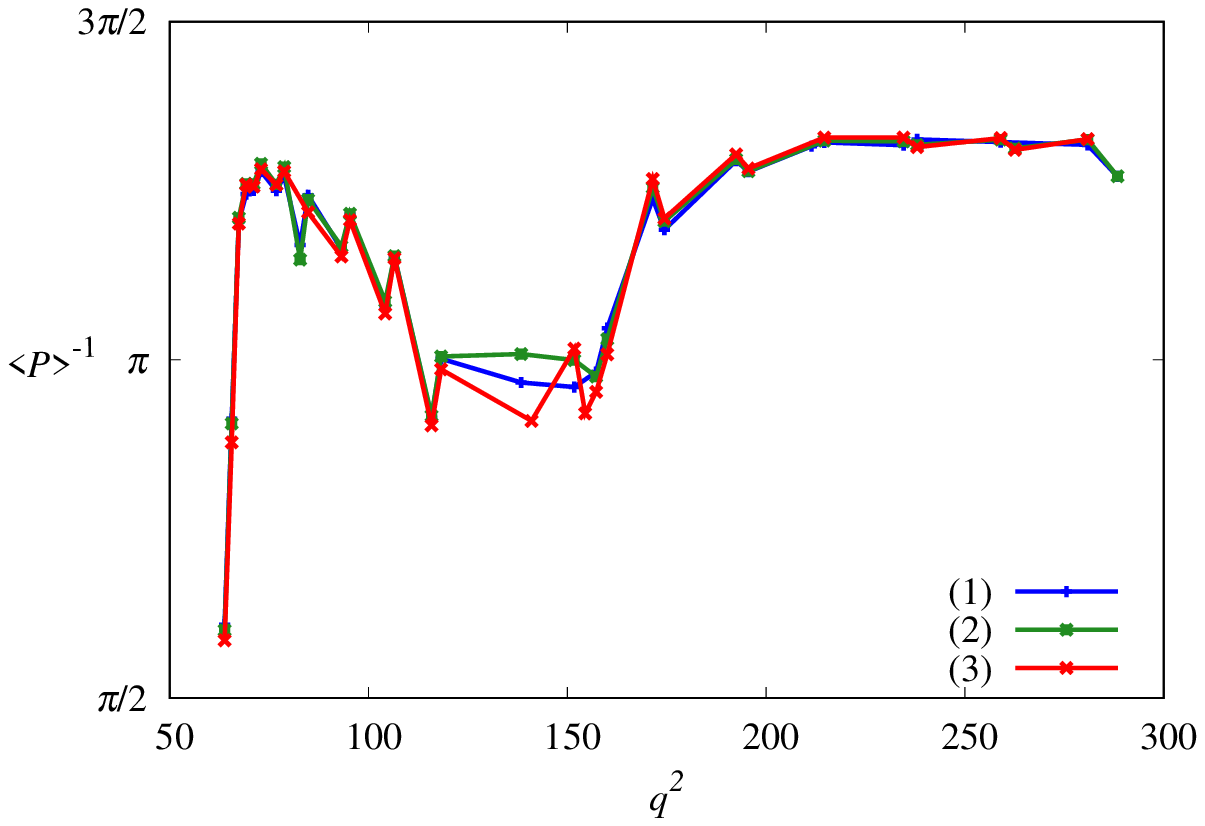}
\includegraphics[width=0.49\textwidth]{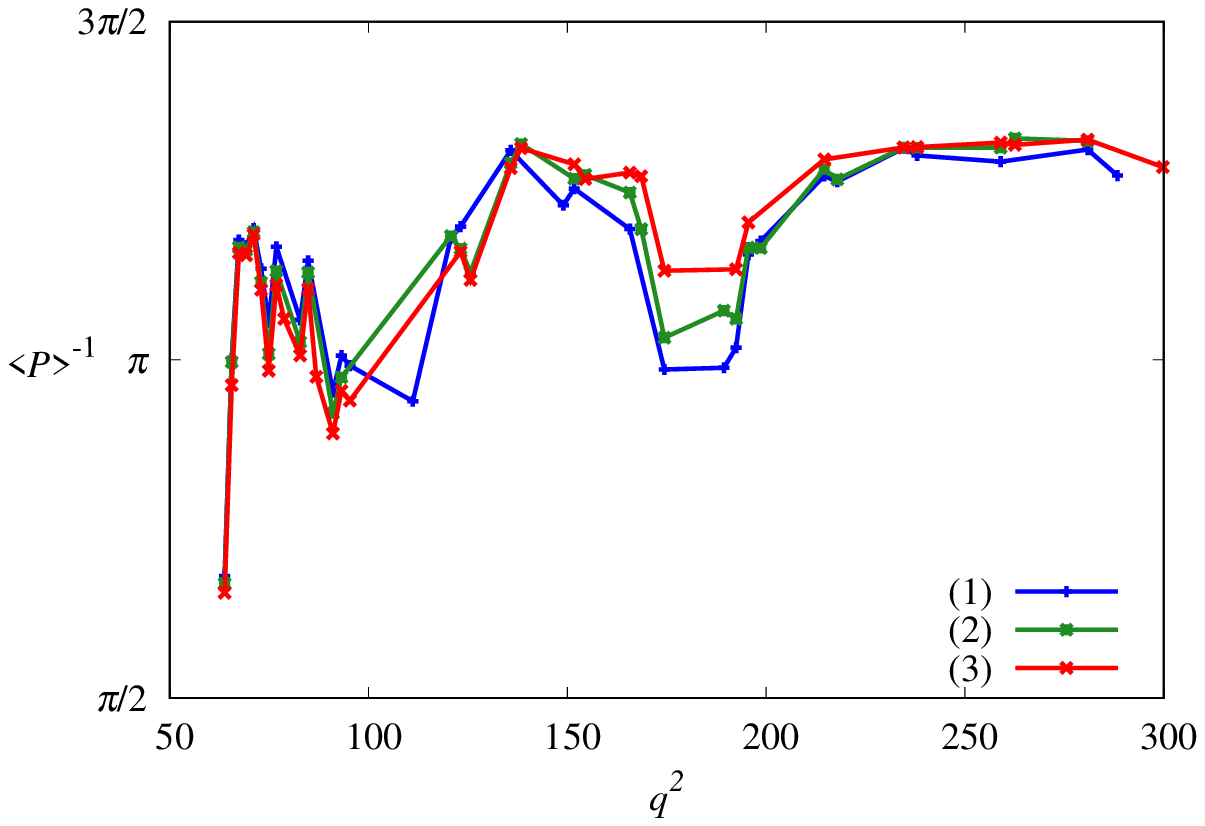}\\
\includegraphics[width=0.49\textwidth]{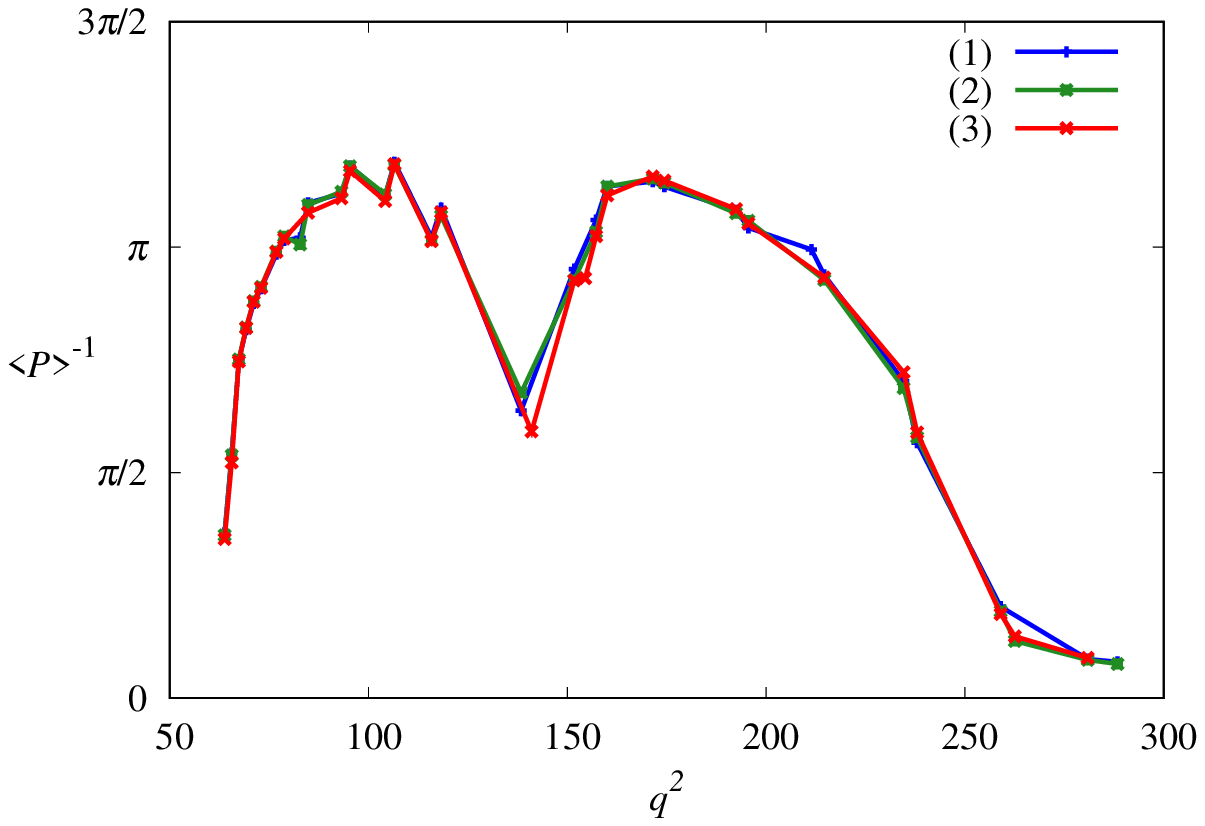}
\includegraphics[width=0.49\textwidth]{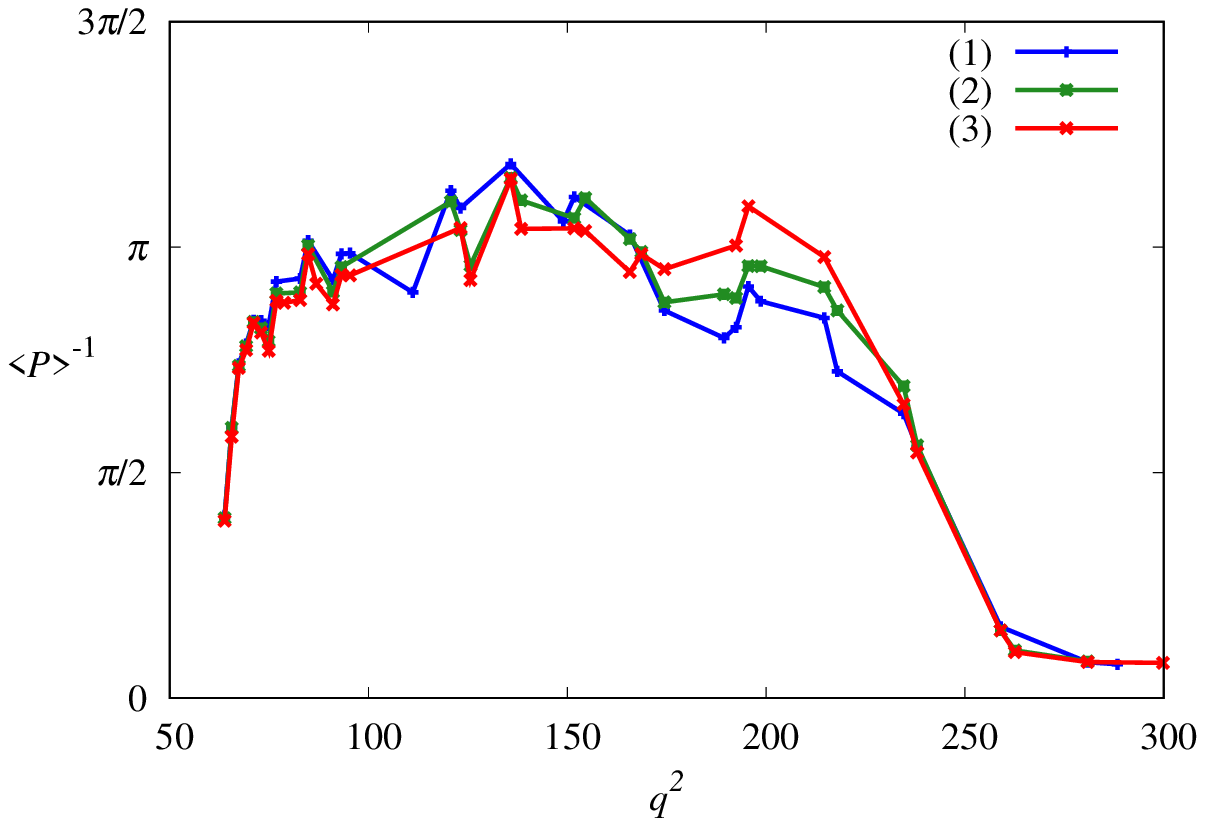}
\caption{Average entropic localisation length and IPR computed over an
ensemble of $N_{r} = 1000$ disorder configurations, for the case $U=15.0$,
$\sqrt{\langle u_{n}^{2} \rangle} = 4.0$ and
$\sqrt{\langle \Delta_{n}^{2} \rangle}=0.007$.
Top panels: average entropic localisation length $\langle l_{e} \rangle$
versus energy $q^{2}$.
Middle panels: Inverse of the average IPR $\langle P \rangle^{-1}$ versus
energy $q^{2}$ for the discrete model~(\ref{tb_model}).
Bottom panels: inverse of the average IPR $\langle P \rangle^{-1}$ versus
energy $q^{2}$ for the continuous model~(\ref{angsch}).
Left panels: one localisation window ($k_{1} = 0.46 \pi$ and $k_{2} = 0.5 \pi$).
Right panels: two localisation windows ($k_{1} = 0.35 \pi$ and $k_{2} = 0.40 \pi$).
\label{av_fig}}
\end{center}
\end{figure}

In our numerical studies, we considered the angular Kronig-Penney
model~(\ref{angsch}) and its tight-binding homologue~(\ref{tb_model})
with $N = 35$ barriers and the same disorder parameters used for their
infinite counterpart in Sec.~\ref{infinite_kp}, i.e., average barrier
strength $U = 15.0$, compositional and structural disorders with respective
intensities $\sqrt{\langle u_{n}^{2} \rangle} = 4.0$ and
$\sqrt{\langle \Delta_{n}^{2} \rangle} =0.007$.
As in Sec.~\ref{infinite_kp}, we considered two cases: in the first one we
generated a single localisation window by setting effective mobility edges
with Bloch vectors $k_{1} = 0.46 \pi$ and $k_{2} = 0.5 \pi$.
In the second case we selected the values $k_{1} = 0.35 \pi$ and
$k_{2} = 0.40 \pi$ and we thus obtained two localisation windows.
We considered positive, absent and negative cross-correlations; as noted
before, in the figures the respective cases are labelled with the numbers~(1),
(2), and (3) and associated to the colours blue, green, and red.
In all cases the averages were computed over an ensemble of
$N_{r} = 1000$ disorder realisations.
We present our numerical data for the average value of the entropic
localisation length $\langle l_{e} \rangle$ in the top
panels of figure~\ref{av_fig};
the inverse of the average of the IPR $\langle P \rangle^{-1}$ is
shown in the middle panels of figure~\ref{av_fig}
for the tight-binding model~(\ref{tb_model})
and in the bottom panels for the Kronig-Penney model~(\ref{angsch}).

When dealing with the entropic localisation length~(\ref{ell})
and the inverse participation ratio~(\ref{ipr}), one should
keep in mind that the average values~(\ref{aell}) and~(\ref{aipr})
do {\em not} provide a complete picture of the behaviour of the
eigenstates of the Kronig-Penney model~(\ref{angsch}).
This is due to the fact that both localisation lengths exhibit
very strong fluctuations from one disorder realisation to the next.
In our numerical study, for example, we found that the typical values of
the standard deviation of $l_{e}$ and $P^{-1}$ varied between $\pi$ and
$2\pi$.
We did not represent the corresponding error bars in figure~\ref{av_fig}
because they were too large and significantly reduced the clarity of the
data.
Even more important, increasing the number of disorder realisations did
not produce a diminution of the standard deviation.
The observed statistical behaviour of the two generalised localisation
lengths agrees with the properties described in the
literature~\cite{izr90,fyo95b,eve08}.

\begin{figure}
\begin{center}
\includegraphics[width=0.49\textwidth]{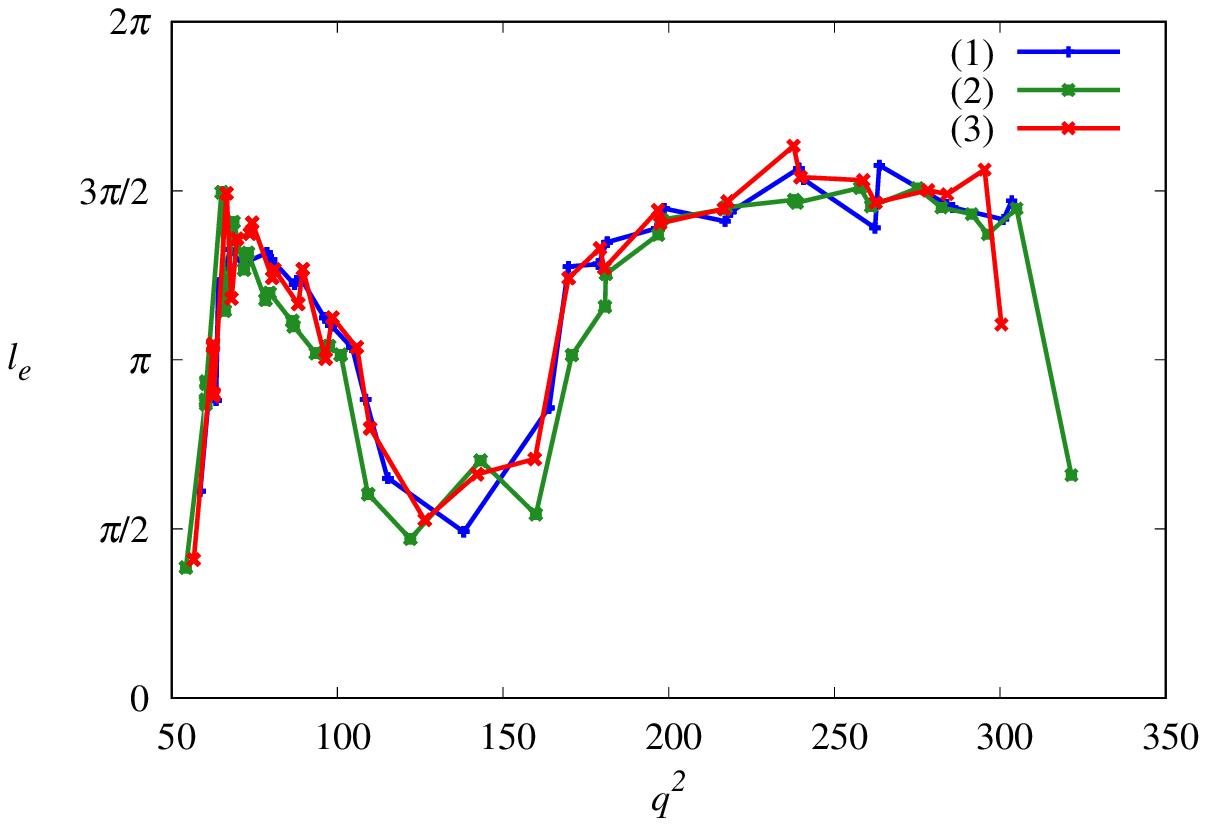}
\includegraphics[width=0.49\textwidth]{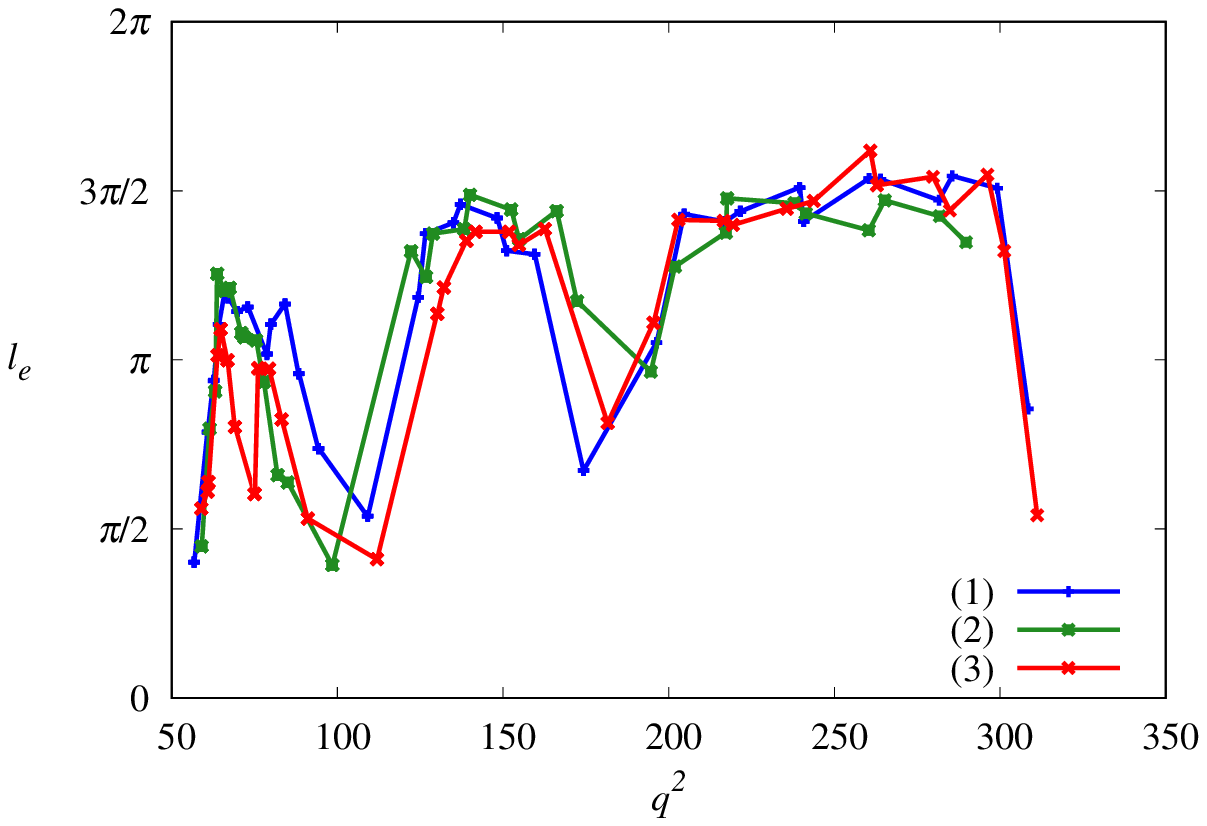}\\
\includegraphics[width=0.49\textwidth]{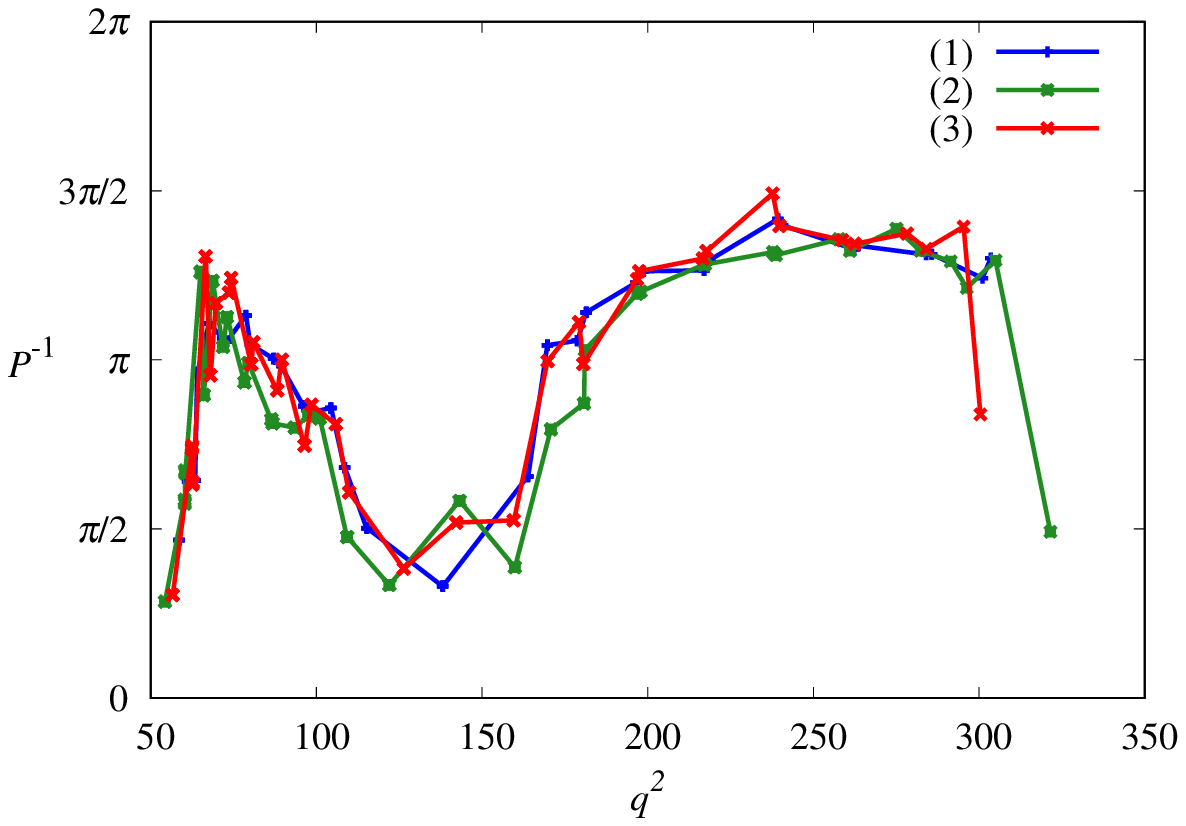}
\includegraphics[width=0.49\textwidth]{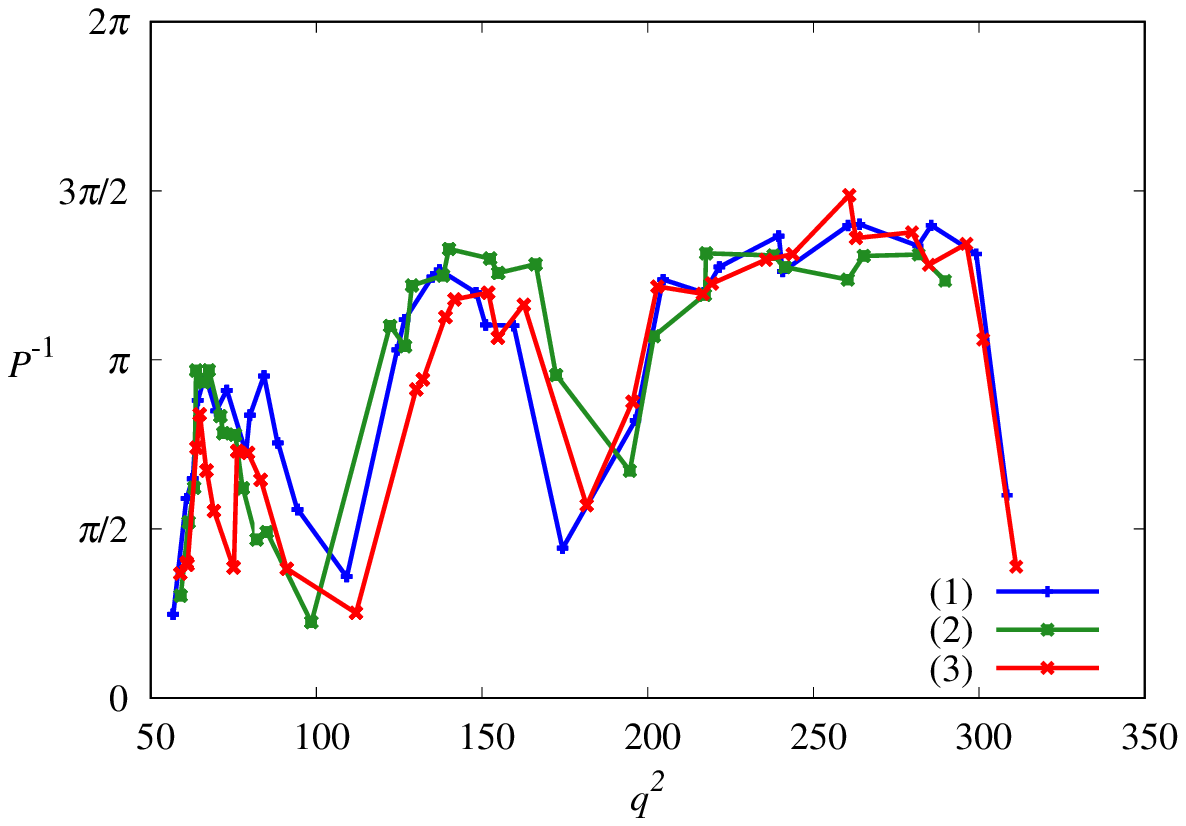}\\
\includegraphics[width=0.49\textwidth]{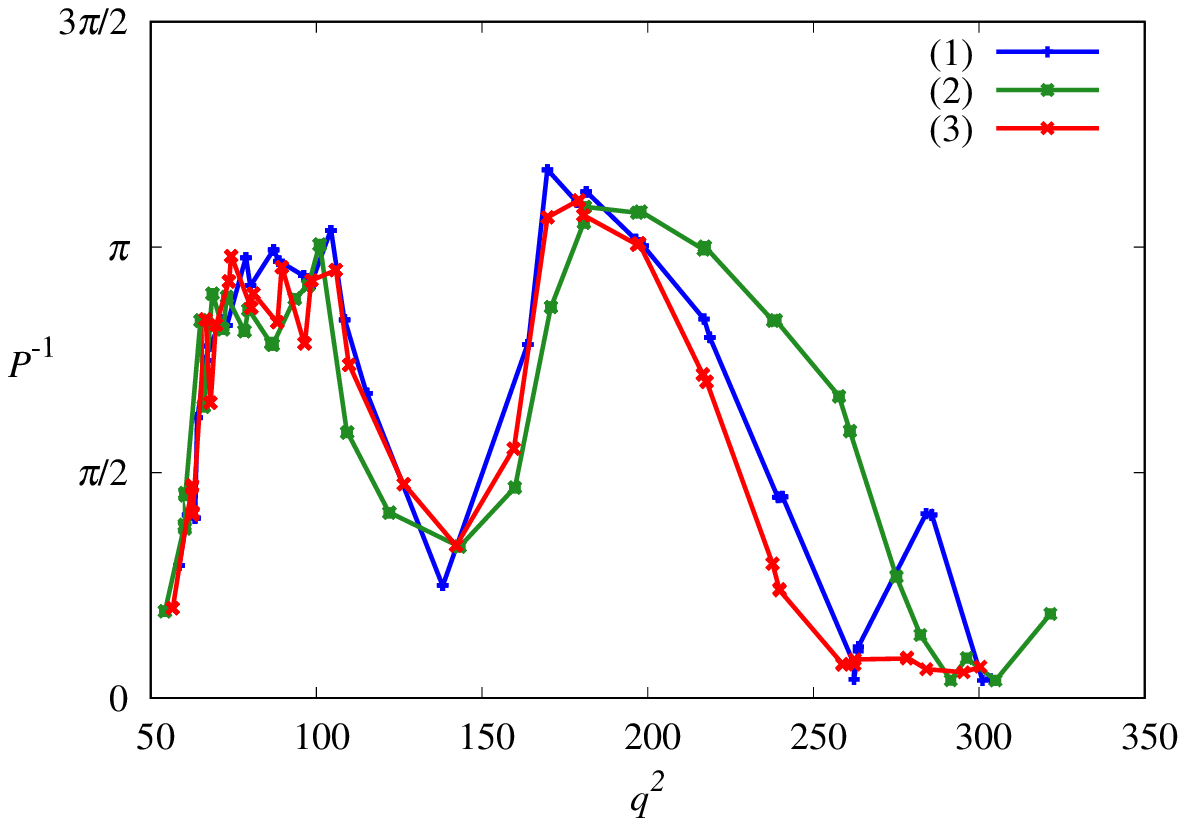}
\includegraphics[width=0.49\textwidth]{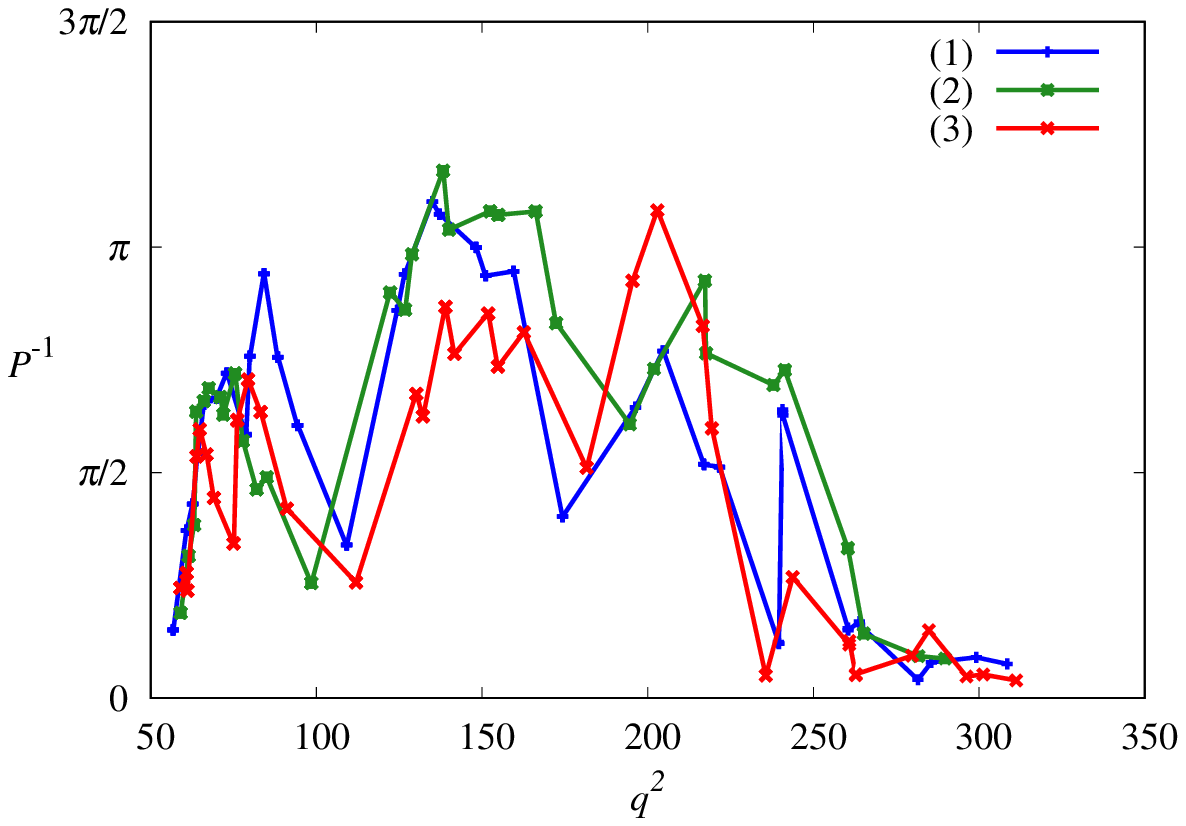}
\caption{As figure~\ref{av_fig} but for a specific disorder realisation.}
\label{opt_fig}
\end{center}
\end{figure}

The strong fluctuations of $l_{e}$ and $P^{-1}$ can be exploited in the
construction of an experimental setup designed to obtain a robust
localisation enhancement.
In fact, one can select a specific disorder realisation which produces
a particularly strong localisation of the angular eigenstates in the
selected energy interval: this is illustrated in the
panels of figure~\ref{opt_fig}
which represent the entropic localisation length and the IPR for such
a single realisation of the disorder.

The numerical results displayed in the figures of this section
clearly show that correlated disorder can produce a significative
enhancement of localisation in predefined energy windows, although the
magnitude of the effect fluctuates considerably over the ensemble of
disorder realisation.
We conclude that the effects of disorder correlations survive (albeit in
an attenuated form) in the tight-binding model~(\ref{tb_model}) and in
the finite Kronig-Penney model~(\ref{angsch}), in spite of the limitations
imposed by the finite size and by the geometry of the system.

\begin{figure}[htb]
\begin{center}
\includegraphics[width=5in]{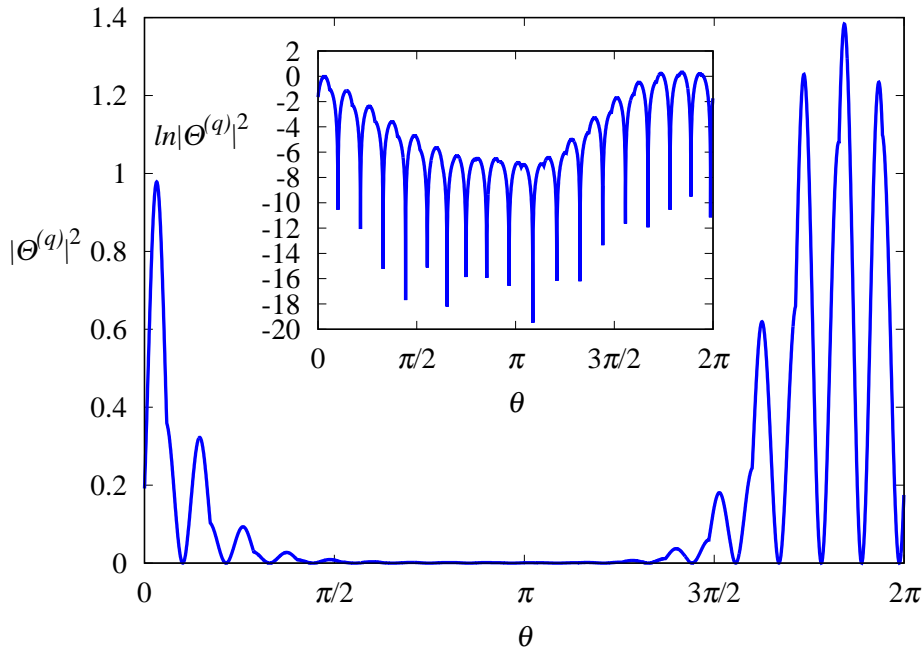}
\caption{Squared modulus of the angular eigenfunction $\Theta^{(q)}(\theta)$
versus the  angle variable $\theta$ with momentum $q \simeq 11.04$ and
energy $q^{2} \simeq 122.0$, for the same disorder realisation as in
figure~\ref{opt_fig}.
\label{localised_wave}}
\end{center}
\end{figure}

As a further illustration of the particularly strong localisation that
can be achieved in specific realisations of the disorder, in
figure~\ref{localised_wave} we show the most localised eigenfunction
of the Kronig-Penney model~(\ref{angsch}) obtained with the
same disorder realisation represented in figure~\ref{opt_fig}.
Specifically, we considered the eigenfunction with momentum $q \simeq 11.045$
and energy $q^{2} \simeq 122.0$ in the first energy band.
The eigenfunction was derived for the case in which structural and
compositional disorder are self-correlated but not cross-correlated.
The energy of this eigenfunction is the closest within the perturbed
spectrum to the energy of the Bloch wave considered in
Sec.~\ref{infinite_kp}. We can consider the wavefunction represented
in figure~\ref{localised_wave} as the counterpart of the wavefunction
pictured in figure~\ref{bloch_wave}. Comparing the two figures, it is
easy to see that correlated disorder manages to produce a rather
good localisation of specific waves.

\subsection{Resonance effects in the continuous Kronig-Penney model}
\label{reson}

So far, in our analysis of the localisation properties of the
Kronig-Penney model~(\ref{angsch}) we have relied on the close
relationship which exists between the model itself and its
tight-binding counterpart~(\ref{tb_model}). In particular, we have
used the correspondence between the continuous eigenfunctions
$\Theta(\theta)$ of the former and the discrete eigenvectors
$\left\{ \Theta_{n} \right\}$ of the latter. This correspondence,
established in Sec.~\ref{angul}, is corroborated by the numerical
data presented in Sec.~\ref{finapkp}, which reveal the roughly parallel
behaviour of the IPR of the eigenstates of the models~(\ref{angsch})
and~(\ref{tb_model}).
However, a careful comparison of the middle panels with
the bottom panels of figure~\ref{av_fig} and \ref{opt_fig}
also shows that the IPR's of the two models behave
differently in the higher part of the energy band. When the energy
approaches the top of the band, the participation ratio of the tight-binding
model~(\ref{tb_model}) keeps high values, indicating that the discrete
eigenstates $\{\Theta_{n}\}$ are extended, whereas the participation ratio
for the Kronig-Penney model~(\ref{angsch}) falls close to zero in the same
energy range, a sign that the spatial extension of the continuous
eigenstates $\Theta(\theta)$ is strongly reduced.
This discrepancy can be explained as a manifestation of resonance
effects occurring in the Kronig-Penney model~(\ref{angsch}) and shows
that, under specific circumstances, the correspondence between this
model and its tight-binding homologue~(\ref{tb_model}) breaks down.

To understand this point, let us consider an eigenvector
$\left\{ \Theta_{n} \right\}$ of the tight-binding model~(\ref{tb_model})
with momentum $q$.
Let us suppose that structural disorder produces a displacement of
the $n$-th and $(n+1)$-th radial barriers such that
\begin{equation}
q \left( \theta_{n+1} - \theta_{n} \right) = \pi + \varepsilon
\label{res_cond}
\end{equation}
with $\varepsilon \to 0$.
For the sake of simplicity, we will assume that only the $n$-th well
satisfies a resonance condition of the form~(\ref{res_cond}).
Substituting the identity~(\ref{res_cond}) in~(\ref{eigenstates}),
one obtains that the eigenstate of the Kronig-Penney model~(\ref{angsch})
with momentum $q$ is equal to
\begin{equation}\fl
\Theta(\theta) = \left\{ \begin{array}{lcl}
\displaystyle
\EuScript{N} \left[ \frac{\Theta_{n+1} - \Theta_{n}}{\varepsilon}
\sin \left[ q \left( \theta - \theta_{n} \right) \right] +
O \left( \varepsilon^{0} \right) \right] & \mbox{ if } &
\theta \in \left[ \theta_{n}, \theta_{n+1} \right] \\
O \left( \varepsilon^{0} \right) & \mbox{ if } &
\theta \notin \left[ \theta_{n}, \theta_{n+1} \right]
\end{array} \right.
\label{res_eig_unnorm}
\end{equation}
The constant $\EuScript{N}$ can be derived from the normalisation
condition~(\ref{norm_cond}); one has
\begin{displaymath}
\EuScript{N} = \sqrt{\frac{2 q}{\pi}}
\frac{\varepsilon}{\Theta_{n+1}-\Theta_{n}} \left[ 1 + O\left(
\varepsilon^{2} \right) \right] .
\end{displaymath}
Substituting this expression in~(\ref{res_eig_unnorm}) one
obtains that the resonant eigenstate has the form
\begin{equation}
\Theta(\theta) = \left\{ \begin{array}{lcl}
\displaystyle
\sqrt{\frac{2 q}{\pi}} \sin \left[ q \left( \theta - \theta_{n} \right) \right]
+ O\left( \varepsilon \right) & \mbox{ if } &
\theta \in \left[ \theta_{n}, \theta_{n+1} \right] \\
O \left( \varepsilon \right) & \mbox{ if } &
\theta \notin \left[ \theta_{n}, \theta_{n+1} \right]
\end{array} \right.
\label{res_eig}
\end{equation}
(\ref{res_eig}) shows that, when the phase of the eigenfunction
increases by an integer multiple of $\pi$ in a potential well, the
mode becomes locked in that well. In other words, the well acts as a
Fabry-P\'{e}rot resonator.

In this case, there is a profound difference between the eigenvectors of
the tight-binding model~(\ref{tb_model}) and the eigenstates of the
continuous Kronig-Penney model~(\ref{angsch}). The discrete eigenvectors
$\{\Theta_{n}\}$ have non-vanishing components in the whole $[0,2\pi]$ range,
whereas the continuous eigenfunctions $\Theta(\theta)$ are significantly
different from zero only in a single well.
When this happens the continuous eigenfunctions $\Theta(\theta)$ are
much more localised than the corresponding eigenvectors $\left\{
\Theta_{n} \right\}$; however, their reduced spatial extension must
be counted as a resonance effect, and not as a true Anderson localisation.
One might say that, in some sense, the eigenvectors of the tight-binding
model~(\ref{tb_model}) offer a better insight on the true localisation
properties of the Kronig-Penney model~(\ref{angsch}) than the eigenfunctions
of the model themselves.

Whether the resonance condition~(\ref{res_cond}) is met or not determines
if resonance effects cooperate with Anderson localisation in shaping the
eigenfunctions of the Kronig-Penney model~(\ref{angsch}).
As an example, let us consider once more the case of a Kronig-Penney
model with $N=35$ barriers, mean field strength $U = 15.0$ and
compositional and structural disorders with strengths
$\sqrt{\langle u_{n}^{2} \rangle} = 4.0$ and
$\sigma_{\Delta} = \sqrt{\langle \Delta_{n}^{2} \rangle} = 0.007$.
In this case the momentum $q$ takes values within the interval
$q \in [q_{\mathrm{min}},q_{\mathrm{max}}]$ with
$q_{\mathrm{min}} \simeq 8.2$ and $q_{\mathrm{max}} \simeq 17.5$
for the first band.
The threshold value of the momentum for the onset of a Fabry-P\'{e}rot
resonance is
\begin{displaymath}
q_{c} \simeq \frac{\pi}{\Delta \theta_{\mathrm{max}}}
\end{displaymath}
where $\Delta \theta_{\mathrm{max}}$ is the largest angular distance
between two consecutive barriers.
In the absence of disorder, one has $\Delta \theta_{\mathrm{max}} =
\alpha = 2\pi/N$ and therefore
\begin{equation}
q_{c} = \frac{\pi}{\alpha} = \frac{N}{2} = 17.5.
\label{qcrit_nodis}
\end{equation}
This shows that the threshold value of the momentum lies right at the
top of the first band and, consequently, resonances are a marginal
phenomenon in the first band of the {\em periodic} Kronig-Penney model.

Things are different when disorder is added, though. In our case
$\sigma_{\Delta}~\ll~\alpha$ and the difference between the
distributions~(\ref{gd}) and~(\ref{tgd}) is irrelevant. Assuming that
each barrier can swing up to $\pm 2 \sigma_{\Delta}$ from its unperturbed
position, one can conclude that the variation of the angular distance
$\Delta \theta_{n}$ lies in the interval $[\alpha - 4 \sigma_{\Delta},
\alpha + 4 \sigma_{\Delta}]$ and that in most cases one has
\begin{displaymath}
\alpha - 3 \sigma_{\Delta} \lesssim \Delta \theta_{n} \lesssim
\alpha + 3 \sigma_{\Delta} .
\end{displaymath}
This gives a critical value of the momentum
\begin{equation}
q_{c} \simeq \frac{\pi}{\alpha + 3 \sigma_{\Delta}} \simeq 15.66
\label{qcrit_dis}
\end{equation}
which is considerably lower than the value~(\ref{qcrit_nodis}) obtained
in the absence of disorder.
The value~(\ref{qcrit_dis}) corresponds to a critical energy
$q_{c}^{2} \simeq 245.5$ which lies well within the first band and
broadly agrees with the observed inflection point where the participation
ratio for the Kronig-Penney model~(\ref{angsch}) starts to decline, see
bottom panels of figure~\ref{av_fig}.

Note that, for resonant states of the form~(\ref{res_eig_unnorm}),
the IPR~(\ref{cont_ipr}) takes the value
\begin{displaymath}
P_{\mathrm{res}}(q) = \frac{3 q}{2 \pi} + O(\varepsilon) ,
\end{displaymath}
which corresponds to a spatial extension
\begin{equation}
P_{\mathrm{res}}^{-1}(q) = \frac{2 \pi}{3 q} + O(\varepsilon) .
\label{res_line}
\end{equation}
We observe that strong resonances which confine the wavefunction to a
single well do not occur for every disorder realisation, therefore
(\ref{res_line}) represents a lower bound for the average participation
ratio.
With the values of the parameters considered in numerical simulations,
we can expect the parameter~(\ref{res_line}) to take values
$P^{-1}_{\mathrm{res}} \simeq 0.12$ to 0.13 for $q$ ranging from 15.6
to 17.5. This is consistent with the asymptotic value $\langle P \rangle^{-1}
\simeq 0.25$ of the average participation ratio obtained in numerical
simulation at the top of the band.
In conclusion, with the present values of the parameters one can expect
the Anderson localisation to be the relevant phenomenon for energies
at the bottom and in the wide middle of the first band, while resonant
states become a dominant feature in the top fringe of the energy band.

\begin{figure}[htb]
\begin{center}
\includegraphics[width=5in]{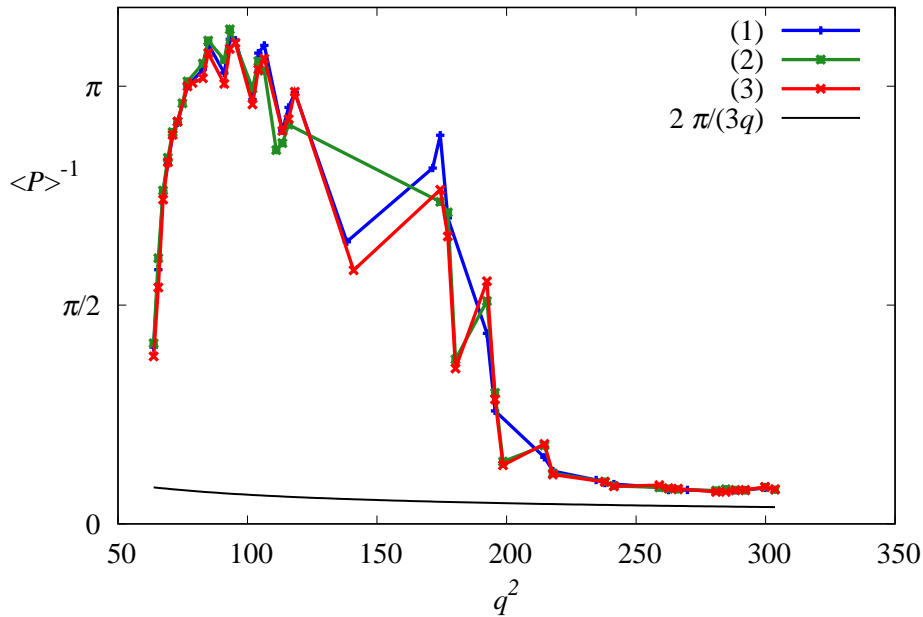}
\caption{Inverse of the average IPR $\langle P \rangle^{-1}$
versus energy $q^{2}$ for the first energy band. The structural disorder
strength is $\sigma_{\Delta} = 0.21$. The black continuous
line represents the limit behaviour given by~(\ref{res_line}).
\label{aver_kp_ipr_sd}}
\end{center}
\end{figure}

The hypothesis that Fabry-P\'{e}rot resonances are the origin of the
drastic reduction of the spatial extension of the wavefunctions of the
Kronig-Penney model~(\ref{angsch}) is further confirmed by the fact that,
if compositional disorder is strengthened, the critical threshold for
the onset of resonant state is lowered.
A stronger structural disorder extends the variation range of the
angular width of the wells; as a consequence, lower values of
$q$ are required to fullfill the resonance condition~(\ref{res_cond}).
Therefore the drop of the participation ratio should occur at lower
energies than is the case for weaker structural disorder.
This is in fact what happens, as can be seen in figure~\ref{aver_kp_ipr_sd},
which represents the inverse of the average IPR as in the bottom panels
of figure~\ref{av_fig}
but for a Kronig-Penney model with {\em stronger} structural disorder, i.e.,
$\sigma_{\Delta} = 0.021$, which is three times the strength of the previous
case; the other parameters of the model were left unchanged.

To conclude this section, we remark that resonances have usually been
neglected in the study of the infinite Kronig-Penney model~(\ref{kpmodel}).
It might be worthwhile to check whether the interplay of localisation and
resonance also play a role in that model.

\subsection{Radial part of the Schr\"{o}dinger equation}
\label{rad_part_schr}

We now turn our attention to the radial part of the wavefunction.
The solutions of~(\ref{radsch}) are linear combinations of
independent Bessel functions of the first and second kind
\begin{displaymath}
R_{q}(r) = c_{1} J_{q}(\sqrt{E} r) + c_{2} Y_{q}(\sqrt{E} r) .
\end{displaymath}
If the particle is confined within the annulus, the wavefunction
must vanish on the inner and outer circles of radii $r_{1}$ and $r_{2}$.
Imposing the boundary conditions
\begin{displaymath}
R_{q}(\sqrt{E} r_{1}) = R_{q}(\sqrt{E} r_{2}) = 0
\end{displaymath}
leads to the solution
\begin{displaymath}
R_{q}(r) = \mathcal{N} \left[ Y_{q}(\sqrt{E} r_{1}) J_{q}(\sqrt{E} r) -
J_{q}(\sqrt{E} r_{1}) Y_{q}(\sqrt{E} r) \right] ,
\end{displaymath}
where $\mathcal{N}$ is a normalisation constant, to be determined with
the condition
\begin{displaymath}
\int_{r_{1}}^{r_{2}} \left| R_{q}(r) \right|^{2} r \mathrm{d} r = 1,
\end{displaymath}
while the values of the energy $E$ can be obtained by solving the
equation
\begin{equation}
Y_{q}(\sqrt{E} r_{1}) J_{q}(\sqrt{E} r_{2}) -
J_{q}(\sqrt{E} r_{1}) Y_{q}(\sqrt{E} r_{2}) = 0 .
\label{crossprod}
\end{equation}
It is convenient to rewrite~(\ref{crossprod}) in the form
\begin{equation}
Y_{q}(x) J_{q}(\varkappa x) -
J_{q}(x) Y_{q}(\varkappa x) = 0
\label{crossprod2}
\end{equation}
with $x = r_{1} \sqrt{E}$ and $\varkappa = r_{2}/r_{1}$.
For any eigenvalue of the momentum $q$, let $x(q,s)$ be the $s$-th zero
of the cross-product of Bessel functions~(\ref{crossprod2}),
with $|x(q,s=1)| \leq |x(q,s=2)| \leq \ldots$
Then the eigenvalues of the total energy of the system can be labelled
with the quantum numbers $(q,s)$ and take the form
\begin{displaymath}
E_{q,s} = \left( \frac{x(q,s)}{r_{1}} \right)^{2} .
\end{displaymath}
On the other hand, for weak disorder each value $q$ of the angular quantum
number can be written as a function of the Bloch wavenumber $k$ and of a band
index $n$; it is therefore convenient to write the energy in the form
\begin{equation}
E(k,n,s) = \left( \frac{x[q(k,n),s]}{r_{1}} \right)^{2} .
\label{spectrum}
\end{equation}

The values $x(q,s)$ can be obtained by solving numerically~(\ref{crossprod2}); for larger values of $s$ one can also
use the asymptotic expansion provided by the McMahon
formula~\cite{mcm94}.

\begin{figure}[htb]
\begin{center}
\includegraphics[width=5in]{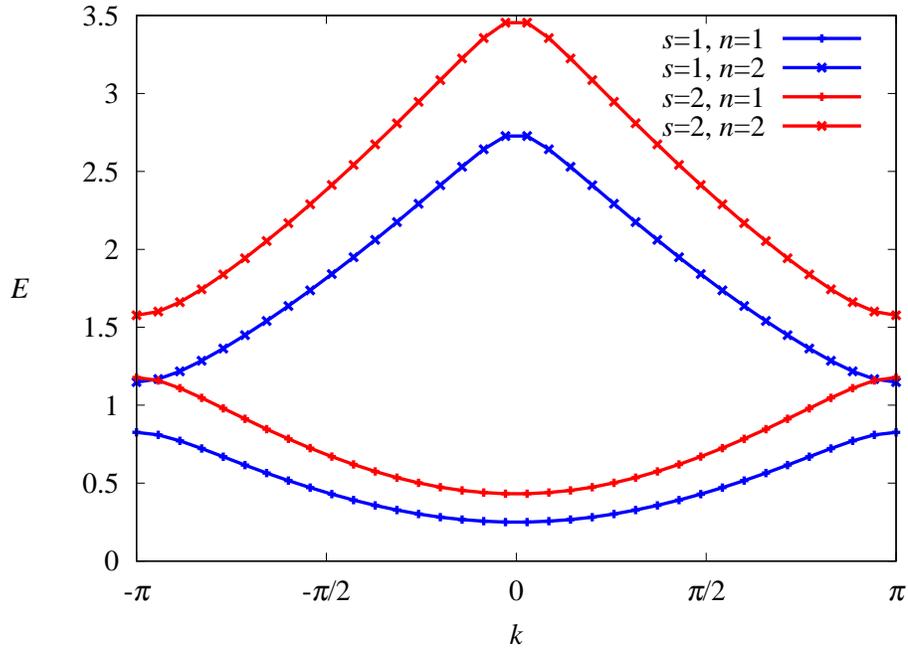}
\caption{Energy levels $E(k,n,s)$ versus $k$ for $n=1,2$ and $s=1,2$. The
energy values are measured in $\mbox{cm}^{-2}$.
\label{spec}}
\end{center}
\end{figure}

In figure~\ref{spec} we plot the energy levels~(\ref{spectrum}) as a function
of the Bloch wavenumber $k$ for the values $n=1,2$ of the band index and
$s=1,2$ of the radial quantum number.
The data were obtained for $r_{1} = 10\,\mbox{cm}$ and $r_{2} = 25\,\mbox{cm}$
and for $U = 15.0$. (The values of $r_{1}$ and $r_{2}$ correspond to the
inner and outer radii of the cavity discussed in~\ref{app:mwc}.)
The computation of the values of $q$ was done in the limit of vanishing disorder.

\begin{figure}
\begin{subfigure}[b]{0.5\textwidth}
\includegraphics[width=\textwidth]{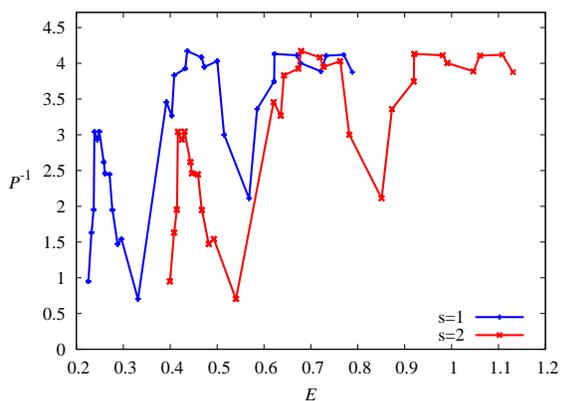}
\caption{Tight-binding model~(\ref{tb_model})}
\end{subfigure}%
\begin{subfigure}[b]{0.5\textwidth}
\includegraphics[width=\textwidth]{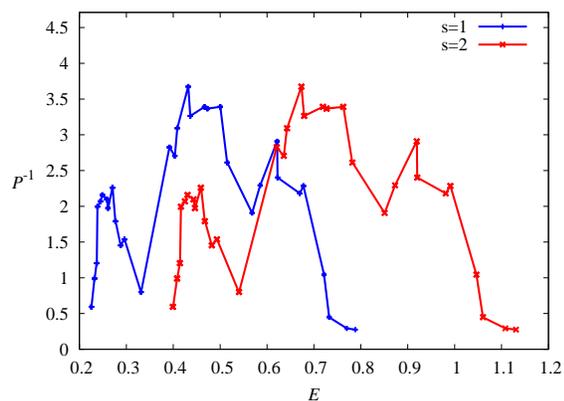}
\caption{Kronig-Penney model~(\ref{kpmodel})}
\end{subfigure}%
\caption{Inverse of the IPR $P^{-1}$ versus energy $E$ (in $\mbox{cm}^{-2}$)
for the discrete model~(\ref{tb_model}) and for the continuous
model~(\ref{kpmodel}) for the first two bands, $s=1$ and $s=2$.
The data correspond to the single disorder
configuration shown in figure~\ref{opt_fig}.}
\label{superposition}
\end{figure}

Figure~\ref{spec} shows that a good deal of overlapping can occur between
energy bands with the same index $n$ but different radial numbers $s$.
This implies that {\em one can localise eigenstates with different radial
numbers within a single window of the energy $E$} of the 2D model~(\ref{2dschr}).
This is what is shown in figure~\ref{superposition} which represents the
behaviour of the participation ratio $P^{-1}$ as a function of the total
energy $E$ (and not of the squared momentum $q^{2}$, as was done in
Secs.~\ref{finapkp} and~\ref{reson}).
The data correspond to the single disorder realisation shown in
figure~\ref{opt_fig}  for the case of self-correlated
disorder without cross-correlations.
We would like to stress that the localisation within the same energy window of
states with different angular quantum numbers is a genuine manifestation of the
{\em two-dimensional} nature of the model~(\ref{2dschr}).

\begin{figure}
\begin{center}
\includegraphics[width=0.49\textwidth]{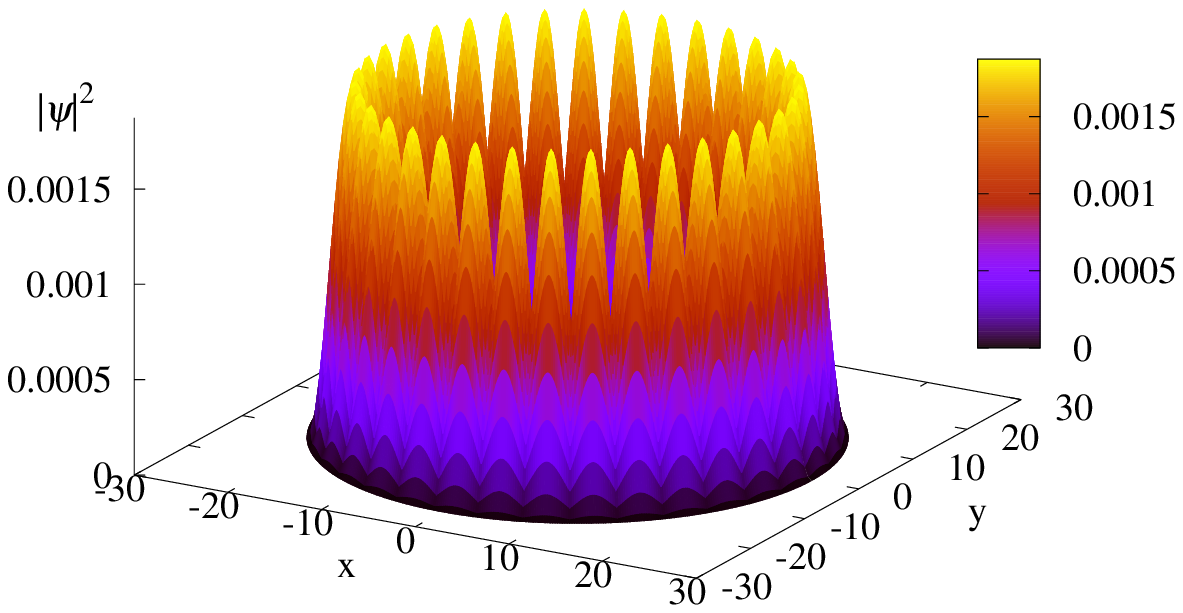}
\includegraphics[width=0.49\textwidth]{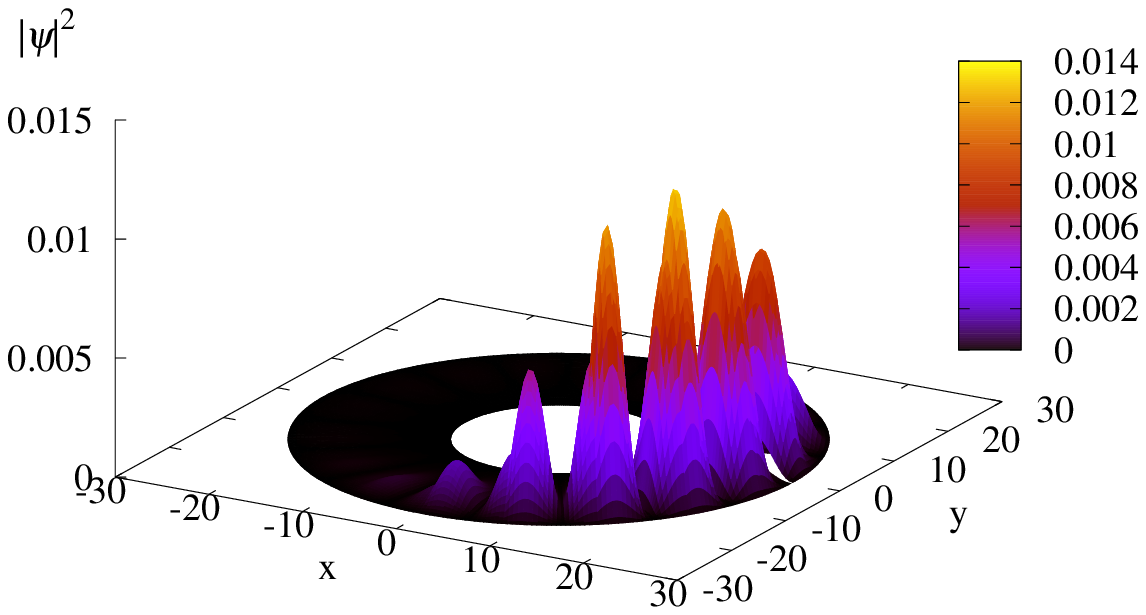}\\
\includegraphics[width=0.49\textwidth]{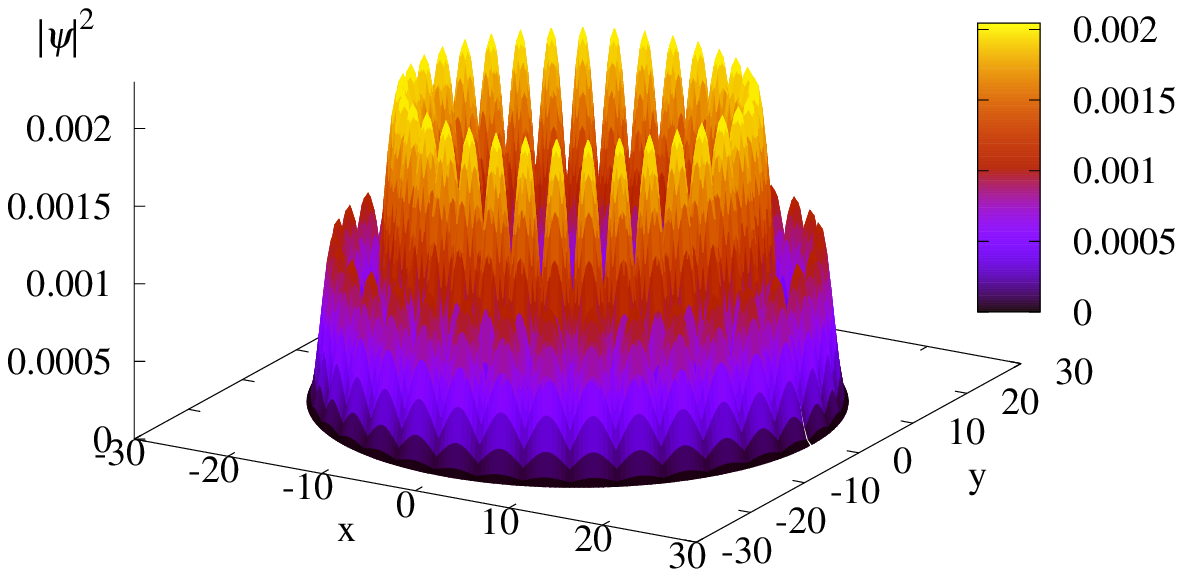}
\includegraphics[width=0.49\textwidth]{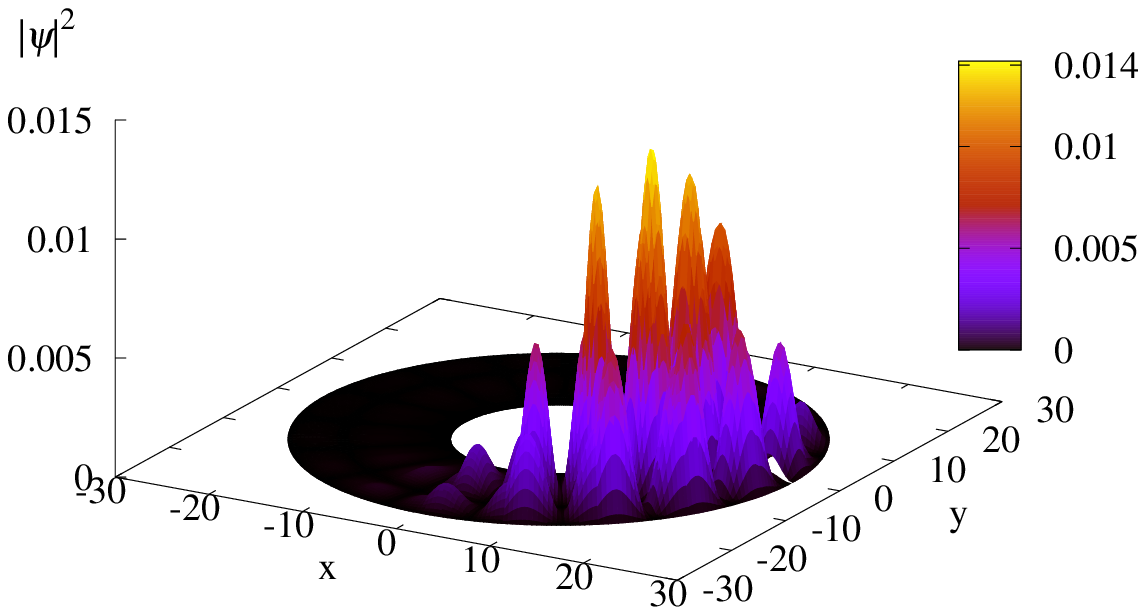}\\
\includegraphics[width=0.49\textwidth]{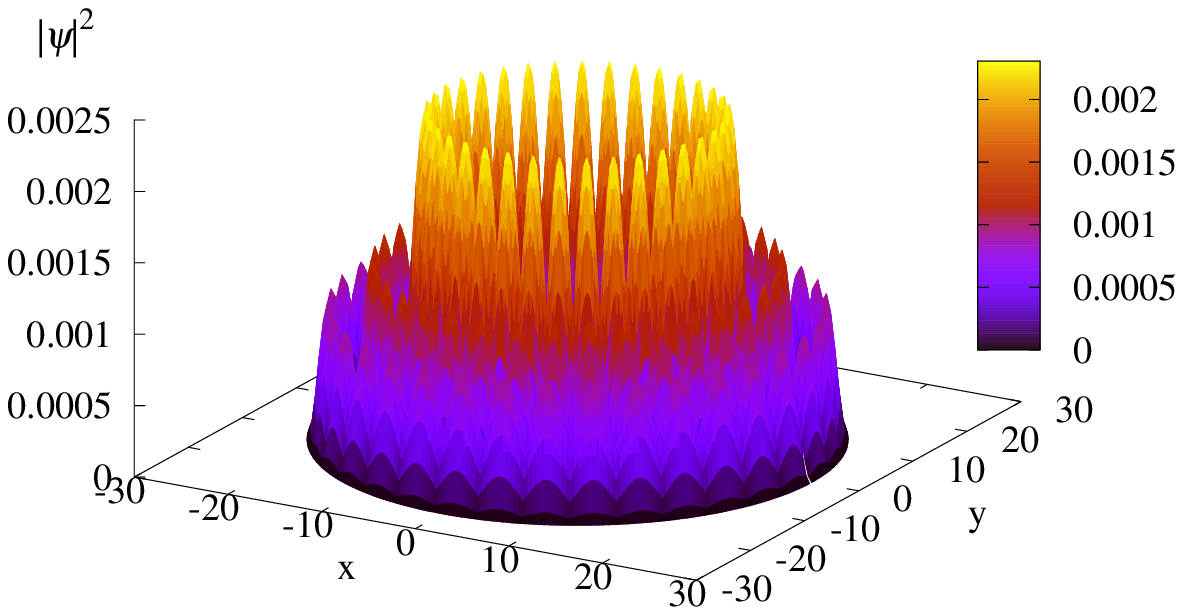}
\includegraphics[width=0.49\textwidth]{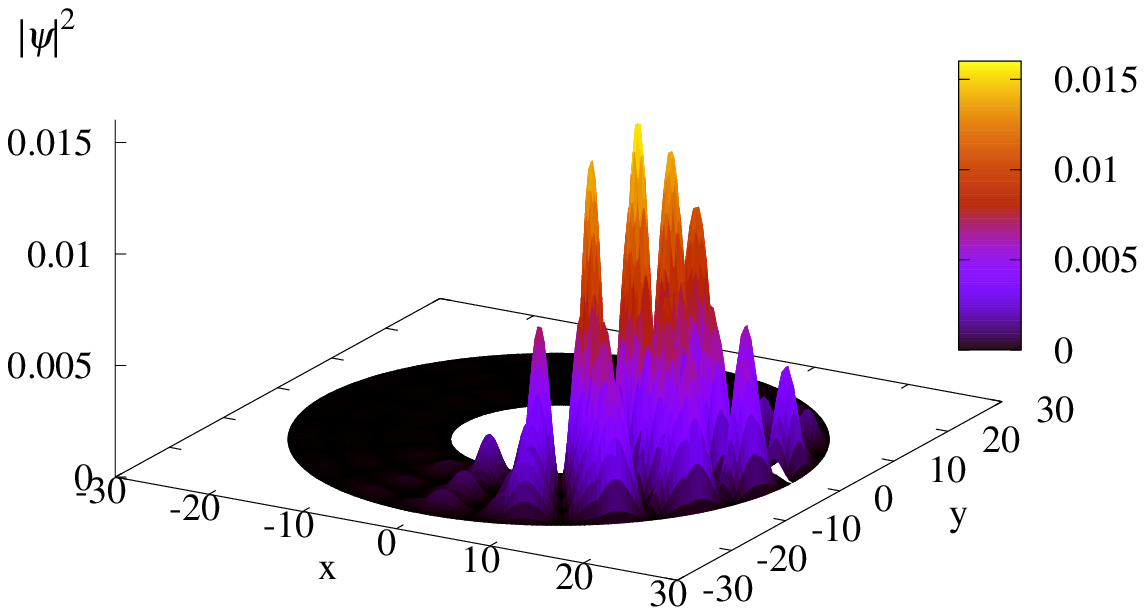}
\caption{Squared modulus $|\psi_{q,s}|^{2}$ of the wavefunction with
quantum numbers $q \simeq 11.04$ and (from top to bottom): $s=1,2,3$.
Left panels: ordered system. Right panels: correlated disordered system.}
\label{wavefun_s}
\end{center}
\end{figure}

To conclude this section, in figure~\ref{wavefun_s}
we show a graphical representation of the disorder-induced
localisation of the complete wavefunctions.
We consider the wavefunctions with quantum numbers $q \simeq 11.04$ and
$s = 1,2,3$; we show the extended wavefunctions in the subfigures on the left
and their localised counterparts in the right subfigures.
The figure~\ref{wavefun_s} complements the
information for the angular part of the wavefunction provided by
figures~\ref{bloch_wave} and~\ref{localised_wave} with the information relative
to the radial part.
The wavefunctions in the three figures differ for the values of the radial
wavenumber~$s$. Note that $s$ is equal to the number of nodes of the radial
part of the wavefunction; therefore the square modulus of the wavefunction has
$s$ circular ridges, whose height decreases for increasing values of the
radial coordinate.

\clearpage
\section{Disorder with radial symmetry}
\label{radial_disorder}

In this section we investigate the second model, with disorder that depends
on the radial coordinate.
This part of the paper is focused on a different 2D problem, namely, a quantum
particle moving on a plane under the influence of a central random potential.

\subsection{Model with radius-dependent disorder}
\label{centr_pot}

For a central random potential $U(r,\theta)=U(r)$ the stationary Schr\"{o}dinger
equation~(\ref{2dschr}) is given by
\begin{equation}
\left[ - \nabla^{2} + U(r) \right] \psi(r,\theta) = E \psi(r,\theta)
\label{2dschr_bis}	
\end{equation}
where $\nabla^{2}$ is the 2D Laplacian~(\ref{2dlap}).
Specifically, we consider a potential of the form
\begin{equation}
U(r) = \sum_{n=1}^{N} U_{n} \delta \left( r - r_{n} \right) .
\label{radpot}	
\end{equation}
We assume that both the positions and the strengths of the
$\delta$-barriers are random variables; the strengths $U_{n}$
are defined by~(\ref{compdis}) with mean value $U=\langle U_n\rangle$
as in the previous model, while the radial positions are given by
\begin{equation}
r_{n} = r_{1} + a(n-1) + a_{n} .
\label{radpos}	
\end{equation}
In~(\ref{radpos}), $a$ represents the average radial distance
between two neighbouring barriers, while $a_{n}$ is the random
radial displacement of the $n$-th barrier from its ``lattice''
position $r_{n}^{(0)} = r_{1} + a (n-1)$. Is it assumed that all
$r_n>0$.
We consider the case of weak disorder, defined by the conditions
\begin{equation}
\begin{array}{cccc}
\langle u_{n}^{2} \rangle \ll U^{2}, &
\langle \Delta_{n}^{2} \rangle E \ll 1, &
\mbox{ and } &
\langle \Delta_{n}^{2} \rangle U \ll 1 ,
\end{array}
\label{weak_rad_dis}
\end{equation}
which coincide with those set by~(\ref{weakdis}) with the substitution
$q^{2} \to E$ and  where the symbol $\Delta_{n}$ now stands for
\begin{displaymath}
\Delta_{n} = a_{n+1} - a_{n} .	
\end{displaymath}

After separating the radial and the angular variables via~(\ref{var_sep}), one obtains that the two parts of the wavefunction
obey the equations
\begin{equation}
\frac{d^{2} \Theta}{d \theta^{2}} + l^{2} \Theta = 0
\label{ang_eq}
\end{equation}
and
\begin{equation}
\frac{d^{2} R}{d r^{2}} + \frac{1}{r} \frac{d R}{d r} + \left[ E -
\sum_{n=1}^{N} U_{n} \delta( r - r_{n}) - \frac{l^{2}}{r^{2}} \right] R
= 0.
\label{rad_eq}
\end{equation}
(\ref{ang_eq}) has solutions
\begin{displaymath}
\Theta_{l}(\theta) = \frac{1}{\sqrt{2 \pi}} e^{i l \theta}	
\end{displaymath}
with $l$ integer.
The functions $\Theta_{l}(\theta)$ are eigenfunctions of the
angular momentum $L = - i \partial/\partial \theta$, which
commutes with the Hamiltonian because of the circular symmetry
of the model.

We are interested in two different but related problems: the
localisation of the quantum states in the model~(\ref{rad_eq})
with an infinite number of circular barriers ($N \to \infty$) and the
transmission properties of a random annulus with a finite number of
barriers.
To analyse these problems we first discuss in Sec.~\ref{2dkp} under
what circumstances a 2D model with circular symmetry can be reduced
to its 1D counterpart.
We then consider the localisation and transmission properties in
Secs.~\ref{2d_lyap} and~\ref{2d_transmission}.

\subsection{The 2D Kronig-Penney model and its 1D analogue}
\label{2dkp}

When $N \to \infty$, the model defined by~(\ref{rad_eq}) represents
a 2D variant of the 1D Kronig-Penney model~(\ref{kpmodel}), with
Schr\"{o}dinger equation
\begin{equation}
\frac{d^{2} R}{d r^{2}} + \frac{1}{r} \frac{d R}{d r} + \left[ E -
\sum_{n=1}^{\infty} U_{n} \delta( r - r_{n}) - \frac{l^{2}}{r^{2}} \right] R
= 0.
\label{rad_eq_infty}
\end{equation}
The analogy with the 1D model~(\ref{kpmodel}) is easier to see if the radial
wavefunction is expressed as
\begin{displaymath}
R(r) = \frac{X(r)}{\sqrt{r}}.
\end{displaymath}
(\ref{rad_eq_infty}) can then be written in the new form
\begin{equation}
\frac{d^{2} X}{d r^{2}} + \left[ E -
\sum_{n=1}^{\infty} U_{n} \delta( r - r_{n}) - \frac{l^{2} - 1/4}{r^{2}}
\right] X = 0.
\label{rad_eq_2}
\end{equation}
Replacing the symbol $r$ for the radial coordinate with $t$ allows one to
interpret~(\ref{rad_eq_2}) as the equation of motion of a classical
dynamical system
\begin{equation}
\ddot{X} + \left[ E -
\sum_{n=1}^{\infty} U_{n} \delta( t - t_{n}) - \frac{l^{2} - 1/4}{t^{2}}
\right] X = 0.
\label{stoc_osc}
\end{equation}
More specifically, (\ref{stoc_osc}) describes the dynamics of a
parametric oscillator whose frequency varies in time under the
influence of two different terms: a deterministic term, due to the
centrifugal potential, which lowers the frequency and tends to vanish
over long time scales, and a stochastic term, which produces a sequence
of abrupt variations of the momentum (``kicks'').

To analyse the dynamics of the stochastic oscillator~(\ref{stoc_osc}), it
is convenient to write the equation of motion in Hamiltonian form.
After introducing the momentum $P = \dot{X}$, one can write~(\ref{stoc_osc})
as
\begin{equation}
\begin{array}{ccl}
\dot{P} & = & \displaystyle
- \left[ E - \frac{l^{2} - 1/4}{t^{2}} -
\sum_{n=1}^{\infty} U_{n} \delta (t - t_{n}) \right] X \\
\dot{X} & = & P \\
\end{array} .
\label{ham_eq}
\end{equation}
If the dynamical equations~(\ref{ham_eq}) are integrated over the interval
$[t_{n}^{-},t_{n+1}^{-}]$ between two kicks, one obtains the Hamiltonian
map
\begin{displaymath}
\left( \begin{array}{c}
X_{n+1} \\
P_{n+1} \\
\end{array} \right) =
{\bf T}_{n}
\left( \begin{array}{c}
X_{n} \\
P_{n} \\
\end{array} \right)
\end{displaymath}
with
\begin{equation}\fl
\begin{array}{ccl}
({\bf T}_{n})_{11} & = & \displaystyle
\frac{\pi}{2} \left\{ \sqrt{E t_{n+1} t_{n}} \left[
J_{l}(\sqrt{E} t_{n+1}) Y_{l}^{\prime}(\sqrt{E} t_{n}) -
Y_{l}(\sqrt{E} t_{n+1}) J_{l}^{\prime}(\sqrt{E} t_{n}) \right] \right. \\
& + & \displaystyle
\frac{1}{2} \sqrt{\frac{t_{n+1}}{t_{n}}}
\left[ J_{l}(\sqrt{E} t_{n+1}) Y_{l}(\sqrt{E} t_{n}) -
Y_{l}(\sqrt{E} t_{n+1}) J_{l}(\sqrt{E} t_{n}) \right] \\
& + & \displaystyle
\left. U_{n} \sqrt{t_{n+1}t_{n}} \left[
Y_{l}(\sqrt{E} t_{n+1}) J_{l}(\sqrt{E} t_{n}) -
J_{l}(\sqrt{E} t_{n+1}) Y_{l}(\sqrt{E} t_{n}) \right] \right\} \\
& & \\
({\bf T}_{n})_{12} & = & \displaystyle
\frac{\pi}{2} \sqrt{t_{n+1} t_{n}}
\left[ Y_{l}(\sqrt{E} t_{n+1}) J_{l}(\sqrt{E} t_{n}) -
J_{l}(\sqrt{E} t_{n+1}) Y_{l}(\sqrt{E} t_{n}) \right] \\
& & \\
({\bf T}_{n})_{21} & = & \displaystyle
\frac{\pi}{2} \left\{ E \sqrt{t_{n+1} t_{n}} \left[
J_{l}^{\prime}(\sqrt{E} t_{n+1}) Y_{l}^{\prime}(\sqrt{E} t_{n}) -
Y_{l}^{\prime}(\sqrt{E} t_{n+1}) J_{l}^{\prime}(\sqrt{E} t_{n})
\right] \right. \\
& + & \displaystyle
\frac{1}{2} \sqrt{E \frac{t_{n+1}}{t_{n}}}
\left[ J_{l}^{\prime}(\sqrt{E} t_{n+1}) Y_{l}(\sqrt{E} t_{n}) -
Y_{l}^{\prime}(\sqrt{E} t_{n+1}) J_{l}(\sqrt{E} t_{n}) \right] \\
& + & \displaystyle
\frac{1}{2} \sqrt{E \frac{t_{n}}{t_{n+1}}} \left[
J_{l}(\sqrt{E} t_{n+1}) Y_{l}^{\prime}(\sqrt{E} t_{n}) -
Y_{l}(\sqrt{E} t_{n+1}) J_{l}^{\prime}(\sqrt{E} t_{n}) \right] \\
& + & \displaystyle
\frac{1}{4\sqrt{t_{n+1} t_{n}}} \left[
J_{l}(\sqrt{E} t_{n+1}) Y_{l}(\sqrt{E} t_{n}) -
Y_{l}(\sqrt{E} t_{n+1}) J_{l}(\sqrt{E} t_{n}) \right] \\
& + & \displaystyle
U_{n} \sqrt{E t_{n+1}t_{n}}
\left[ Y_{l}^{\prime}(\sqrt{E} t_{n+1}) J_{l}(\sqrt{E} t_{n}) -
J_{l}^{\prime}(\sqrt{E} t_{n+1}) Y_{l}(\sqrt{E} t_{n}) \right] \\
& + & \displaystyle
\left. \frac{1}{2} U_{n} \sqrt{E\frac{t_{n}}{t_{n+1}}}  \left[
Y_{l}(\sqrt{E} t_{n+1}) J_{l}(\sqrt{E} t_{n}) -
J_{l}(\sqrt{E} t_{n+1}) Y_{l}(\sqrt{E} t_{n}) \right] \right\} \\
& & \\
({\bf T}_{n})_{22} & = & \displaystyle
\frac{\pi}{2} \left\{ \sqrt{E t_{n+1} t_{n}} \left[
Y_{l}^{\prime}(\sqrt{E} t_{n+1}) J_{l}(\sqrt{E} t_{n}) -
J_{l}^{\prime}(\sqrt{E} t_{n+1}) Y_{l}(\sqrt{E} t_{n}) \right] \right. \\
& + & \displaystyle
\left. \frac{1}{2} \sqrt{\frac{t_{n}}{t_{n+1}}} \left[
Y_{l}(\sqrt{E} t_{n+1}) J_{l}(\sqrt{E} t_{n}) -
J_{l}(\sqrt{E} t_{n+1}) Y_{l}(\sqrt{E} t_{n}) \right] \right\} \\
\end{array}
\label{t_matrix}
\end{equation}
In~(\ref{t_matrix}) the symbols $J_{l}$ and $Y_{l}$ represent Bessel
functions of the first and second kind.

Furstenberg's theorem~\cite{fur60,fur63,fur71} ensures that the Lyapunov exponent
\begin{equation}
\lambda = \lim_{N \to \infty} \frac{1}{N} \ln \left| \left| \mathbf{T}_{N}
\cdots \mathbf{T}_{1} \mathbf{v}_{0} \right| \right|
\label{furst}
\end{equation} 
exists with probability 1 and assumes a positive value, independent of the
disorder realisation, for every initial condition
\begin{displaymath}
\mathbf{v}_{0}~=~(X_{0},P_{0})~\neq~(0,0) .
\end{displaymath}
On the other hand, Borland's conjecture allows one to identify
the Lyapunov exponent~(\ref{furst}) with the inverse of the localisation
length $l_{\mathrm{loc}}$ for the circular Kronig-Penney
model~(\ref{rad_eq})~\cite{bor63}.
When considering the right-hand side of~(\ref{furst}), it is important to
observe that the limit is determined by the behaviour of the evolution
matrix~(\ref{t_matrix}) for $n \gg 1$. This can be seen as follows.
Let $\bar{N}$ be a large but finite value of the index $n$.
The value of $\lambda$ in~(\ref{furst}) does not depend on the election of
the initial vector $\mathbf{v}_{0}$; it is therefore possible to consider a
vector
\begin{equation}
\mathbf{v}_{0} = \mathbf{T}_{1}^{-1} \mathbf{T}_{2}^{-1} \cdots
\mathbf{T}_{\bar{N}}^{-1} \mathbf{w}_{0} ,
\label{v0}
\end{equation}
with arbitrary $\mathbf{w}_{0} \neq \mathbf{0}$.
Substituting the vector~(\ref{v0}) in~(\ref{furst}), one obtains that the
Lyapunov exponent can be obtained with the limit
\begin{equation}
\lambda = \lim_{N \to \infty} \frac{1}{N} \ln \left| \left| \mathbf{T}_{N}
\cdots \mathbf{T}_{\bar{N}} \mathbf{w}_{0} \right| \right|
\label{furst2}
\end{equation}
with $\mathbf{w}_{0}$ selected at will and $\bar{N}$ arbitrarily large, although
finite.

With these considerations in mind, we can expand the right-hand side
of~(\ref{t_matrix}) in powers of $1/t_{n}$ and obtain
\begin{equation}
{\mathbf T}_{n} = {\mathbf T}_{n}^{(0)} -
\frac{1}{t_{n}^{2}} {\mathbf T}_{n}^{(2)} + \ldots
\label{t_exp}
\end{equation}
The terms in the right-hand side of~(\ref{t_exp}) can be further expanded
in the disorder strength, which can be measured with the parameter
\begin{displaymath}
\sigma = \sqrt{ \left\langle \left( \frac{u_{n}}{\sqrt{E}} + U \Delta_{n} \right)^{2}
\right\rangle}	.
\end{displaymath}
In this way one obtains a double expansion in powers of $\sigma$ and $1/t_{n}$,
\begin{equation}
\mathbf{T}_{n} =
\sum_{k,p} \sigma^{k} \frac{1}{t_{n}^{p}} {\mathbf{T}}_{n}^{(k,p)} =
{\mathbf{T}}_{n}^{(0,0)} + \sigma {\mathbf{T}}_{n}^{(1,0)} +
\sigma^{2} {\mathbf{T}}_{n}^{(2,0)} + \ldots -
\frac{1}{t_{n}^{2}} {\mathbf{T}}_{n}^{(0,2)} + \ldots
\label{double_exp}
\end{equation}
The matrices in the right-hand side of the expansions~(\ref{t_exp})
and(\ref{double_exp}) can be worked out and are given in~\ref{expansion_matrices};
their explicit form, however, is not particularly relevant for the present
considerations. The key point is that expansion~(\ref{double_exp}) makes it possible
to conclude that the contribution of the centrifugal potential is asymptotically
``drowned'' by the noisy terms. More precisely, the random terms overshadow
the centrifugal ones as soon as
\begin{equation}
t_{n} \gtrsim \frac{l}{\sqrt{\sigma}} .
\label{asympt_cond}
\end{equation}

We can now select the integer $\bar{N}$ in~(\ref{furst2}) in such a
way that the condition~(\ref{asympt_cond}) is fulfilled for all the
matrices on the right-hand side of~(\ref{furst2}).
This leads to the important result that the localisation length in
the model~(\ref{rad_eq_infty}) {\em does not depend} on the centrifugal
potential.
This might have been guessed from a direct examination of~(\ref{rad_eq_infty}),
but the fact that the random potential is a succession of $\delta$-barriers makes
it somewhat tricky to estimate quantitatively the impact of the random part of the
potential with respect to the centrifugal part. This difficulty is overcome with
the use of the transfer matrix approach adopted in this section.

After dropping the second term in the right-hand side of the
expansion~(\ref{t_exp}), one is left with the evolution matrix
$\mathbf{T}_{n}^{(0)}$, which has the {\em same form} of the evolution
matrix for the 1D Kronig-Penney model~(\ref{kpmodel})~\cite{her08}.
We are thus led to conclude that the 2D model~(\ref{rad_eq_infty}) and its
1D counterpart~(\ref{kpmodel}) share essential features, such as the
band structure and the localisation length.
This important result is not restricted to 2D models with random
potentials of the form~(\ref{radpot}), but is actually valid for any
central random potential which falls off as $U(r) \sim 1/r^{\alpha}$
with $\alpha < 2$. Conversely, if the central potential decays faster
than the centrifugal potential, the latter cannot be neglected in
the computation of the Lyapunov exponent, and the result need not
be the same as in the 1D case.

To sum up, we have reached the essential result that, in the weak disorder
case, the Lyapunov exponent for the 2D model~(\ref{rad_eq_infty}) is given by
the 1D formula~(\ref{invloc}), with the substitution of $a$ for $\alpha$ and
where $q$ must now be interpreted as the square root of the energy $E$
rather than as an angular quantum number.
In the next section, after discussing how to compute the Lyapunov exponent,
we shall see that the analytical formula~(\ref{invloc}) matches well the
numerically-obtained values of the Lyapunov exponent, thus corroborating
the conclusions of this section.

\subsection{The Lyapunov exponent}
\label{2d_lyap}

The Lyapunov exponent of the model~(\ref{rad_eq_infty}) can be computed
using the transfer matrix technique. It is convenient to put the transfer
matrix in the running-wave representation, rather than in the
standing-wave representation used in the previous section.
For this reason, we write the solution of~(\ref{rad_eq_infty}) in
the potential wells in terms of Hankel functions of the
first and second kind. We assume that the particle is in a state of
angular momentum $l$; this defines the order of the Hankel functions.
Using the symbols $A_{n}$ and $B_{n}$ for the amplitudes of the
outward- and inward-directed waves, we can write the radial wavefunction
within the $n$-th potential well as
\begin{equation}
\begin{array}{ccc}
R(r) = A_{n} H_{l}^{(1)}(\sqrt{E} r) + B_{n} H_{l}^{(2)}(\sqrt{E} r) &
\mbox{ for } & r_{n} < r < r_{n+1}	.
\end{array}
\label{nth_well}
\end{equation}

By integrating (\ref{rad_eq_infty}) across the $n$-th barrier, one
can connect the amplitudes of the waves in the $(n-1)$-th and $n$-th wells.
One obtains the map
\begin{equation}
\left( \begin{array}{c}
A_{n} \\
B_{n} \\
\end{array} \right) =
\mathbf{M}_{n}
\left( \begin{array}{c}
A_{n-1} \\
B_{n-1} \\
\end{array} \right) ,
\label{running_wave_tmatrix}
\end{equation}
where $\mathbf{M}_{n}$ is the transfer matrix in the running-wave
representation with elements
\begin{equation}
\begin{array}{ccl}
(\mathbf{M}_{n})_{11} & = & \displaystyle
1 - i \frac{\pi}{4} r_{n} U_{n} H_{l}^{(1)}(\sqrt{E} r_{n})
H_{l}^{(2)}(\sqrt{E} r_{n}) ,\\
(\mathbf{M}_{n})_{12} & = & \displaystyle
-i \frac{\pi}{4} r_{n} U_{n}
\left[ H_{l}^{(2)}(\sqrt{E} r_{n}) \right]^{2} ,\\
(\mathbf{M}_{n})_{21} & = & \displaystyle
i \frac{\pi}{4} r_{n} U_{n}
\left[ H_{l}^{(1)}(\sqrt{E} r_{n}) \right]^{2} ,\\
(\mathbf{M}_{n})_{22} & = & \displaystyle
1 + i \frac{\pi}{4} r_{n} U_{n} H_{l}^{(1)}(\sqrt{E} r_{n})
H_{l}^{(2)}(\sqrt{E} r_{n}) .\\
\end{array}	
\label{m_matrix}
\end{equation}
Note that $\det \mathbf{M}_{n} = 1$.

The Lyapunov exponent can be defined as~\cite{mar08}
\begin{equation}
\lambda = \lim_{N \to \infty} \frac{1}{N} \sum_{n=1}^{N} \ln |R_{n}| ,
\label{num_lyap}
\end{equation}
where $R_{n}$ is the (complex) ratio of the amplitudes of the outgoing wave
in the $(n-1)$-th and $n$-th wells,
\begin{equation}
R_{n} = \frac{A_{n}}{A_{n-1}} .	
\label{ratio_n}
\end{equation}
The ratio~(\ref{ratio_n}) obeys the recursive relation
\begin{equation}\fl
\begin{array}{ccl}
R_{n} & = & \displaystyle
1 - i \frac{\pi}{4} r_{n} U_{n}
H_{l}^{(1)}(\sqrt{E} r_{n}) H_{l}^{(2)}(\sqrt{E} r_{n}) +
\frac{r_{n} U_{n} \left[ H_{l}^{(2)} (\sqrt{E} r_{n}) \right]^{2}}
{r_{n-1} U_{n-1} \left[ H_{l}^{(2)} (\sqrt{E} r_{n-1}) \right]^{2}} \\
& \times & \displaystyle
\left\{ 1 + i \frac{\pi}{4} r_{n-1} U_{n-1} H_{l}^{(1)}(\sqrt{E} r_{n-1})
H_{l}^{(2)}(\sqrt{E} r_{n-1}) - \frac{1}{R_{n-1}}  \right\} \\
\end{array}
\label{r_map}	
\end{equation}
which is obtained after eliminating the amplitudes $B_{n}$ from the
map~(\ref{running_wave_tmatrix}).
Eqs.~(\ref{num_lyap}) and~(\ref{r_map}) make possible an efficient numerical
computation of the inverse localisation length $\lambda$.

\begin{figure}
\begin{center}
\includegraphics[width=0.49\textwidth]{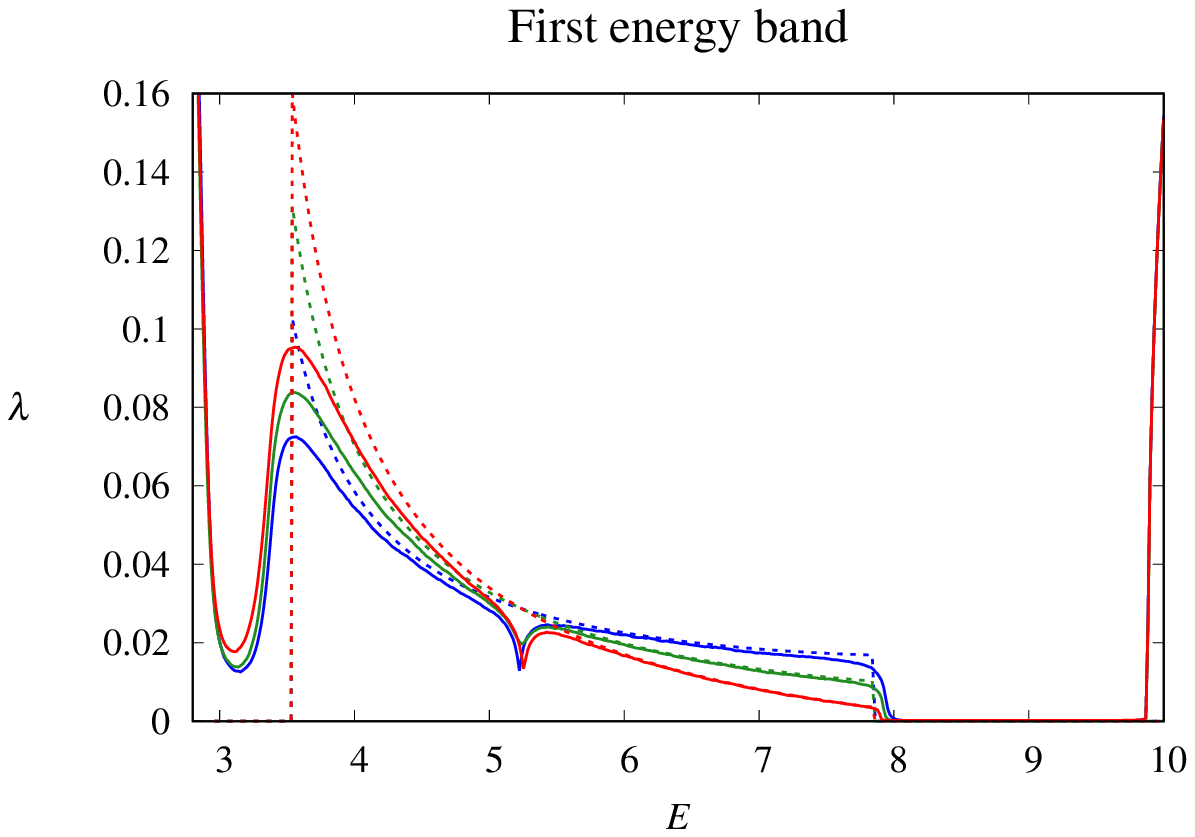}
\includegraphics[width=0.49\textwidth]{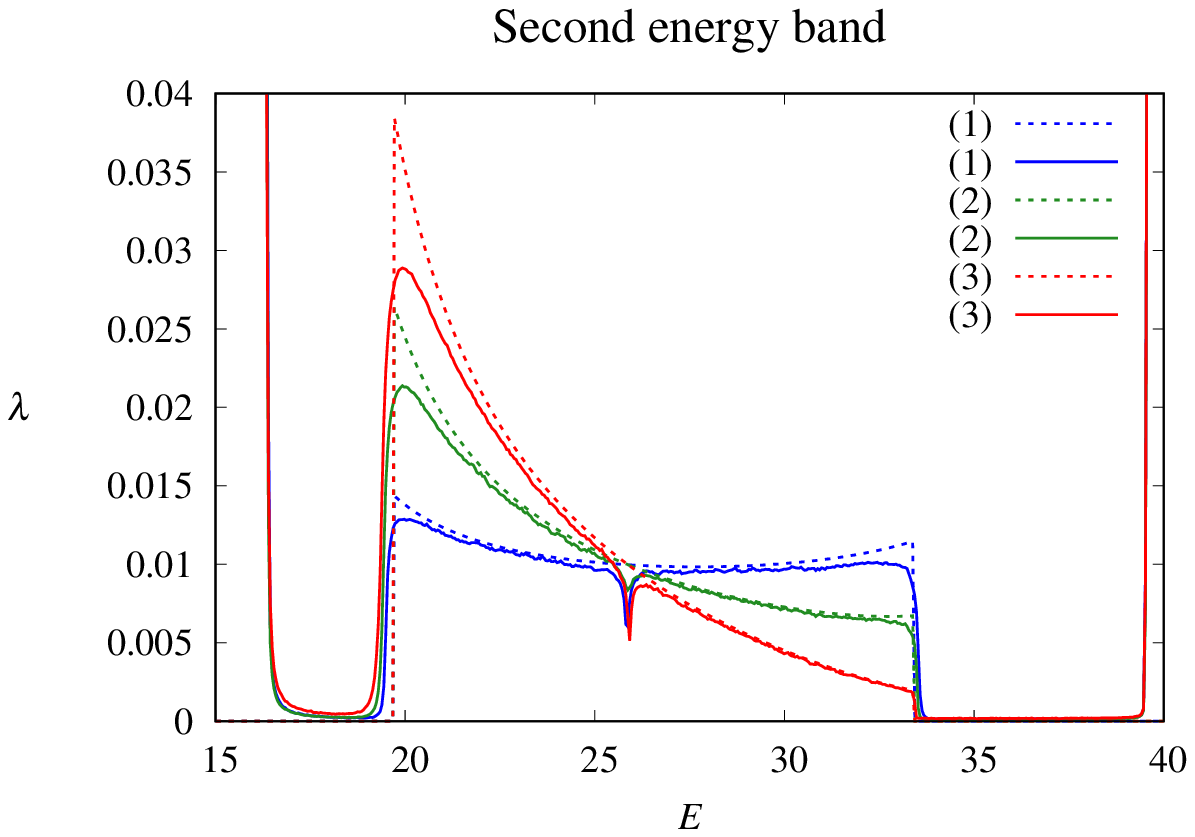}\\
\includegraphics[width=0.49\textwidth]{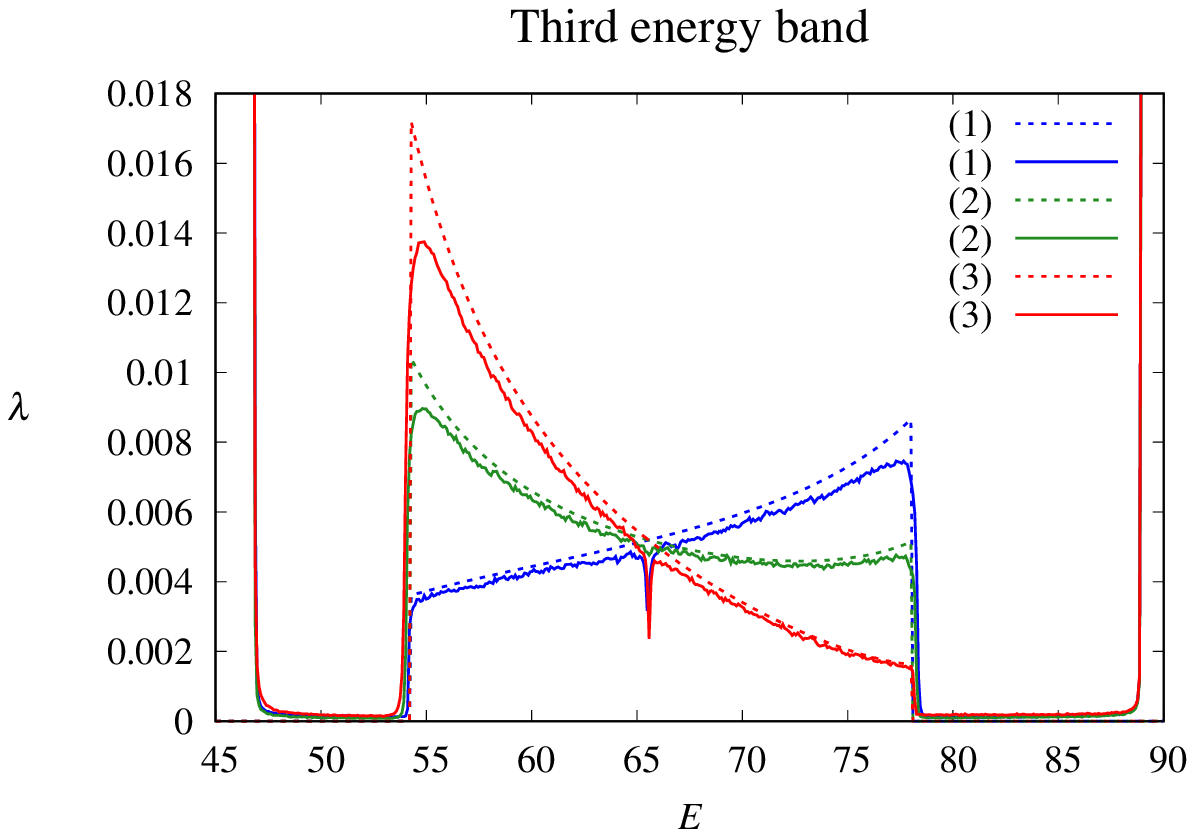}
\includegraphics[width=0.49\textwidth]{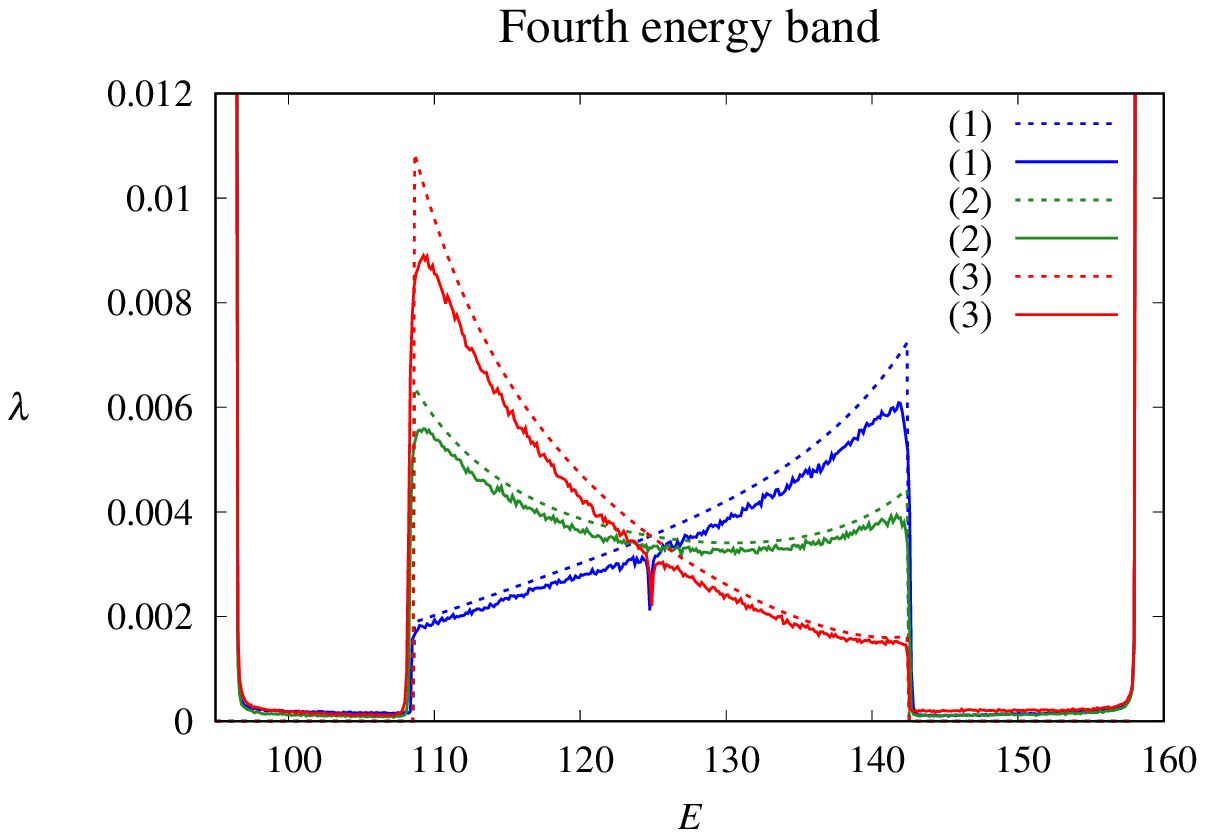}\\
\caption{Lyapunov exponent $\lambda$ (in $\mbox{cm}^{-1}$) versus energy $E$ (in
$\mbox{cm}^{-2}$) for the first 4 bands.
The dotted lines correspond to the analytical expression~(\ref{invloc}); the solid lines
represent the numerical values obtained using Eqs.~(\ref{num_lyap}) and~(\ref{r_map}).
The data were obtained for the 2D Kronig-Penney model~(\ref{rad_eq_infty}) with the
following parameters: $U=4.0\,\mbox{cm}^{-1}$, $r_{1} = 5\,\mbox{cm}$, and $a = 1\,\mbox{cm}$.
Disorder strength: $\sqrt{\langle u_n^2\rangle} = 1.0\,\mbox{cm}^{-1}$ and
$\sqrt{\langle \Delta_n^2\rangle} = 0.02\,\mbox{cm}$.
The mobility edges were set at $k_{1} = \pi/4$ and $k_{2} = \pi/2$. The data labelled with
1 (blue), 2 (green), and 3 (red) correspond to the cases of positive, absent, and negative
cross-correlations.
The rise of the Lyapunov exponent at the lower and upper parts of the windows corresponds
to the band structure of the periodic system.
\label{radial_lyap}}
\end{center}
\end{figure}

\begin{figure}
\begin{center}
\includegraphics[width=5in]{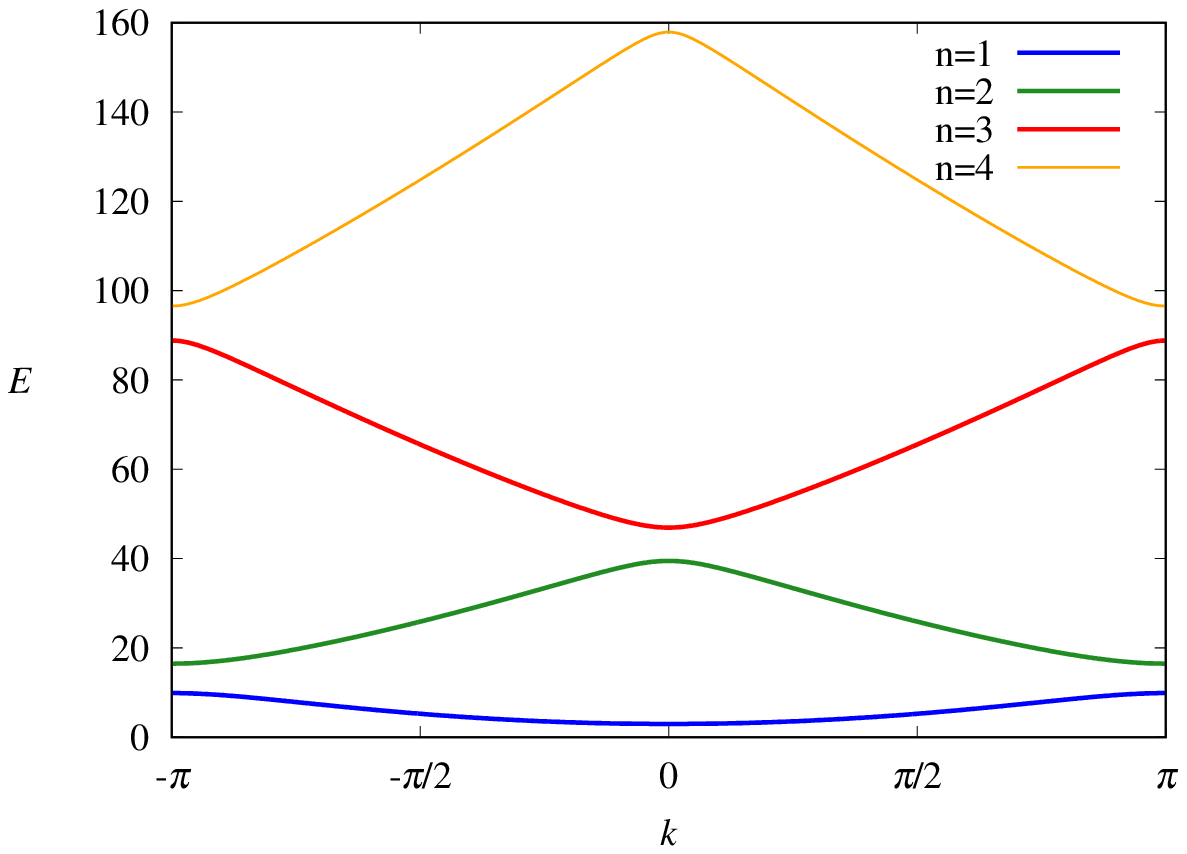}
\caption{Energy $E$ (in $\mbox{cm}^{-2}$) versus the Bloch vector $k$ (dimensionless).
The first four bands are represented, with the band index taking the values $n = 1,2,3,4$.
The data were obtained for $U=4.0\,\mbox{cm}^{-1}$ and $a = 1\,\mbox{cm}$.
\label{rad_spectrum}}
\end{center}
\end{figure}

In figure~\ref{radial_lyap} we compare the numerical values of the Lyapunov
exponent~(\ref{num_lyap}) with the analytical expression~(\ref{invloc}).
The represented data were obtained for the 2D Kronig-Penney model~(\ref{rad_eq_infty})
with $U=4.0\,\mbox{cm}^{-1}$, $r_{1} = 5\,\mbox{cm}$, and $a = 1\,\mbox{cm}$.
The values of the disorder strength are $\sqrt{\langle u_n^2\rangle}~=~1.0\,\mbox{cm}^{-1}$
and $\sqrt{\langle \Delta_n^2\rangle}~=~0.02\,\mbox{cm}$.
The mobility edges were set at $k_{1} = \pi/4$ and $k_{2} = \pi/2$.
Note that self-correlations of the disorder are responsible for the effective mobility
edges; cross-correlations between structural and compositional disorder, on the
other hand, allows one to shift the region of strongest localisation towards the
higher or lower energies of the localisation window.
We considered a particle with angular momentum $l=0$, although this choice is not
especially relevant, as discussed below.
As can be seen from figure~\ref{radial_lyap}, the numerical data generally match the
theoretical predictions for the Lyapunov exponent in a satisfactory way.
A discrepancy occurs for $k = \pi/2$, where the Kronig-Penney model exhibits an anomaly
which is the equivalent of the band-centre anomaly in the Anderson model, and where the
expression~(\ref{invloc}) for the Lyapunov exponent fails as Thouless formula does
at the band centre~\cite{her10b}.
The band structure also corresponds to the theoretical expectations: this can be seen by
considering that the points where the numerically computed Lyapunov exponent starts a rapid
increase coincide with the band limits obtained by solving (\ref{rotangle}) for the
1D Kronig-Penney model~(\ref{kpmodel}).
We show the band structure derived from~(\ref{rotangle}) in figure~\ref{rad_spectrum}.

\begin{figure}
\begin{center}
\includegraphics[width=5in]{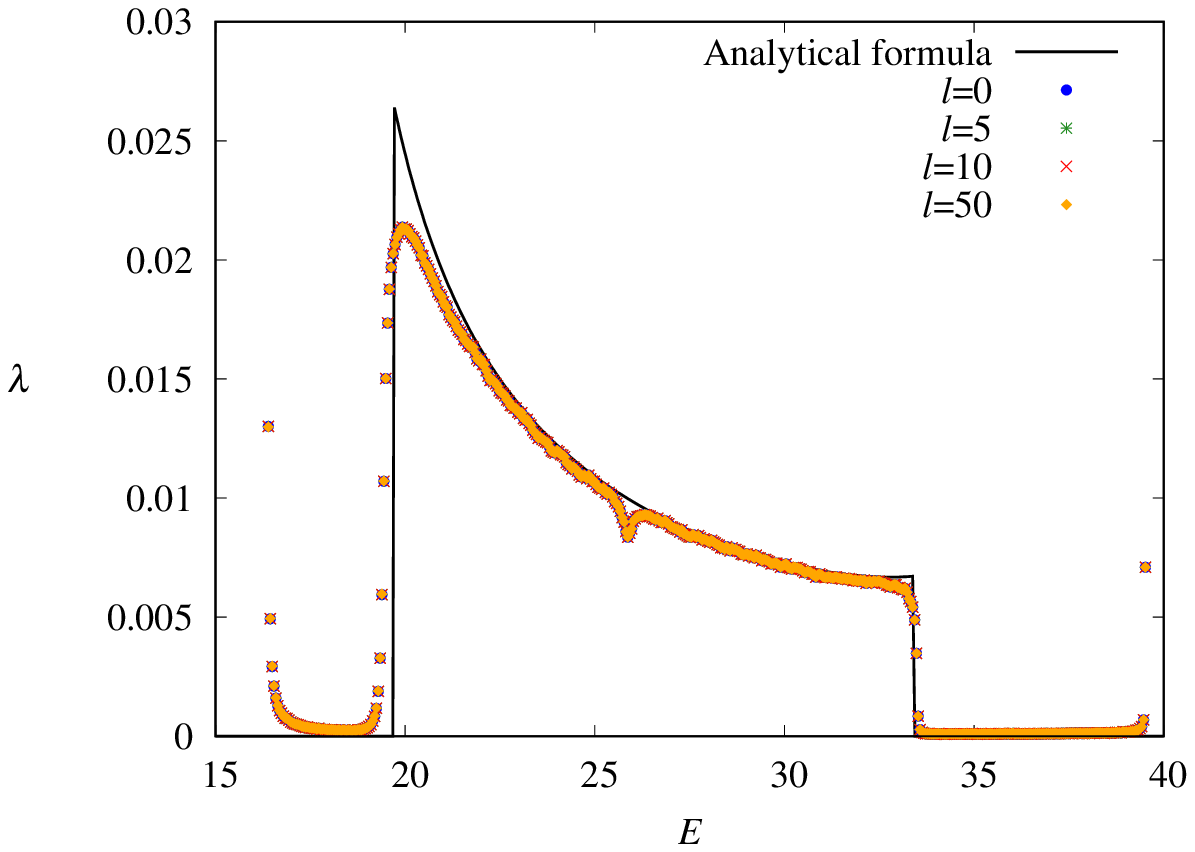}
\caption{Lyapunov exponent $\lambda$ (in $\mbox{cm}^{-1}$) versus energy $E$ (in
$\mbox{cm}^{-2}$) for several values of the angular momentum $l$.
The considered values of the energy correspond to the second energy band.
The data were obtained for the 2D Kronig-Penney model~(\ref{rad_eq_infty}) with the
following parameters: $U=4.0\,\mbox{cm}^{-1}$, $r_{1} = 5\,\mbox{cm}$, and $a = 1\,\mbox{cm}$.
Disorder strength: $\sqrt{\langle u_n^2\rangle} = 1.0\,\mbox{cm}^{-1}$ and
$\sqrt{\langle \Delta_n^2\rangle} = 0.02\,\mbox{cm}$.
The mobility edges were set at $k_{1} = \pi/4$ and $k_{2} = \pi/2$. Structural and
compositional disorder were not cross-correlated.
\label{rad_lyap_l}}
\end{center}
\end{figure}

The data in figure~\ref{radial_lyap} correspond to an {\it s}-wave, but
the results do not change if we consider a particle in a different angular state.
This is shown in figure~\ref{rad_lyap_l}, which represents the numerical values of
the Lyapunov exponent in the second energy band for four values of the angular
momentum ranging from $l=0$ to $l=50$. The data were obtained with the same
values of the parameters used in figure~\ref{radial_lyap}.
The complete collapse of the data corresponding to different values of $l$ confirms
that the Lyapunov exponent does not depend on the angular momentum and shows that the
centrifugal potential plays a negligible role in the localisation of the wavefunction.

In mathematical terms, this irrelevance is confirmed by considering the
behaviour of the ratio $R_{n}$ in the limit of large $n$.
In fact, for $n \gg 1$ one can use the asymptotic expansions
for the Hankel functions
\begin{displaymath}
\begin{array}{ccl}
H_{l}^{(1)}(z) & = & \displaystyle
\sqrt{\frac{2}{\pi z}} e^{i \left( z - \frac{\pi}{2} l - \frac{\pi}{4} \right)}
\left[ 1 + i \frac{4 l^{2} -1}{8} \frac{1}{z} + \ldots \right] \\
H_{l}^{(2)}(z) & = & \displaystyle
\sqrt{\frac{2}{\pi z}} e^{-i \left( z - \frac{\pi}{2} l - \frac{\pi}{4} \right)}
\left[ 1 - i \frac{4 l^{2} -1}{8} \frac{1}{z} + \ldots \right] \\
\end{array}	
\end{displaymath}
and write the map~(\ref{r_map}) in the simplified form
\begin{equation}
R_{n} \simeq - \frac{U_{n}}{U_{n-1}} e^{-i2\sqrt{E} \Delta_{n-1}} \frac{1}{R_{n-1}}
+ 1 - \frac{i}{2} \frac{U_{n}}{q} + \frac{U_{n}}{U_{n-1}} \left( 1 + \frac{i}{2}
\frac{U_{n-1}}{\sqrt{E}} \right) ,
\label{asymp_r_map}
\end{equation}
which does not show any dependence on the value of the angular momentum.

The asymptotic map~(\ref{asymp_r_map}) can also be used to confirm that the 1D
model~(\ref{kpmodel}) and in its 2D counterpart~(\ref{rad_eq_infty}) have the
same band structure.
To see this point, let us consider an eigenstate of the Schr\"{o}dinger
equation~(\ref{2dschr_bis}) in the $n$-th potential well having the form
\begin{displaymath}
\begin{array}{ccc}
\displaystyle
\psi_{\mathrm{out}} (r, \theta) = A_{n} H_{l}^{(1)} \left( \sqrt{E} r \right)
e^{il \theta} & \mbox{ for } & r_{n} < r < r_{n+1} .
\end{array}
\end{displaymath}
The amplitude ratio $R_{n}$ can then be written
\begin{displaymath}
R_{n} = \frac{A_{n}}{A_{n-1}} = \frac{\psi_{\mathrm{out}}(r_{n}, \theta)}
{\psi_{\mathrm{out}}(r_{n-1}, \theta)}
\frac{H_{l}^{(1)}(\sqrt{E} r_{n-1})}{H_{l}^{(1)}(\sqrt{E} r_{n})} .
\end{displaymath}
For $n \gg 1$ one can use the asymptotic expansion of the Hankel function and
write
\begin{equation}
R_{n} \simeq \frac{\psi_{\mathrm{out}}(r_{n}, \theta)}
{\psi_{\mathrm{out}}(r_{n-1}, \theta)} e^{-i \sqrt{E} \Delta_{n}}.
\label{ratio_out}
\end{equation}

In the absence of disorder, and assuming that the centrifugal potential can be
neglected for $r_{n} \to \infty$, one can expect that Bloch's theorem applies
and that increasing the radial coordinate of a lattice step $a$ should produce
a change of phase in the wavefunction:
\begin{displaymath}
\frac{\psi^{\mathrm{out}}(r_{n}, \theta)}{\psi_{\mathrm{out}}
(r_{n-1}, \theta)} \simeq e^{ik} ,
\end{displaymath}
where $k$ is the Bloch wavevector.
Substituting this expression in~(\ref{ratio_out}) and taking into account
that without structural disorder $\Delta_{n} = a$, one obtains
\begin{equation}
R_{n} \simeq e^{i \left( k - \sqrt{E} a \right)} .
\label{ratio_nodis}
\end{equation}
On the other hand, in the absence of disorder the asymptotic map~(\ref{asymp_r_map})
reduces to the form
\begin{equation}
R_{n} e^{i \sqrt{E} a} \simeq - \frac{1}{\displaystyle R_{n-1} e^{i \sqrt{E} a}}
+ 2 \cos \left( \sqrt{E} a \right) + \frac{U}{\sqrt{E}} \sin \left( \sqrt{E} a \right).
\label{r_map_nodis}
\end{equation}
Substituting the expression~(\ref{ratio_nodis}) in~(\ref{r_map_nodis}), one
recovers~(\ref{rotangle}) with $q$ and $\alpha$ replaced by $\sqrt{E}$ and $a$.
We have thus reached again the conclusion that the 1D model~(\ref{kpmodel}) and
its 2D equivalent~(\ref{rad_eq}) have the same band structure.

\subsection{The transmission coefficient}
\label{2d_transmission}

\begin{figure}[htb]
\begin{center}
\includegraphics[width=3.5in,height=2.5in]{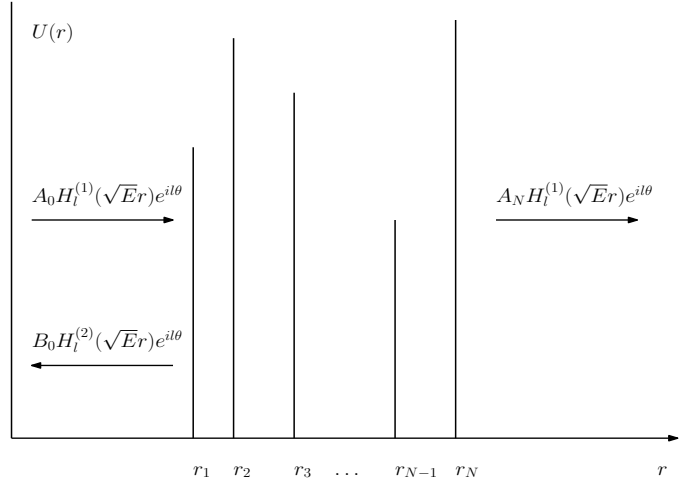}
\caption{Schematic representation of the impinging, reflected, and
transmitted waves.
\label{2dtrans}}
\end{center}
\end{figure}

We now consider the transmission properties of an annulus with a finite
number $N$ of circular barriers. As can be seen from (\ref{rad_eq}) in this
case the plane of motion is divided in three regions: a central disk with
$r < r_{1}$ where the quantum particle is free, the annular region with
$r_{1} \leq r \leq r_{N}$ where the particle is scattered by the $N$
barriers, and the plane beyond the last barrier with $r > r_{N}$, where the
particle is free again.
In the inner disk the solution can be written as a superposition of an
incident and a reflected wave,
\begin{displaymath}
\begin{array}{ccc}
R(r) = A_{0} H_{l}^{(1)}(\sqrt{E} r) + B_{0} H_{l}^{(2)}(\sqrt{E} r) &
\mbox{ for } & r < r_{1} .
\end{array}
\end{displaymath}
while in the outer space only a transmitted wave is present
\begin{displaymath}
\begin{array}{ccc}
R(r) = A_{N} H_{l}^{(1)}(\sqrt{E} r) & \mbox{ for } &
r_{N} < r .
\end{array}
\end{displaymath}
Within the potential wells of the annulus, the solution is given
by~(\ref{nth_well}).
The situation is graphically represented in figure~\ref{2dtrans}.
The amplitudes of the waves on both sides of the annulus can be linked
with the total transfer matrix
\begin{equation}
\mathcal{M}_{N} = \mathbf{M}_{N} \mathbf{M}_{N-1} \cdots \mathbf{M}_{1}
\label{tot_transf_matrix}
\end{equation}
with the individual transfer matrices defined by~(\ref{m_matrix}).
One has
\begin{equation}
\left( \begin{array}{c}
A_{N} \\
0 \\
\end{array} \right) =
\mathcal{M}_{N}
\left( \begin{array}{c}
A_{0} \\
B_{0} \\
\end{array} \right) .
\label{total_transfer}
\end{equation}

We define the transmission coefficient as the ratio of the
transmitted and incident probability currents
\begin{equation}
T_{N} = \frac{\displaystyle \int_{0}^{2 \pi} \left. \left(
\mathbf{j}_{\mathrm{out}} \cdot \hat{r} \right) \right|_{r=r_{N}}
r_{N} \mathrm{d} \theta }
{\displaystyle \int_{0}^{2 \pi} \left. \left( \mathbf{j}_{\mathrm{in}}
\cdot \hat{r} \right) \right|_{r=r_{1}} r_{1} \mathrm{d} \theta }
\label{def_trans_coef}
\end{equation}
with
\begin{displaymath}
\mathbf{j} = -\frac{1}{2m} \left[ \psi^{\ast} \hat{\mathbf{p}} \psi +
\psi \left( \hat{\mathbf{p}} \psi \right)^{\ast} \right] .
\end{displaymath}
In~(\ref{def_trans_coef}) we integrate the probability current
around the inner and outer rings of the annulus to compensate for
the $1/\sqrt{r}$ decrease of the amplitude of the circular waves.
In physical terms, this corresponds to using circular detectors to
measure the intensity of the total impinging and outgoing waves.
Taking into account that the incident and transmitted waves have the form
\begin{displaymath}
\psi_{\mathrm{in}}(r,\theta) = A_{0} H_{l}^{(1)}(\sqrt{E}r) e^{il \theta}
\end{displaymath}
and
\begin{displaymath}
\psi_{\mathrm{out}}(r,\theta) = A_{N} H_{l}^{(1)}(\sqrt{E}r) e^{il \theta} ,
\end{displaymath}
one obtains that the transmission coefficient~(\ref{def_trans_coef}) is
equal to the squared ratio of the amplitudes of the outgoing and incoming
waves
\begin{equation}
T_{N} = \left| \frac{A_{N}}{A_{0}} \right|^{2} .
\label{trans_coef}
\end{equation}
Using~(\ref{total_transfer}) and the fact that $\det \mathcal{M}_{N} = 1$,
one can express the transmission coefficient~(\ref{trans_coef}) as
\begin{equation}
T_{N} = \frac{1}{\left| \left(\mathcal{M}_{N} \right)_{22} \right|^{2}} .
\label{trans_coef_2}
\end{equation}
(\ref{trans_coef_2}) can be used to compute numerically the transmission
coefficient across the random annulus.

From the analytical point of view, knowledge of the inverse localisation
length~(\ref{invloc}) is sufficient to determine the behaviour of the
transmission coefficient in the localised regime: one has
\begin{equation}
\ln T_{N} \simeq - 2 \frac{r_{N} - r_{1}}{l_{\mathrm{loc}}}
\label{loc_reg}
\end{equation}
for $(r_{N} - r_{1}) \gg l_{\mathrm{loc}}$.
In the metallic and intermediate regimes, i.e., for
$(r_{N}~-~r_{1})~\lesssim~l_{\mathrm{loc}}$, the Lyapunov exponent
does not determine the transmission coefficient completely.
However, one can invoke the single-parameter scaling theory which,
in the formulation of Pichard~\cite{Pic86}, postulates that the conductance
$g$ of a 1D system of size $x$ can be expressed as a function
$f(x/l_{\mathrm{loc}})$ of the ratio of the finite sample size to the
localisation length in the infinite model. Pichard showed that the ansatz
$g(x) = f(x/l_{\mathrm{loc}})$ is equivalent to the ordinary formulation
$d \ln g/d\ln L = \beta(g)$ of the SPS hypothesis. His argument can be
transposed without modification to the present case with the width
$r_{N} - r_{1}$ of the annulus replacing the length $x$ of the random
sample (although the specific form of the function $f$ need not be the
same as in the 1D case studied in~\cite{Pic86}).
We can conclude that transmission across the random annulus depends on the
ratio of the ring width $r_{N} - r_{1}$ to the localisation length
$l_{\mathrm{loc}}$, i.e.,
\begin{equation}
T_{N} = F \left( \frac{r_{N} - r_{1}}{l_{\mathrm{loc}}} \right) .
\label{sps_t}
\end{equation}
The single-parameter scaling theory leaves unspecified the form of the
function $F(x)$; however (\ref{sps_t}) still allows one to make
significant predictions: in particular, one can conclude that the
transmission coefficient approaches a unitary value whenever
$\lambda \to 0$.

\begin{figure}
\begin{center}
\includegraphics[width=0.49\textwidth]{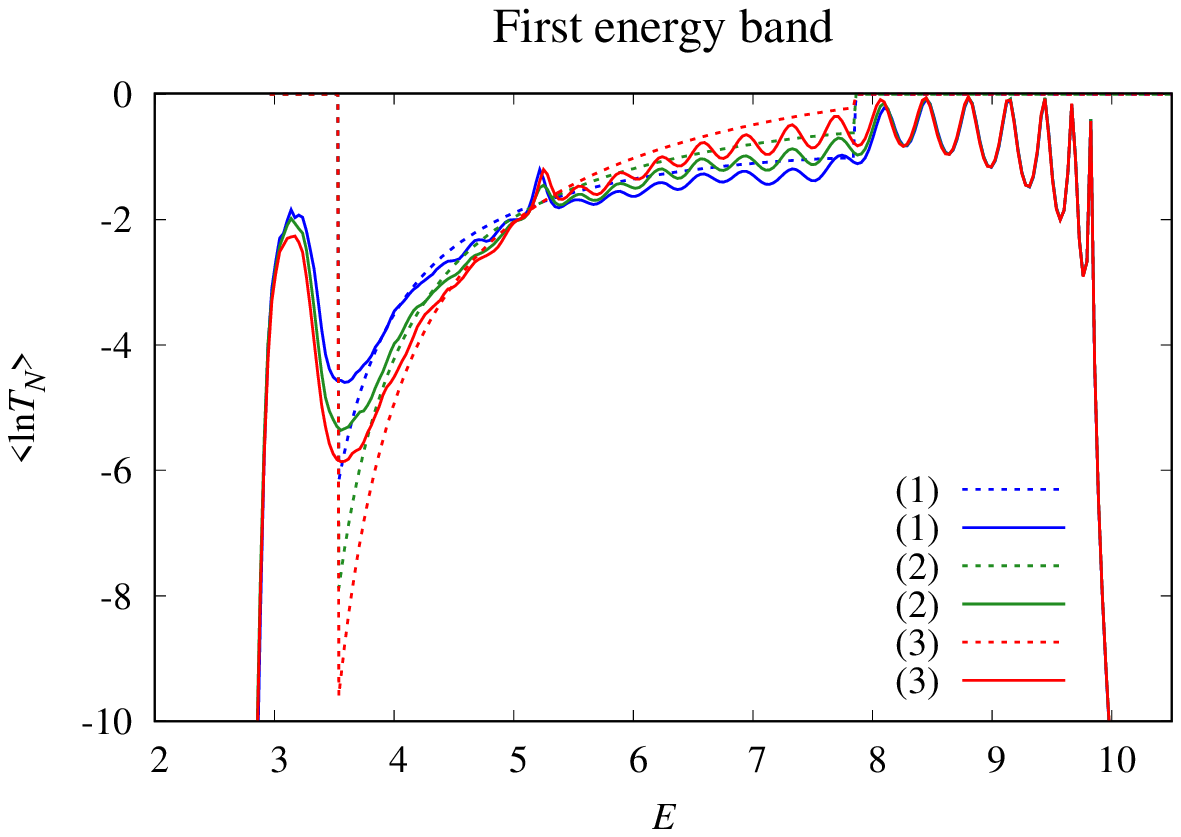}
\includegraphics[width=0.49\textwidth]{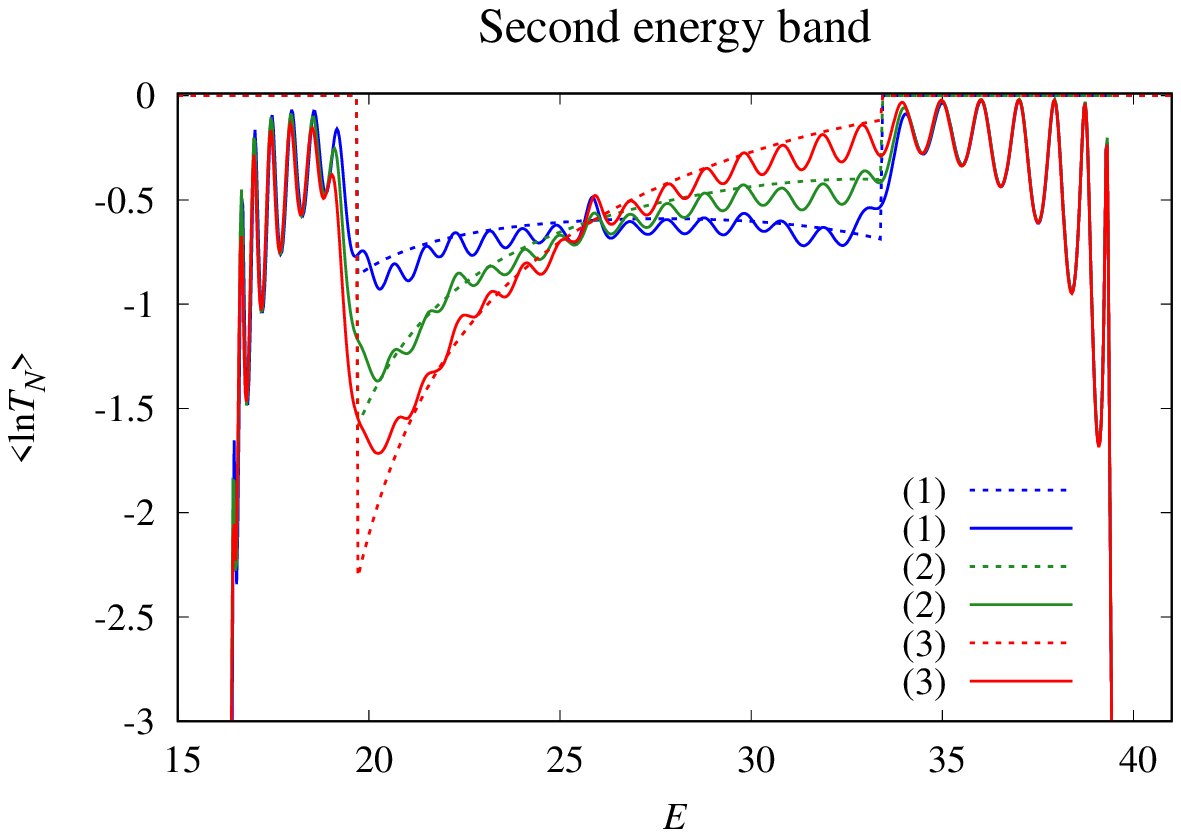}\\
\includegraphics[width=0.49\textwidth]{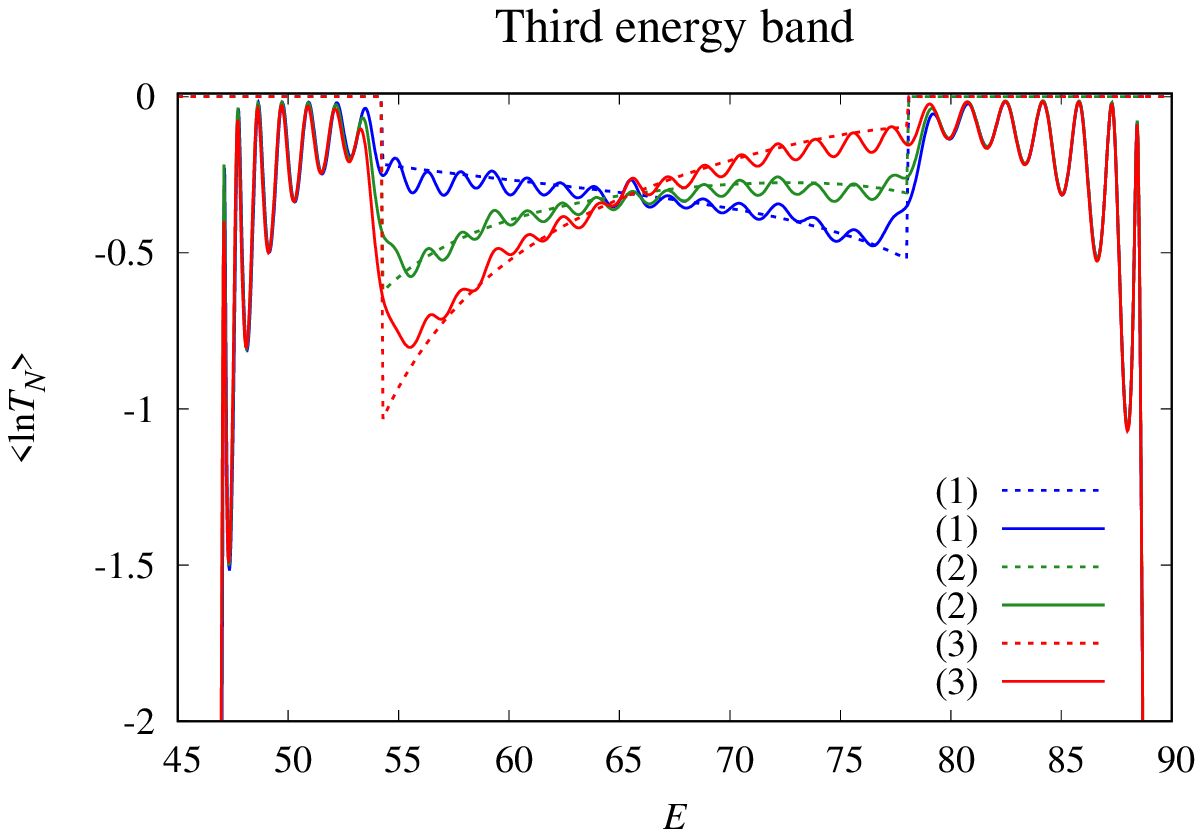}
\includegraphics[width=0.49\textwidth]{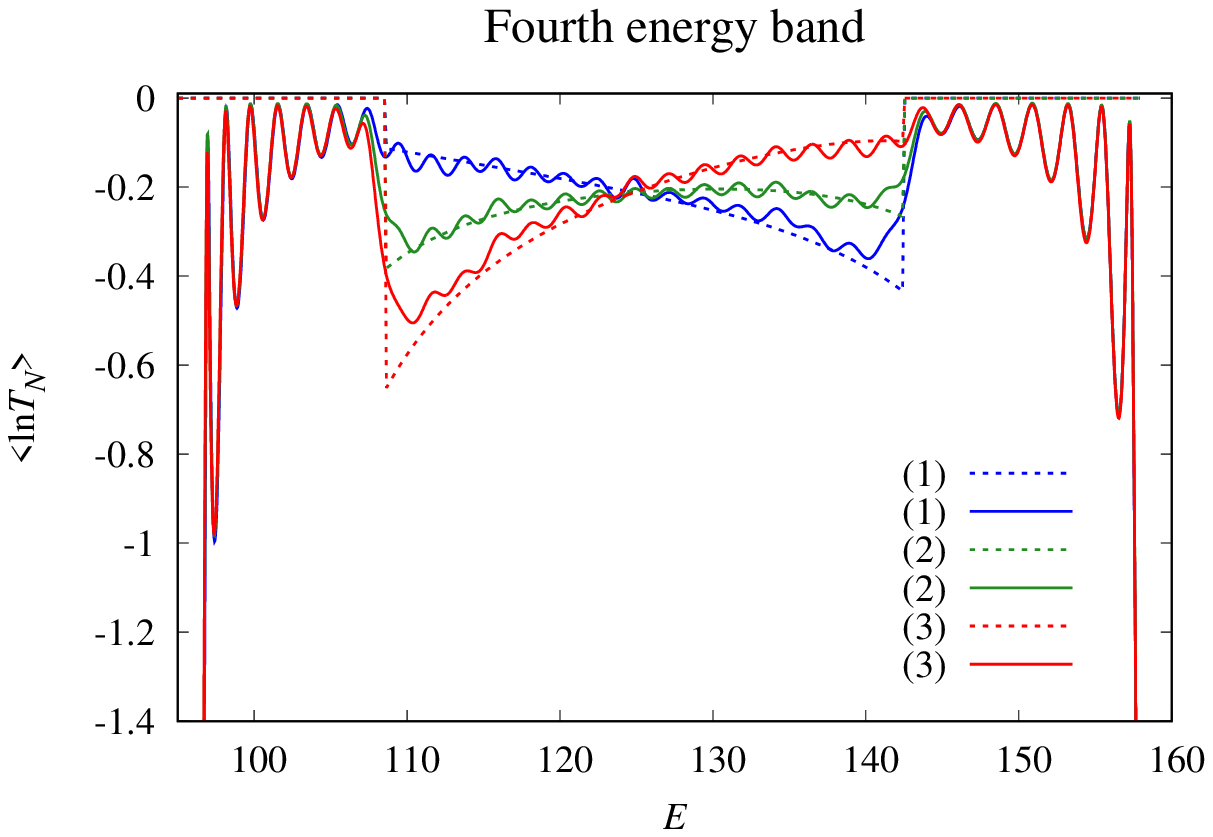}\\
\caption{Average of the logarithm of the transmission coefficient
$\langle \ln T_{N} \rangle$ versus energy $E$ (in $\mbox{cm}^{-2}$) for $N =30$
barriers. Each panel represents one of the first four energy bands.
The dotted lines correspond to the analytical expression~(\ref{loc_reg}); the solid lines
correspond to the numerical evaluation of~(\ref{trans_coef_2}). The average was
computed over an ensemble of $N_{r} = 1000$ disorder realisations.
The data were obtained for the 2D Kronig-Penney model~(\ref{rad_eq}) with the
following parameters: $U=4.0\,\mbox{cm}^{-1}$, $r_{1} = 5\,\mbox{cm}$, and $a = 1\,\mbox{cm}$.
Disorder strength: $\sqrt{\langle u_n^2\rangle} = 1.0\,\mbox{cm}^{-1}$ and
$\sqrt{\langle \Delta_n^2\rangle} = 0.02\,\mbox{cm}$.
The mobility edges were set at $k_{1} = \pi/4$ and $k_{2} = \pi/2$. The data labelled with
1 (blue), 2 (green), and 3 (red) correspond to the cases of positive, absent, and negative
cross-correlations.
\label{trans_n30}}
\end{center}
\end{figure}

\begin{figure}
\begin{center}
\includegraphics[width=0.49\textwidth]{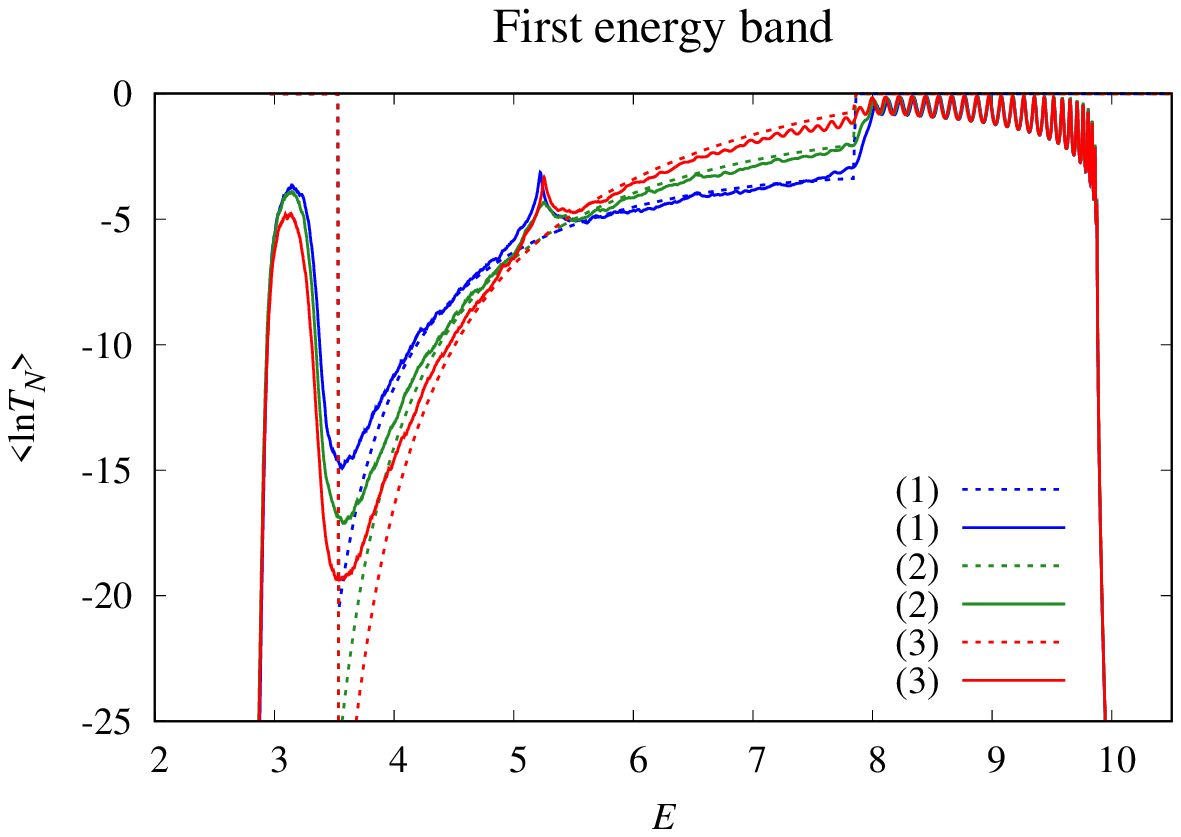}
\includegraphics[width=0.49\textwidth]{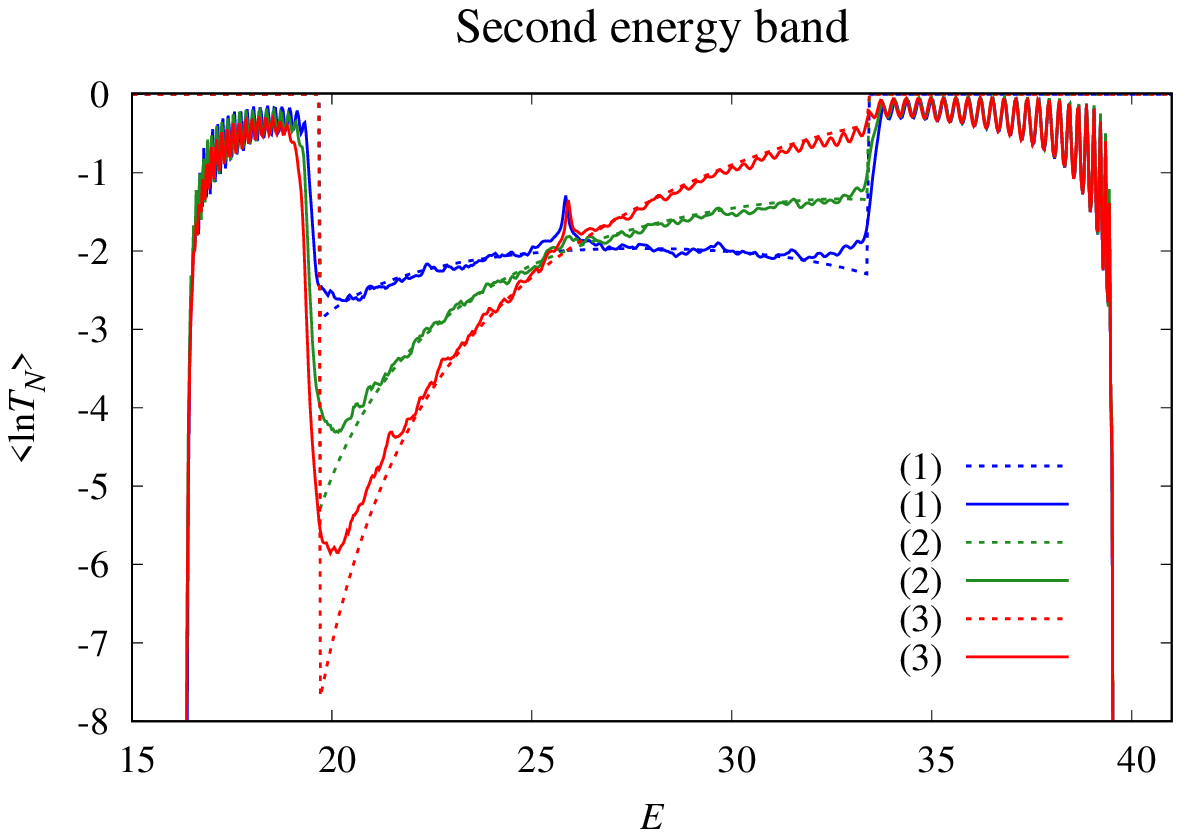}\\
\includegraphics[width=0.49\textwidth]{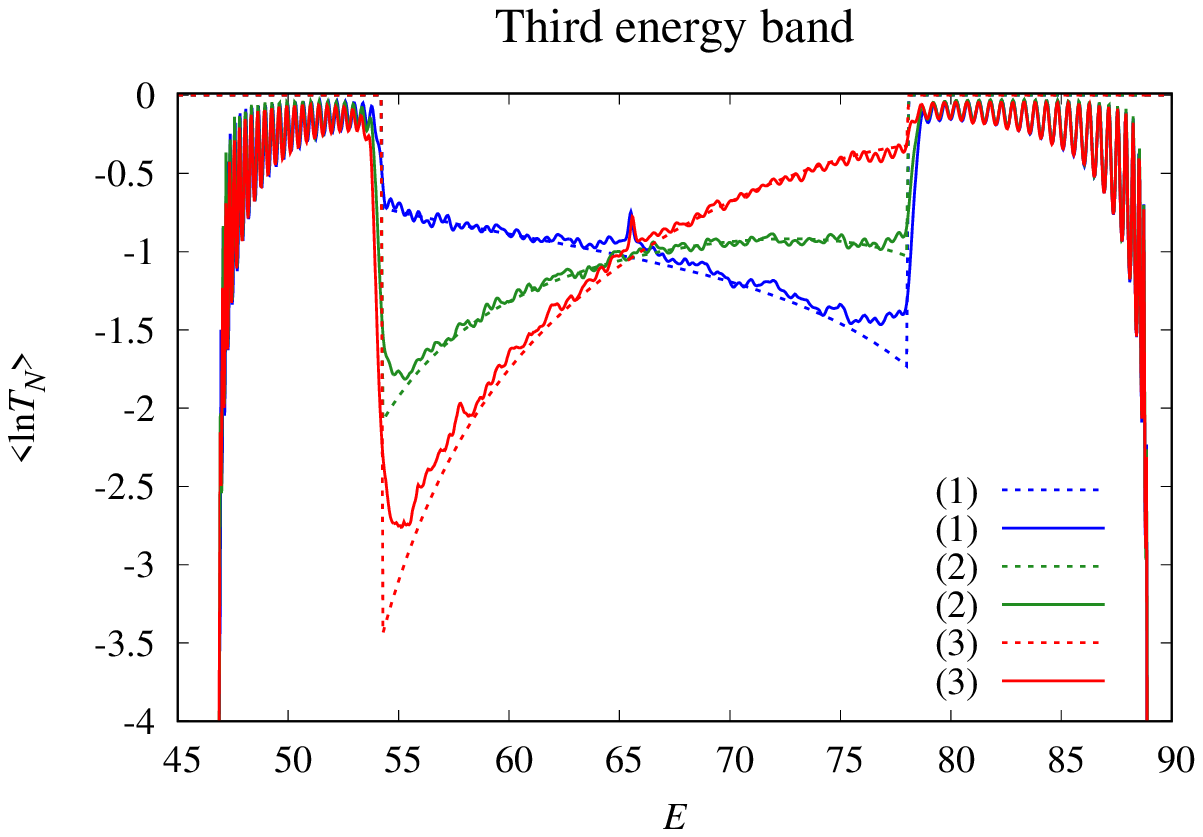}
\includegraphics[width=0.49\textwidth]{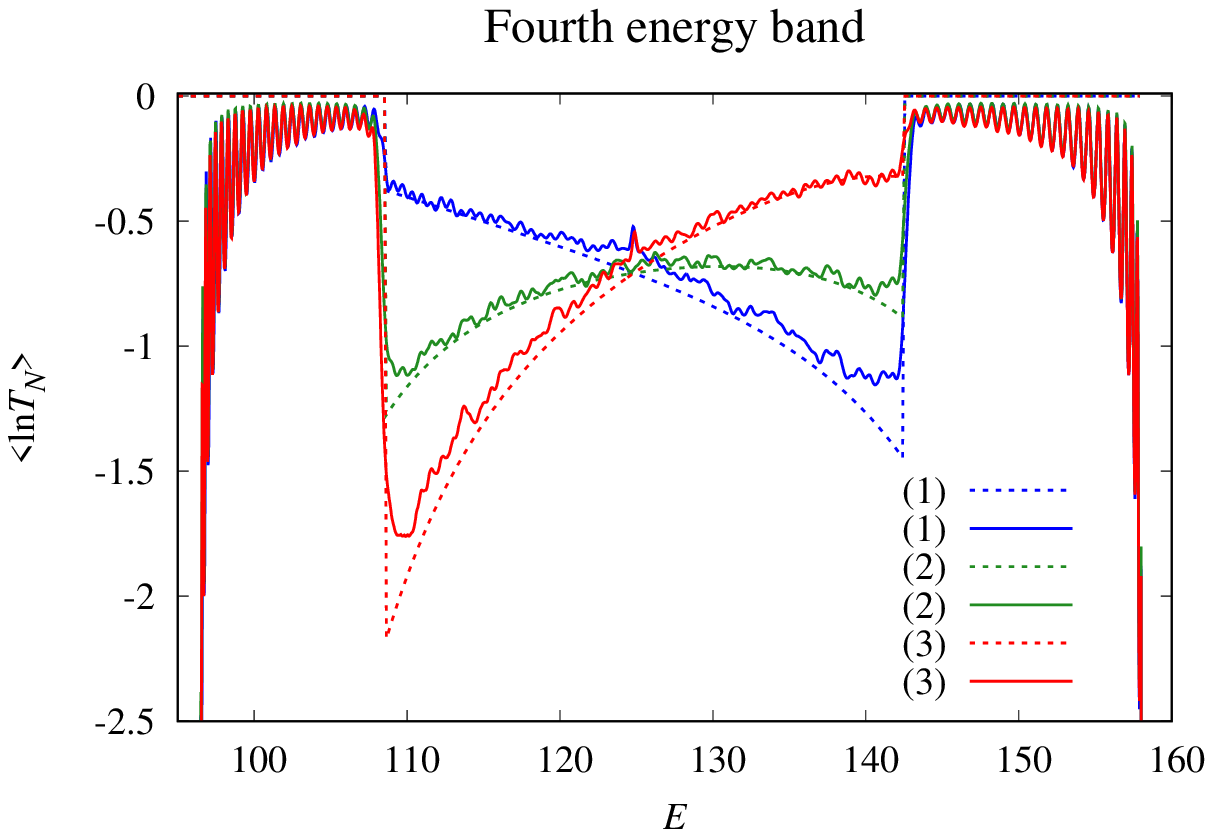}\\
\caption{As figure~\ref{trans_n30} but for $N =100$ barriers.
\label{trans_n100}}
\end{center}
\end{figure}

\begin{figure}
\begin{center}
\includegraphics[width=0.49\textwidth]{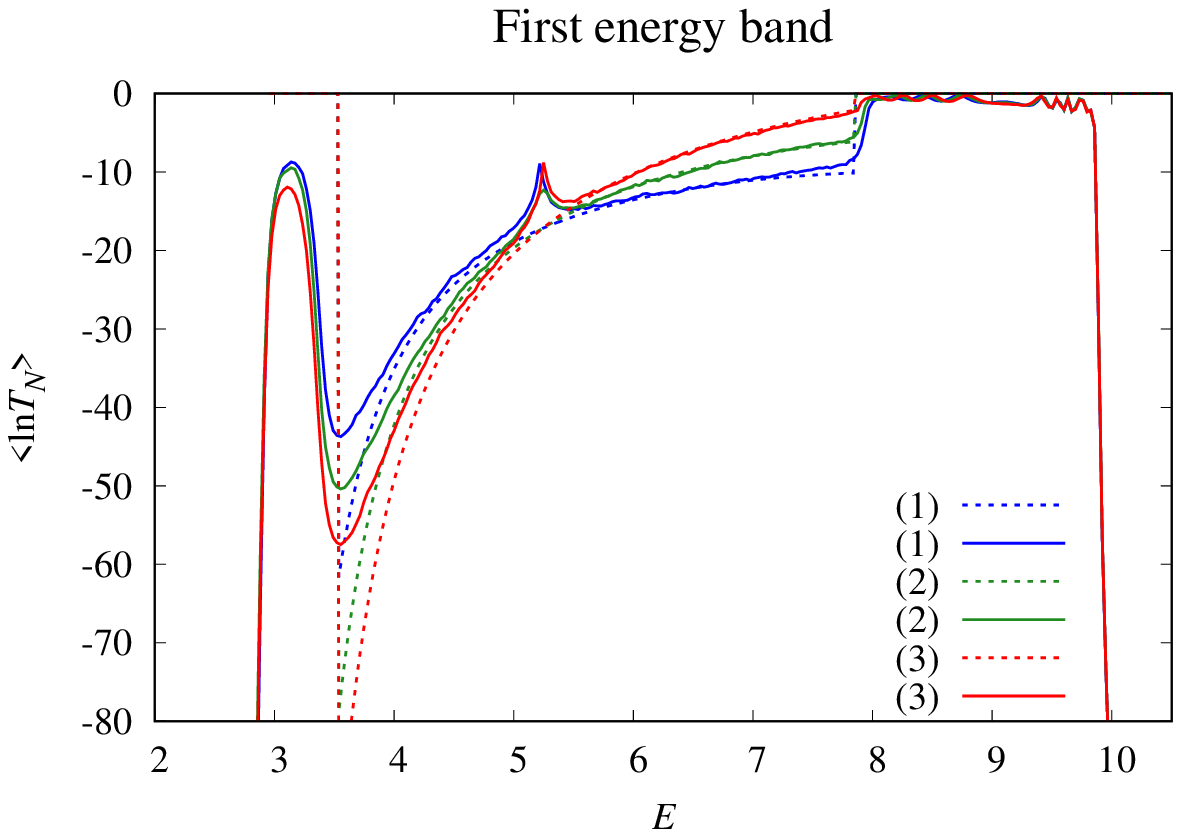}
\includegraphics[width=0.49\textwidth]{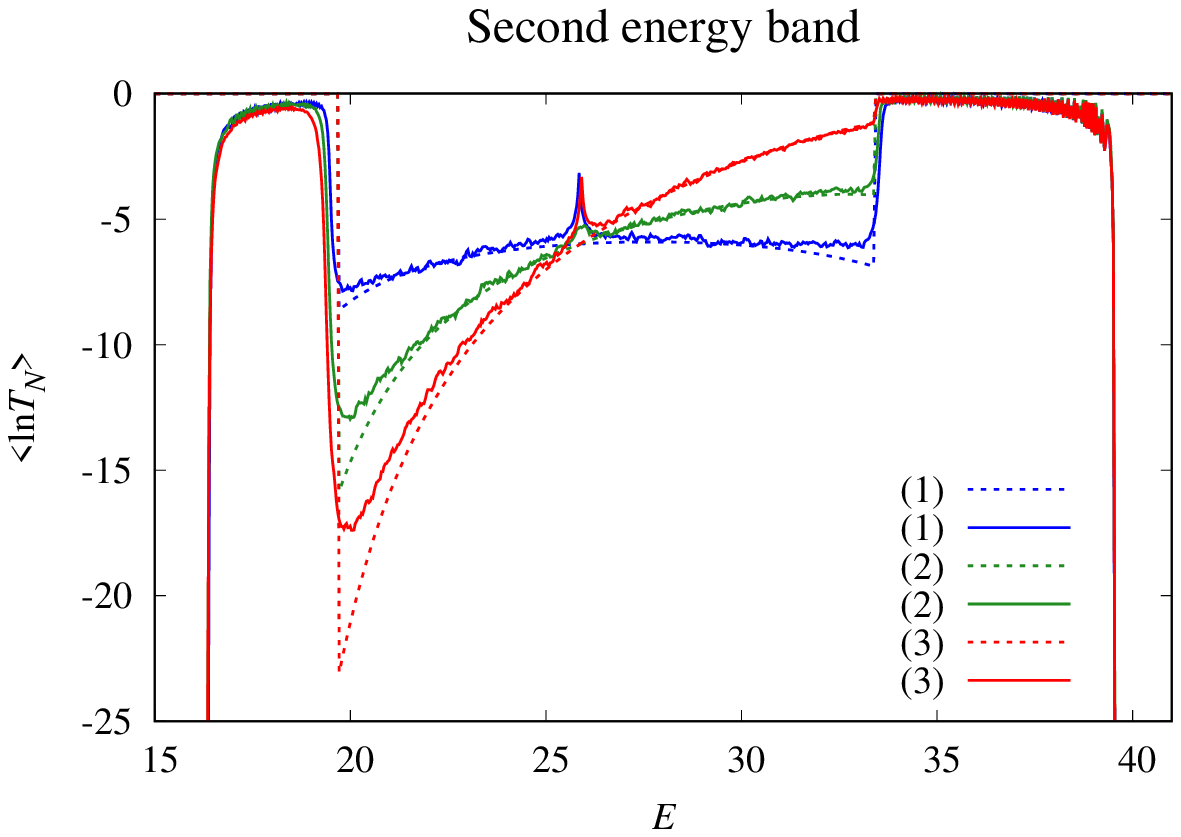}\\
\includegraphics[width=0.49\textwidth]{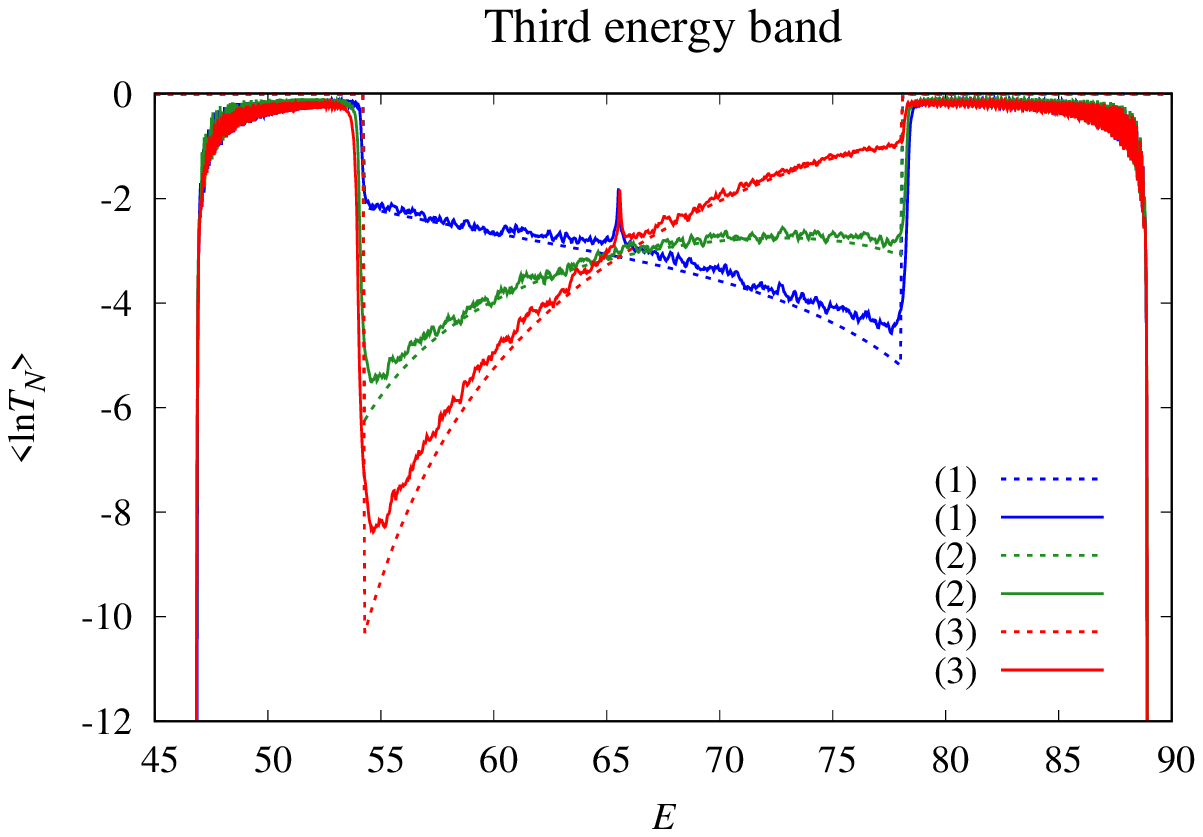}
\includegraphics[width=0.49\textwidth]{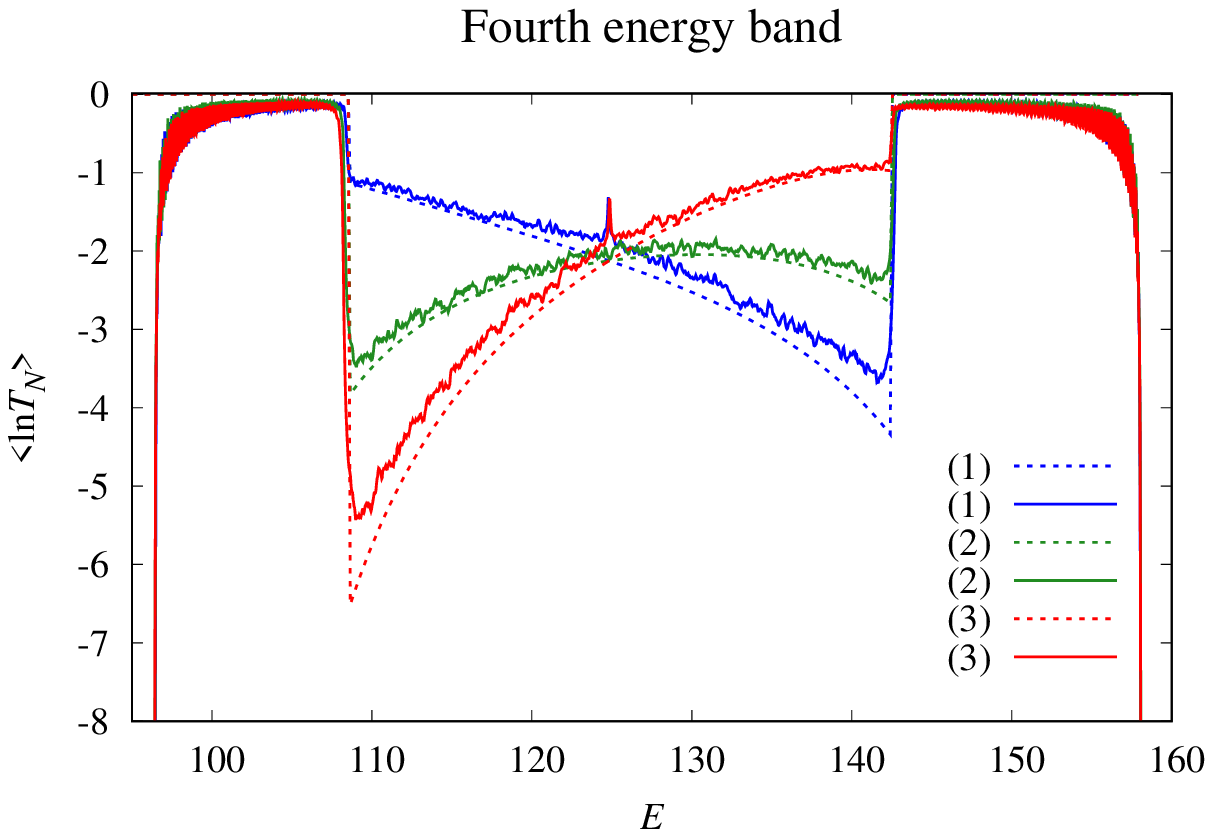}\\
\caption{As figure~\ref{trans_n30} but for $N =300$ barriers.
\label{trans_n300}}
\end{center}
\end{figure}

In figures~\ref{trans_n30},~\ref{trans_n100}, and~\ref{trans_n300} we compare
the data obtained by numerical evaluation of the transmission
coefficient~(\ref{trans_coef_2}) with the analytical formula~(\ref{loc_reg}).
We determined the transmission coefficient for the Kronig-Penney
model~(\ref{rad_eq}) with a finite number $N$ of barriers; we considered
the values $N=30$, $N=100$, and $N=300$.
We selected the same values of the parameters already used in
Sec.~\ref{2d_lyap}, i.e., $U=4.0\,\mbox{cm}^{-1}$, $r_{1} = 5\,\mbox{cm}$, and
$a = 1\,\mbox{cm}$. We set the disorder strengths at
$\sqrt{\langle u_n^2\rangle} = 1.0\,\mbox{cm}^{-1}$ and
$\sqrt{\langle \Delta_n^2\rangle} = 0.02\,\mbox{cm}$ and we put
mobility edges at $k_{1} = \pi/4$ and $k_{2} = \pi/2$.
The numerical data were obtained for an {\it s}-wave, but the results do
not change if different values of the angular momentum $l$ are selected.
In order to obtain a better understanding of the behaviour of the transmission
coefficient as a function of the energy of the incoming wave, we averaged
$\ln T_{N}$ over an ensemble of $N_{r} = 1000$ disorder realisations, thus
eliminating the fluctuations associated to each individual realisation of
the disorder.

The numerical data show that formula~(\ref{loc_reg}) describes the behaviour
of the transmission coefficient well when the number of barriers is large ($N \sim 300$);
the formula works reasonably well also in the case in which the number of
barriers is reduced ($N \sim 30$) or intermediate ($N \sim 100$), although in these cases
(\ref{loc_reg}) fails to reproduce the oscillations of $\ln T_{N}$ with $E$.

Figures~\ref{trans_n30},~\ref{trans_n100}, and~\ref{trans_n300} clearly show that
the use of correlated disorder to effectively suppress transmission in predefined
energy windows is not restricted to 1D models but works well also in 2D systems.
As expected the windows where transmission is inhibited correspond to the energy
intervals where correlated disorder enhances the localisation and increases the
Lyapunov exponent.
As a consequence, transmission is permitted (in fact, it is enhanced) only in the
windows where the Lyapunov exponent tends to zero. We can conclude that the use of
correlated disorder makes possible to generate small transmission bands for which the
annulus becomes almost completely transparent.
By moving the mobility edges, one can thus select the energy of the transmitted
waves and use the annulus as a band filter.
\clearpage

\section{Conclusions}
\label{conclu}

In this work we have studied the localisation and transport properties of
2D~models with Hamiltonians that are separable in polar coordinates and
have random potentials that depend on a single variable, $r$ or $\theta$.
Broadly speaking, our analysis confirms what one would intuitively expect for
such models: mathematically, the separability condition allows one to study
these 2D systems in terms of models of lower dimensionality; physically,
localisation occurs on the coordinate associated with the random part of the
potential.
These conclusions, however, come with several caveats and putting them on a
firm footing required a precise assessment of the analogies and differences
which exist between the 2D models under study and their 1D counterparts.

In the case of models with angular disorder, we found that the geometry of
the 2D problem sets unavoidable constraints on the 1D associated model: first
and foremost, the 1D model has a {\em finite} size and its wavefunctions must
satisfy periodic boundary conditions. In ordinary 1D models the finite size of
the random sample can be neglected if disorder is sufficiently strong to
produce localisation on a length scale smaller than the system size.
This possibility is partially precluded in the present case, because the
geometry of the 2D model imposes severe constraints on the strength of
structural disorder. A good degree of localisation can be achieved even for
weak disorder, however, if long-range correlations of the random potential
are appropriately used to this effect.
Mathematically, the finite size of the system complicates the analysis of its
eigenfunctions: it is not possible to study the tails of the eigenstates by
considering the asymptotic behaviour of products of transfer matrices because
one deals with a finite and limited number of transfer matrices.
To determine the localisation properties of the 2D model we had to
resort to the direct evaluation of the eigenstates of the associated tight-binding
model which is the 1D discrete counterpart of the original 2D system.
In so doing, however, we had to take into account that the correspondence
between discrete and continuous models can be tricky when resonance phenomena
come into play. In fact, we found that a striking feature of the 2D model
with random angular potential lies in the fact that its eigenstates are shaped
not only by localisation, but also by Fabry-P\'{e}rot resonances which occur
when the ratio of the angular momentum to the angular width of a potential well
is an integer multiple of $\pi$.

In the case of 2D models with a central random potential, we have shown that
the localisation of the radial part of the wavefunction occurs exactly as in
the 1D case provided that the random potential falls off away from the force
centre more slowly than the centrifugal potential.
This circumstance can be exploited to reproduce in 2D models the
localisation-delocalisation transitions that are generated in 1D systems
by specific long-range correlations of the disorder. In this way one can
design 2D filters which allow the transmission of waves within predefined
frequency windows, thus overcoming the limitations of the 1D geometry of
previous devices of this kind.

As a final remark, we would like to point out that in this paper we restricted
our attention to 2D models. The present approach, however, can be applied also
to 3D models with similar features; we plan to extend our results to this class
of system in a future paper.

\section*{Acknowledgements}

L.~T.\ gratefully acknowledges the warm hospitality and support of the Institut
de Physique de Nice, where much of this work was done. L.~T.\ also acknowledges
the support of a CONACyT sabbatical fellowship and of the CIC-UMSNH 2018-2019
grant.

\appendix
\setcounter{section}{0}

\section{Application to microwave cavities}
\label{app:mwc}

In this appendix we discuss how the model~(\ref{2dschr}) can be applied
to different physical problems beyond that of a quantum particle in an annulus.
Specifically, we show how a slight modification of (\ref{2dschr}) can be
reinterpreted  as the Helmholtz equation describing the electric field of a
transverse magnetic (TM) mode in a microwave cavity shaped as a ring cake tin.

In fact, one can consider a microwave cavity vertically bounded by two
coaxial cylinders (of radii $r_{1}$ and $r_{2}$) and closed at the bottom and
at the top by two horizontal plates separated by a distance $d_{0}$.
It is  natural to introduce cylindrical coordinates and to set the $z$-axis
along the cylinder axis.
If one assumes that the top and bottom plates of the cavity are flat, the
electric field of the $m$-th TM mode can be written in the form
\begin{displaymath}
\mathbf{E}_{TM}(r,\theta,z) =
E_{z}(r,\theta)
\sin \left( \frac{m \pi}{d_{0}} z \right) \mathbf{\hat{z}}
\end{displaymath}
with the amplitude $E_{z}(r,\theta)$ satisfying the wave equation
\begin{equation}
\left[ -\nabla^{2} + \left( \frac{m \pi}{d_{0}} \right)^{2} \right]
E_{z}(r,\theta) = \frac{\omega^{2}}{c^{2}} E_{z}(r,\theta)
\label{tmmode}
\end{equation}
where $\nabla^{2}$ is the 2D Laplacian~(\ref{2dlap}).

If the top and bottom plates are not perfectly flat, so that the distance
$d(r,\theta)$ between them is not constant, (\ref{tmmode}) ceases to be
exact.
It remains approximately satisfied, however, if the distance $d(r,\theta)$
changes slowly with the position.
In this case it is convenient to rewrite (\ref{tmmode}) in the form
\begin{equation}\fl
\left\{ -\nabla^{2} + \left[ \left( \frac{m \pi}{d(r,\theta)} \right)^{2}
- \left( \frac{m \pi}{d_{\mathrm max}}\right)^{2} \right] \right\}
E_{z}(r,\theta) = \left[ \frac{\omega^{2}}{c^{2}} -
\left( \frac{m \pi}{d_{\mathrm max}}\right)^{2}  \right] E_{z}(r,\theta) ,
\label{tmmode2}
\end{equation}
where $d_{\mathrm max}$ represents the maximum distance between the top and
bottom plates. The term $(m \pi/d(r,\theta))^{2}$ in the left-hand side of
(\ref{tmmode2}) plays the role of a potential; from this point of
view, subtracting the constant term $(m \pi/d_{\mathrm{max}})^{2}$ is
equivalent to setting the potential equal to zero on the bottom plate
of the cavity.

Comparing Eqs.~(\ref{2dschr}) and~(\ref{tmmode2}), one can see that they
have identical forms and that the same mathematical solutions describe
the wavefunction $\psi(r,\theta)$ and the longitudinal electric field
$E_{z}(r,\theta)$.
We observe that the energy of the quantum particle in model~(\ref{2dschr})
and the frequency of the electric field in the cavity~(\ref{tmmode2}) are
connected as follows
\begin{displaymath}
E = \frac{\omega^{2}}{c^{2}} - \left( \frac{m \pi}{d_{\mathrm{max}}} \right)^{2};
\end{displaymath}
the potential $U(r,\theta)$ and the distance $d(r,\theta)$, on the other
hand, are related by the identity
\begin{equation}
U(r,\theta) = m^{2} \pi^{2} \left[ \frac{1}{d^{2}(r,\theta)}
- \frac{1}{d_{\mathrm max}^{2}} \right] .
\label{pot_corresp}
\end{equation}

(\ref{pot_corresp}) shows how the potential $U(r,\theta)$ in the
quantum mechanical model~(\ref{2dschr}) can be reproduced with an
appropriate topographical shape of the bottom plate in a microwave
cavity. Obviously, there is no way to reproduce physically the
$\delta$-barriers in the potential~(\ref{deltabar}); the latter,
however, should be seen as a schematic representation of a real
potential of the form
\begin{displaymath}
U_{\varepsilon}(r,\theta) = \frac{1}{r^{2}}
\sum_{n=1}^{N} U_{n} \eta_{\varepsilon}(\theta - \theta_{n})
\end{displaymath}
with
\begin{displaymath}
\eta_{\varepsilon}(\theta) = \left\{ \begin{array}{ccc}
\displaystyle \frac{1}{2 \varepsilon} & \mbox{ if } &
\theta \in [-\varepsilon, \varepsilon] \\
0 & \mbox{ if } & \theta \notin [-\varepsilon, \varepsilon] \\
\end{array} \right.
\end{displaymath}
where $\varepsilon$ is a small positive parameter.
A potential of this kind can be reproduced by inserting appropriately
shaped wedges in a ring microwave cavity.

We observe that the theoretical predictions obtained with the use of
the 1D Kronig-Penney model with $\delta$-barriers have worked remarkably
well when applied to quasi-1D waveguides with physical barriers of various
shapes and kinds (screws in~\cite{kuh00}, bars in~\cite{die12a}).
In light of the past experience, we expect that the use of models
with $\delta$-barriers should produce reasonable results even for 2D
cavities with rectangular wedges, provided that the width of the barriers
is sufficiently smaller than $1/k_{\theta}$, where $k_{\theta}$ is the
wavenumber in the angular direction.
We would like to emphasise as well that in the weak-disorder limit the
localisation length essentially depends on the binary correlator of
the random potential, rather than on the potential itself. Although
specific details, like the value of the Lyapunov exponent, the exact
frequencies for which mobility edges appear and the spatial location of
the localisation centres do depend on the form of the potential, the
general features of the spectral shaping of the localisation length should
still be observable even when the $\delta$-barrier potential does not
represent too closely the actual potential in the experimental setup,
assuming that the disorder correlations are the same.

\section{Generation of self- and cross-correlated random sequences}
\label{random_sequences}

In this appendix we summarise how one can generate two random sequences
$\{u_{n}\}$ and $\{\Delta_{n}\}$ with pre-defined self- and
cross-correlations.
As a first step, one generates two uncorrelated sequences
$\{ X_{n}^{(1)}\}$ and $\{ X_{n}^{(2)}\}$ of independent variables.
The two white noises are then cross-correlated via the transformation
\begin{displaymath}
\begin{array}{ccl}
Y_{n}^{(1)} & = & X_{n}^{(1)} \cos \eta + X_{n}^{(2)} \sin \eta \\
Y_{n}^{(2)} & = & X_{n}^{(1)} \sin \eta + X_{n}^{(2)} \cos \eta , \\
\end{array}
\end{displaymath}
where $\eta$ is the parameter which determines the degree of cross-correlation
of the $Y_{n}^{(1)}$ and $Y_{n}^{(2)}$ sequences.
Finally one filters the cross-correlated white noises $Y_{n}^{(1)}$ and
$Y_{n}^{(2)}$ so that they acquire the desired self-correlation properties.
The operation is carried out with the convolution products
\begin{equation}
\begin{array}{ccl}
u_{n} & = & \displaystyle
\sum_{l=-\infty}^{\infty} c^{(1)}_{l} Y_{n-l}^{(1)} \\
\Delta_{n} & = & \displaystyle
\sum_{l=-\infty}^{\infty} c^{(2)}_{l} Y_{n-l}^{(2)} \\
\end{array},
\label{convol}
\end{equation}
in which the coefficients $c^{(1)}_{n}$ and $c^{(2)}_{n}$ are derived by
the pre-defined power spectra $W_{1}$ and $W_{2}$ with the help of the
following identities
\begin{displaymath}
\begin{array}{ccl}
c^{(1)}_{n} & = & \displaystyle
\frac{2}{\pi} \int_{0}^{\pi/2} \sqrt{\langle u_{l}^{2} \rangle W_{1}(x)}
\cos \left( 2 n x \right) \mathrm{d}x \\
c^{(2)}_{n} & = & \displaystyle
\frac{2}{\pi} \int_{0}^{\pi/2} \sqrt{\langle \Delta_{l}^{2} \rangle W_{2}(x)}
\cos \left( 2 n x \right) \mathrm{d}x .\\
\end{array}
\end{displaymath}

\section{Explicit form of the matrix terms in expansions~(\ref{t_exp})
and~(\ref{double_exp})}
\label{expansion_matrices}

In this appendix we provide the explicit form of the matrices that
appear in the expansions~(\ref{t_exp}) and~(\ref{double_exp}).
In the rhs of~(\ref{t_exp}), the zero-th order term $\mathbf{T}_{n}^{(0)}$ has
matrix elements
\begin{displaymath}
\begin{array}{ccl}
\left({\mathbf T}_{n}^{(0)} \right)_{11} & = & \displaystyle
\cos \left[ \sqrt{E} \left( a + \Delta_{n} \right)\right] +
\frac{U_{n}}{\sqrt{E}} \sin \left[ \sqrt{E} \left( a + \Delta_{n} \right)\right] \\
\left({\mathbf T}_{n}^{(0)} \right)_{12} & = & \displaystyle
\frac{1}{\sqrt{E}} \sin \left[\sqrt{E} \left( a + \Delta_{n} \right) \right] \\
\left({\mathbf T}_{n}^{(0)} \right)_{21} & = & \displaystyle
-\sqrt{E} \sin \left[\sqrt{E} \left( a + \Delta_{n} \right) \right]
+ U_{n} \cos \left[ \sqrt{E} \left( a + \Delta_{n} \right) \right] \\
\left({\mathbf T}_{n}^{(0)} \right)_{11} & = & \displaystyle
\cos \left[ \sqrt{E} \left( a + \Delta_{n} \right) \right] \\
\end{array}
\end{displaymath}
while the matrix elements of the second-order term $\mathbf{T}_{n}^{(2)}$ are
\begin{displaymath}
\begin{array}{ccl}
\left( {\mathbf T}_{n}^{(2)} \right)_{11} & = & \displaystyle
-\frac{l^{2}-1/4}{2\sqrt{E}} \left\{ \left[ \frac{1}{E} U_{n} - a - \Delta_{n} \right]
\sin [\sqrt{E} (a + \Delta_{n})] \right. \\
& + & \displaystyle \left.
\frac{1}{\sqrt{E}} U_{n} (a + \Delta_{n}) \cos [\sqrt{E} (a + \Delta_{n})] \right\} \\
\left( {\mathbf T}_{n}^{(2)} \right)_{12} & = & \displaystyle
\frac{l^{2}-1/4}{2\sqrt{E}} \left\{
- \frac{1}{E} \sin \left[\sqrt{E} \left( a + \Delta_{n} \right) \right] \right. \\
& + & \displaystyle \left. \frac{1}{\sqrt{E}} (a + \Delta_{n})
\cos \left[\sqrt{E} \left( a + \Delta_{n} \right) \right] \right\} \\
\left( {\mathbf T}_{n}^{(2)} \right)_{21} & = & \displaystyle
-\frac{l^{2}-1/4}{2\sqrt{E}} \left\{ \left[ 1 + U_{n} (a + \Delta_{n}) \right]
\sin \left[\sqrt{E} \left( a + \Delta_{n} \right) \right] \right. \\
& - & \displaystyle \left.
\sqrt{E} (a + \Delta_{n}) \cos \left[\sqrt{E}
\left( a + \Delta_{n} \right) \right] \right\} \\
\left( {\mathbf T}_{n}^{(2)} \right)_{21} & = & \displaystyle
- \frac{l^{2}-1/4}{2\sqrt{E}} (a + \Delta_{n})
\sin \left[\sqrt{E} \left( a + \Delta_{n} \right) \right] .\\
\end{array}
\end{displaymath}

In expansion~(\ref{double_exp}), the unperturbed term is equal to
\begin{displaymath}
{\mathbf{T}}_{n}^{(0,0)} = \left(
\begin{array}{cc}
\displaystyle
\cos(\sqrt{E}a) + \frac{U}{\sqrt{E}} \sin(\sqrt{E}a) &
\displaystyle
\frac{1}{\sqrt{E}}\sin(\sqrt{E}a) \\
-\sqrt{E} \sin(\sqrt{E}a) + U \cos(\sqrt{E}a) & \cos(\sqrt{E}a) \\	
\end{array}
\right) .
\end{displaymath}
The dominant corrections due to disorder are the random matrices
$\mathbf{T}_{n}^{(1,0)}$ and $\mathbf{T}_{n}^{(2,0)}$, with elements
\begin{displaymath}
\begin{array}{ccl}
\left( {\mathbf{T}}_{n}^{(1,0)} \right)_{11} & = & \displaystyle
\frac{1}{\sigma} \left[ \Delta_{n} U \cos(\sqrt{E}a) +
\left( \frac{u_{n}}{\sqrt{E}} - \sqrt{E} \Delta_{n} \right) \sin(\sqrt{E}a) \right] \\
\left( {\mathbf{T}}_{n}^{(1,0)} \right)_{12} & = & \displaystyle
\frac{\Delta_{n}}{\sigma} \cos(\sqrt{E}a) \\
\left( {\mathbf{T}}_{n}^{(1,0)} \right)_{21} & = & \displaystyle
\frac{1}{\sigma} \left[ \left( u_{n} - E \Delta_{n} \right) \cos(\sqrt{E}a) -
\sqrt{E} U \sin ( \sqrt{E} a) \right] \\
\left( {\mathbf{T}}_{n}^{(1,0)} \right)_{22} & = & \displaystyle
-\sqrt{E} \frac{\Delta_{n}}{\sigma} \sin(\sqrt{E} a) \\
\end{array}
\end{displaymath}
and
\begin{displaymath}\fl
\begin{array}{ccl}
\left( {\mathbf{T}}_{n}^{(2,0)} \right)_{11} & = & \displaystyle
\frac{1}{\sigma^{2}} \left[ \left( u_{n} \Delta_{n} - \frac{1}{2} E \Delta_{n}^{2} \right)
\cos (\sqrt{E} a) - \frac{1}{2} U \sqrt{E} \Delta_{n}^{2} \sin (\sqrt{E} a ) \right] \\
\left( {\mathbf{T}}_{n}^{(2,0)} \right)_{12} & = & \displaystyle
-\frac{1}{2} \sqrt{E} \frac{\Delta_{n}^{2}}{\sigma^{2}} \sin ( \sqrt{E} a) \\
\left( {\mathbf{T}}_{n}^{(2,0)} \right)_{21} & = & \displaystyle
\frac{1}{\sigma^{2}} \left[ -\frac{1}{2} E U  \Delta_{n}^{2} \cos (\sqrt{E} a) +
\sqrt{E} \left( E \Delta_{n}^{2} - u_{n} \Delta_{n} \right) \sin ( \sqrt{E} a ) \right] \\
\left( {\mathbf{T}}_{n}^{(2,0)} \right)_{22} & = & \displaystyle
-\frac{1}{2} E \frac{\Delta_{n}^{2}}{\sigma^{2}} \cos (\sqrt{E} a ) \\
\end{array}
\end{displaymath}
The leading correction due to the centrifugal potential, on the other hand, is
proportional to the matrix $\mathbf{T}_{n}^{(0,2)}$, whose elements are
\begin{displaymath}
\begin{array}{ccl}
\left( {\mathbf{T}}_{n}^{(0,2)} \right)_{11} & = & \displaystyle
\frac{l^{2}-1/4}{2\sqrt{E}} \left[
\frac{U}{E} \sin (\sqrt{E} a) - \frac{U}{\sqrt{E}} a \cos (\sqrt{E} a) \right] \\
\left( {\mathbf{T}}_{n}^{(0,2)} \right)_{12} & = & \displaystyle
\frac{l^{2}-1/4}{2\sqrt{E}} \left[
-\frac{1}{E} \sin (\sqrt{E} a) + \frac{1}{\sqrt{E}} a \cos ( \sqrt{E} a) \right] \\
\left( {\mathbf{T}}_{n}^{(0,2)} \right)_{21} & = & \displaystyle
\frac{l^{2}-1/4}{2\sqrt{E}} \left[
(1 + U a) \sin (\sqrt{E} a) + \sqrt{E} a \cos (\sqrt{E} a) \right] \\
\left( {\mathbf{T}}_{n}^{(0,2)} \right)_{22} & = & \displaystyle
- \frac{l^{2}-1/4}{2\sqrt{E}}  a \sin (\sqrt{E} a ) .\\
\end{array}
\end{displaymath}
\newpage

\newcommand{\newblock}{}


\end{document}